%% file: daSilvaetal2015_aph.tex
\documentclass{aa}

\usepackage{natbib}
\usepackage{txfonts}
\usepackage{longtable}
\usepackage{color}

\bibpunct{(}{)}{;}{a}{}{,}

\begin{document}

\def\RG{$R_{\rm G}$}
\def\kms{km\,s$^{-1}$}
\def\teff{$T_{\rm eff}$}
\def\logg{$\log{g}$}
\def\loggf{$\log{gf}$}
\def\vt{$\upsilon_t$}
\def\logP{$\log{P}$}
\def\potex{$\chi_{\rm ex}$}
\def\dexkpc{dex\,kpc$^{-1}$}

\titlerunning{On the neutron-capture elements across the Galactic thin disk using Cepheids}
\authorrunning{...}

\title{On the neutron-capture elements across the Galactic thin disk using Cepheids
\thanks{Based on spectra collected with the UVES spectrograph available at
the ESO Very Large Telescope (VLT), Cerro Paranal, Chile (ESO Proposals:
081.D-0928(A), PI: S. Pedicelli; 082.D-0901(A), PI: S. Pedicelli;
089.D-0767(C), PI: K. Genovali).}$^,$
\thanks{Tables ... are only
available in electronic form at the CDS via anonymous ftp to
cdsarc.u-strasbg.fr (130.79.128.5) or via
http://cdsweb.u-strasbg.fr/cgi-bin/qcat?J/A+A/}}

\author{
R. da Silva\inst{1,2,3} \and
B. Lemasle\inst{4} \and
G. Bono\inst{2,3} \and
K. Genovali\inst{2} \and
A. McWilliam\inst{5} \and
S. Cristallo\inst{6,7} \and
M. Bergemann\inst{8} \and \\
R. Buonanno\inst{2,6} \and
M. Fabrizio\inst{1,6} \and
I. Ferraro\inst{3} \and
P. Fran\c cois\inst{9,10} \and
G. Iannicola\inst{3} \and
L. Inno\inst{8} \and
C.D. Laney\inst{11,12} \and \\
R.-P. Kudritzki\inst{13,14,15} \and
N. Matsunaga\inst{16} \and
M. Nonino\inst{17} \and
F. Primas\inst{18} \and
N. Przybilla\inst{19} \and
M. Romaniello\inst{18,20} \and \\
F. Th\'evenin\inst{21} \and
M.A. Urbaneja \inst{19}
}

\institute{
ASI Science Data Center, Via del Politecnico snc, 00133 Rome, Italy
\and Dipartimento di Fisica, Universit\`a di Roma Tor Vergata, via della Ricerca
Scientifica 1, 00133 Rome, Italy
\and INAF--Osservatorio Astronomico di Roma, via Frascati 33, 00078 Monte 
Porzio Catone, Rome, Italy
\and Anton Pannekoek Institute for Astronomy, University of Amsterdam,
Science Park 904, PO Box 94249, 1090 GE, Amsterdam, The Netherlands
\and  The Observatories of the Carnegie Institute of Washington, 
813 Santa Barbara Street, Pasadena, CA 91101, USA
\and INAF--Osservatorio Astronomico di Teramo, via Mentore Maggini s.n.c.,
64100 Teramo, Italy
\and INFN Sezione Napoli, Napoli, Italy
\and Max-Planck-Institut f\"ur Astronomy, D-69117, Heidelberg, Germany
\and GEPI, Observatoire de Paris, CNRS, Universit\'e Paris Diderot, Place
Jules Janssen, 92190 Meudon, France
\and UPJV, Universit\'e de Picardie Jules Verne, 33 rue St. Leu,
80080 Amiens, France
\and Department of Physics and Astronomy, N283 ESC, Brigham Young
University, Provo, UT 84601, USA
\and South African Astronomical Observatory, PO Box 9, Observatory 7935,
South Africa
\and Institute for Astronomy, University of Hawaii, 2680 Woodlawn Drive,
Honolulu, HI 96822, USA
\and Max-Planck-Institute for Astrophysics, Karl-Schwarzschild-Str.1,
D-85741 Garching, Germany
\and University Observatory Munich, Scheinerstr. 1, D-81679 Munich, Germany
\and Department of Astronomy, School of Science, The University of Tokyo,
7-3-1 Hongo, Bunkyo-ku, Tokyo 113-0033, Japan
\and INAF--Osservatorio Astronomico di Trieste, via G. B. Tiepolo 11,
34143, Trieste, Italy
\and European Southern Observatory, Karl-Schwarzschild-Str. 2,
85748 Garching bei M\"unchen, Germany
\and Institute for Astro- and Particle Physics, University of Innsbruck,
Technikerstr. 25/8, A-6020 Innsbruck, Austria
\and Excellence Cluster Universe, Boltzmannstr. 2, 85748, Garching bei
M\"unchen, Germany
\and Laboratoire Lagrange, CNRS/UMR 7293, Observatoire de la C\^ote d'Azur,
Bd de l'Observatoire, CS 34229, 06304 Nice, France
}

\date{Received / accepted}

%
%

\abstract{

We present new accurate abundances for five neutron-capture
(Y, La, Ce, Nd, Eu) elements in 73 classical Cepheids located across the
Galactic thin disk. Individual abundances are based on high spectral
resolution ($R$ $\sim$ 38\,000) and high signal-to-noise ratio
(S/N $\sim$ 50-300) spectra collected with UVES at ESO VLT for the
DIONYSOS project. 
Taking account for similar Cepheid abundances provided either
by our group (111 stars) or available in the literature, we end 
up with a sample of 435 Cepheids covering a broad range in iron 
abundances ($-$1.6 < [Fe/H] < 0.6). We found, using homogeneous 
individual distances and abundance scales, well defined gradients 
for the above elements. However, the slope of the
light s-process element (Y) is at least a factor of two steeper 
than the slopes of heavy s- (La, Ce, Nd) and r- (Eu) process elements.
The s to r abundance ratio ([La/Eu]) of Cepheids shows a well defined
anticorrelation with of both Eu and Fe. On the other
hand, Galactic field stars attain an almost constant value and only when
they approach solar iron abundance display a mild enhancement in La.
The [Y/Eu] ratio shows a mild evidence of a correlation with Eu and,
in particular, with iron abundance for field Galactic stars.
We also investigated the s-process index -- [hs/ls] -- and we found a well
defined anticorrelation, as expected, between [La/Y] and iron abundance.
Moreover, we found a strong correlation between [La/Y] and [La/Fe] and, 
in particular, a clear separation between Galactic and Sagittarius red 
giants. Finally, the comparison between predictions for low-mass asymptotic 
giant branch stars and the observed [La/Y] ratio indicate a very good 
agreement over the entire metallicity range covered by Cepheids. However, 
the observed spread, at fixed iron content, is larger than predicted by 
current models.
} 


\maketitle

%
%
\section{Introduction}
\label{intro}

The use of classical Cepheids as solid distance indicators dates
back to more than  one century ago
\citep{Leavitt1908,LeavittPickering1912}. The evidence that
individual distances could be estimated on the basis of the
pulsation period and of the mean magnitude made Cepheids also very
popular stellar tracers. However, the use of classical Cepheids as
tracers of young stellar populations is more recent and dates back
to the seminal investigations by \citet{KraftSchmidt1963}. To
constrain the geometry, the rotation, and the density distribution
of the Galactic thin disk they used distances and radial
velocities of a sample of 267 Cepheids. The above empirical
evidences were soundly supplemented by pioneering evolutionary and
pulsation investigations suggesting that the pulsation period of
classical Cepheids is tightly anti-correlated with their age
\citep{Kippenhahnetal1969,MeyerHofmeister1969}.

During the last twenty years, classical Cepheids have been the
cross road of several  photometric and spectroscopic
investigations. Thanks to the new optical -- OGLE-IV
\citep{Udalskietal2015} -- and NIR -- IRIS
\citep{MivilleDeschenesLagache2005}, VVV \citep{Minnitietal2010}
-- photometric surveys classical Cepheids have been identified and
characterized in the thin disk and in the different components of
the Galactic spheroid hosting young-stellar populations: {\it i)}
inner disk
(\citealt{Matsunagaetal2013}; Inno et al., in prep.); {\it
ii)} nuclear bulge \citep{Matsunagaetal2011}; {\it iii)} beyond
the nuclear bulge \citep{Feastetal2014,Dekanyetal2015}; {\it iv)}
outer disk \citep[i.e.,][]{Metzgeretal1992,Pontetal2001}. Thus
suggesting that they can provide solid constraints on the impact
that environment has on the recent star formation episodes of the
Galactic thin disk.

The use of high-resolution spectrographs revealed that classical
Cepheids are also excellent tracers of the chemical enrichment of
intermediate-mass stars across the thin disk
\citep{Kraft1965,ContiWallerstein1969}. In spite of these
indisputable advantages in using classical Cepheids as stellar
tracers, the first detailed investigation of the thin disk iron
gradient dates back to \citet{Harris1981,HarrisPilachowski1984}.
More recent theoretical and empirical investigations revealed that classical
Cepheids, being yellow and red giants (RGs) and supergiants, display in
their spectra hundreds of iron lines
\citep{Andrievskyetal2002a,Andrievskyetal2002b,Andrievskyetal2002c,Andrievskyetal2004,Yongetal2006,Lucketal2006,Lemasleetal2007,Lemasleetal2008,Pedicellietal2010,Lucketal2011,LuckLambert2011,Genovalietal2013,Genovalietal2014},
dozens of $\alpha$-element
\citep[e.g.,][]{Andrievskyetal2004,LuckLambert2011,Lemasleetal2013,Genovalietal2015},
and iron peak lines
\citep[e.g.,][]{Andrievskyetal2004,LuckLambert2011} together with
a well defined continuum. They are also excellent laboratories to
constrain the impact of non-LTE effects in the abundance of both
CNO \citep{Lucketal2011,Martinetal2015} and Na
\citep{Takedaetal2013,Genovalietal2015} in RGs.
Cepheids have also been identified to constrain the abundance of
lithium in young stellar populations \citep{Kovtyukhetal2005,LuckLambert2011}.

The above investigations move along a well defined path concerning
the elemental abundance analysis and the chemical enrichment
history of the thin disk. In this context neutron capture elements
play a crucial role, since they trace the yields of a broad range
of stellar structures \citep[see, e.g.,][]{Snedenetal2008}. They
are typically split in two different groups, the heavy elements
formed either via slow neutron-capture (``s-process'') or rapid
neutron capture (``r-process''), i.e., either
slow or rapid compared to the $\beta$-decay time scale.

The ``main s-process'' is considered to occur in asymptotic giant
branch (AGB) stars during recurrent thermal pulses
\citep{Gallinoetal1998,Bussoetal1999}. On the other hand, the
``weak s-process'', takes place in massive fast evolving primitive
stars and produce elements with atomic mass number smaller than
A = 90 \citep{Raiterietal1993,Pignatarietal2010}.
The astrophysical sites of the r-process elements are even
more complex. Indeed, the recent literature on the nucleosynthesis
of r elements is quite rich. It has been suggested that they can
be produced by core-collapse supernovae (SNe) with a mass of the order
of 20~M$_\odot$
\citep{Thielemannetal2011,Wanajo2013,TsujimotoNishimura2015}; by
core-collapse SNe of very massive stars \citep[25 $\le$
M/M$_\odot \le$ 45,][]{Boydetal2012}; by electron capture
SNe in intermediate-mass stars \citep[8--10~M$_\odot$,][]{WoosleyHeger2015};
and by more complex astrophysical mechanisms such as the neutrino driven
wind ensuing to the merging of neutron stars
\citep[e.g.,][]{WanajoJanka2012,Bergeretal2013}.

The abundance of s- and r-process elements in Galactic classical
Cepheids have already been discussed in several recent papers
\citep[][]{Andrievskyetal2004,Lucketal2011,LuckLambert2011,Lemasleetal2013}.
However, we still lack a homogeneous and detailed analysis of weak
and main s-process elements in the Galaxy and their dependence on the iron
abundance -- a detailed analysis of s and r elements of RGs in the 
Sagittarius dwarf galaxy has been performed by
\citet[][hereinafter M13]{McWilliametal2013}. Moreover, and even more
importantly, we still lack a quantitative analysis of the spatial
abundance pattern of s- and r-process elements and, in particular,
the possible occurrence of an age dependence.

In this investigation we focus our attention on the abundance of
one light (Y) and three heavy (La, Ce, Nd) dominated
s-process elements, and a single r-process dominated
element (Eu). The abundances of 73 Galactic Cepheids of our sample
were complemented with the abundances of 363 Cepheids available in
the literature.

The structure of the paper is the following. In \S~2 we discuss
the spectra we collected together with the approach we adopted for
data reduction and analysis. In this section we also mention the
different samples of Cepheid abundances, based on high-resolution
spectra, available in the literature and the approach we adopted
to provide a homogeneous metallicity scale. In \S~3 we discuss the
radial gradients of [element/H] ratios for neutron-capture elements and
their comparison with iron and $\alpha$-element gradients. In this
section we also discuss the radial gradients of [element/Fe]
ratios and their age dependence. In \S~4 we present s- and 
r-process element abundances and compare their distribution
with dwarf and giant stars available in the literature. The
differentiation between weak and main s-process elements
and the comparison with the literature are also discussed
in this section. The summary of the results of the current
investigation are given in \S~5 together with a brief outline of
the future development of this project.

%
%
\section{Observations, data reduction and analysis}
\label{obs}

\subsection{Spectroscopic data}
\label{spec_data}

In this work we used the same high-resolution ($R$ $\sim$ 38\,000)
and high signal-to-noise ratio (S/N) spectra reported in
\citet[][hereinafter G14]{Genovalietal2014} in our determination of
iron abundances and atmospheric parameters, and in
\citet[][hereinafter G15]{Genovalietal2015} in our study of
$\alpha$-element abundances. A total of 122 spectra of 75 Galactic
Cepheids were collected with the UVES spectrograph at the ESO VLT
(Cerro Paranal, Chile) using two different instrument settings:
$i)$ with the former one we collected 80 spectra of 74 stars in
the wavelength ranges of $\sim$3760--4985~\AA,
$\sim$5684--7520~\AA, and $\sim$7663--9458~\AA; $ii)$ with the
latter one we collected 42 spectra of a control sample of 11
Cepheids in the wavelength ranges of $\sim$4786--5750~\AA\ and
$\sim$5833--6806~\AA. For more details on the instrumental
settings used we refer the reader to the G14 and G15 papers.

\begin{figure*}
\centering
\resizebox{0.9\hsize}{!}{\includegraphics[angle=-90]{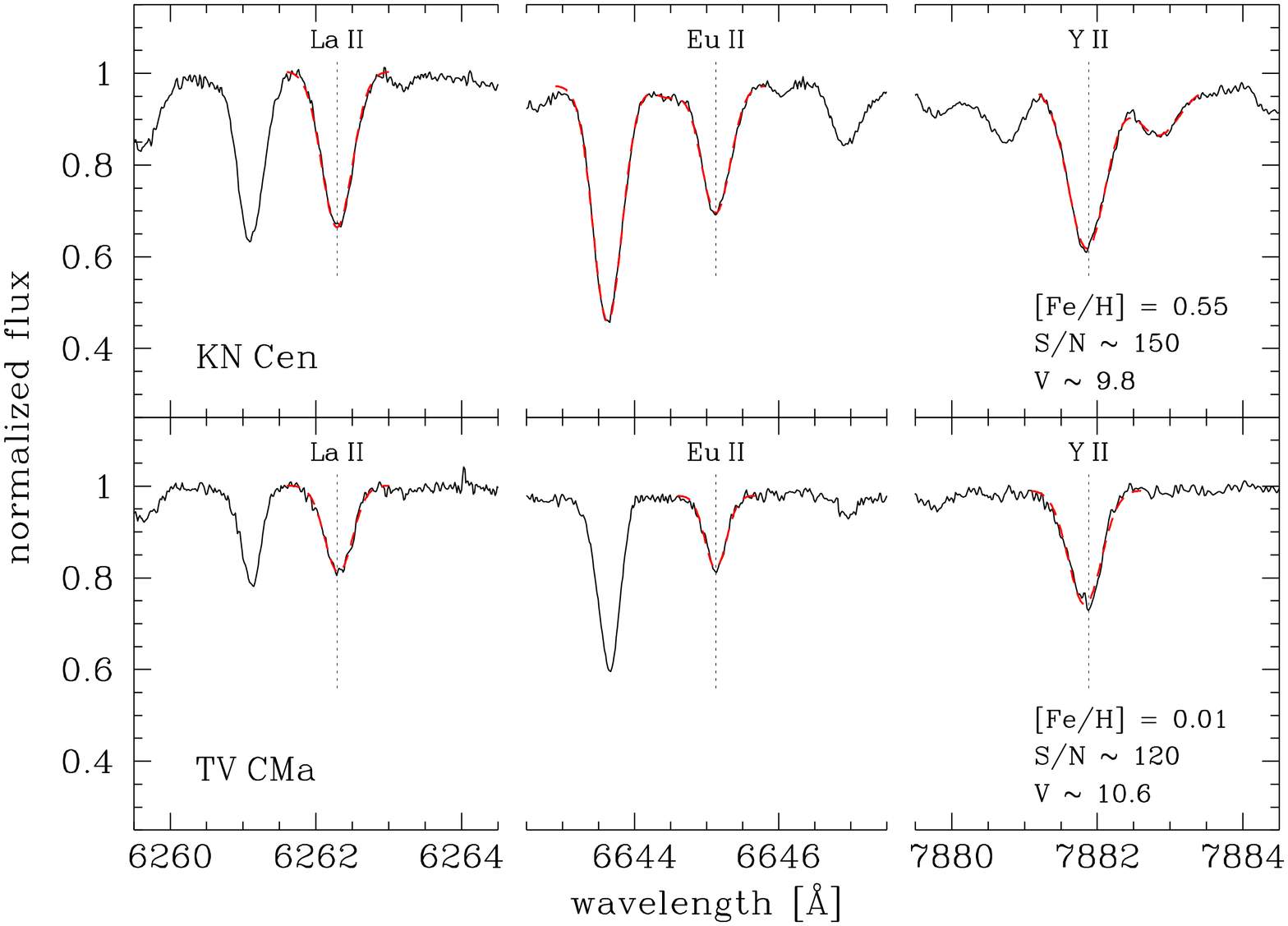}}
\caption{High-resolution ($R \sim 38\,000$) UVES spectrum of KN\,Cen and TV\,CMa. The apparent visual magnitude and the S/N in the spectral range $\lambda \sim5650-7500$~\AA\ are also labeled. The vertical dashed lines display some of the spectral lines (\ion{La}{ii} 6262.29, \ion{Eu}{ii} 6645.13, \ion{Y}{ii} 7881.88~\AA) adopted to estimate the abundances.}
\label{spectra}
\end{figure*}

In the same way as done in the papers mentioned above, we used the 11 Cepheids (V340\,Ara, AV\,Sgr, VY\,Sgr, UZ\,Sct, Z\,Sct, V367\,Sct, WZ\,Sgr, XX\,Sgr, KQ\,Sco, RY\,Sco, V500\,Sco) as a control sample. For these stars we have from four to six spectra each, collected with both the instrumental configurations (with the exception of V500\,Sco, which has 4 spectra collected only with the second instrument setting). For the same reasons reported in G15, the stars BB\,Gem and GQ\,Ori (both observed using the first instrument setting) were not included in the current analysis, thus we are left with 73 stars. The S/N are typically better 
than $\sim$100 per extracted pixel for all the \'echelle orders in 
the case of the first instrumental configuration (see examples in 
Fig.~\ref{spectra}), and ranges from $\sim$50 to roughly 300 for the 
second one. All the spectra were reduced using the ESO UVES pipeline 
Reflex v2.1 \citep{Ballesteretal2011}.

\subsection{Atmospheric parameters and abundances}
\label{meth}

We adopted the same iron abundances and atmospheric parameters derived by G14. The iron abundances are based on the equivalent width (EW) of about 100-200 \ion{Fe}{i} and about 20-40 \ion{Fe}{ii} lines, the number of lines depending on the spectral range used. The number of lines also varies according to the metallicity and to the spectral type of the star (at the time of the observation). To determine the atmospheric parameters, we set a limit of EW $<$ 120~m\AA\ to remain in the linear part of the curve of growth. For the objects where the number of weak lines was too small, we increased the limit to 180~m\AA . This slightly increases the uncertainties affecting the correlated atmospheric parameters, namely the effective temperature (\teff) and the microturbulent velocity (\vt). For more details on the impact that typical uncertainties on \teff, surface gravity (\logg), and \vt\ have on the iron abundances, see Table~2 of G14.

The \teff\ of individual spectra was estimated using the line depth ratio 
(LDR) method and calibrations derived by \citet{KovtyukhGorlova2000}. We 
adopted these calibrations because those provided by \citet{Kovtyukh2007} 
are not publicly available. Note that the difference in the temperature 
scale between the two calibrations is quite modest
\citep[see Fig. 4 in][]{Kovtyukh2007}. The latter calibrations have been 
criticized by \citet{Lyubimkovetal2010} and by \citet{Luck2014}, suggesting 
an overestimate for effective temperatures hotter than 6500--6800~K. In 
passing, we note that in our sample only a minor fraction, six out of 73 
Cepheids, has an effective temperature hotter than 6300~K. We also note that 
the estimated values of \teff\ were validated by verifying that the
\ion{Fe}{i} abundances do not depend on the excitation potential (\potex), 
i.e., the slope of [\ion{Fe}{i}/H] vs. \potex\ should be as close to zero.

The \logg\ was derived through the ionization equilibrium between
\ion{Fe}{i} and \ion{Fe}{ii} lines, and \vt\ was derived by minimizing the 
slope in the [\ion{Fe}{i}] vs. EW plot. This means that the \logg\ value is 
changed until the Fe I and Fe II lines provide the same abundance, while the 
\vt\ value is changed until the dependence of the derived abundances on the 
EWs is removed. Indeed, weak and strong lines are supposed to provide the
same elemental abundances. In this context it is worth mentioning that we 
are dealing with radial variables and the quoted physical parameters (\teff,
\logg, \vt) undergo cyclic variation along the pulsation cycle. The internal 
consistency of the adopted values was validated by G14 using calibrating 
Cepheids, i.e., objects for which we have from four to six spectra.

Concerning the abundance of the neutron-capture elements Y, La, Ce, Nd, and
Eu, we used the linelist provided by \citet{Lemasleetal2013}, with the same
atomic parameters (\potex\ and \loggf) listed in their Table~A.1, but with
small differences in the number of lines. We used six \ion{Y}{II} lines 
(5119.12, 5289.81, 5402.77, 5509.91, 5728.89, and 7881.88~\AA) instead of 
seven (the line at 6795.41~\AA\ was not used because the abundances derived 
using this line are systematically smaller than those derived using the 
other six lines). We also used the six lines for \ion{La}{II} (5114.56, 
5290.82, 5805.77, 6262.29, 6390.48, and 6774.27~\AA), three lines for 
\ion{Ce}{II} (4562.37, 5518.49, and 6043.39~\AA) instead of four (the 
abundance provided by the line at 4486.91~\AA\ was very often discrepant to 
the values provided by the other lines, therefore we did not used it), the 
six \ion{Nd}{II} lines (4959.12, 5092.79, 5130.59, 5181.17, 5431.52, and 
6740.08~\AA), and the two ones for \ion{Eu}{II} (6437.64 and 6645.13~\AA).

As previously done in G15 for the $\alpha$ elements, the EWs for these 
neutron-capture elements were measured using the {\it Automatic Routine for 
line Equivalent widths in stellar Spectra} \citep[ARES,][]{Sousaetal2007}, 
and double-checked using the {\it splot} task of IRAF\,\footnote{{\it Image 
Reduction and Analysis Facility}, distributed by the National Optical
Astronomy Observatories (NOAO), USA.}. Again, the internal dispersion is 
smaller than 6~m\AA\ and there is no evidence of systematics.

The abundances were derived with the {\it calrai} spectrum synthesis 
package, originally developed by \citet{Spite1967} and regularly improved 
since then. The package allow us to compute synthetic spectra by 
interpolating over a large grid of hydrostatic, LTE, and plane-parallel or 
spherical stellar atmospheres models \citep[MARCS,][]{Gustafssonetal2008}. 

For all the elements studied here, we assumed the standard solar abundances 
provided by \citet{Grevesseetal1996}, namely A(Fe)$_\odot$ = 7.50,
A(Y)$_\odot$ = 2.24, A(La)$_\odot$ = 1.17, A(Ce)$_\odot$ = 1.58,
A(Nd)$_\odot$ = 1.50, and A(Eu)$_\odot$ = 0.51.
Note that recent spectroscopic estimates of solar abundances by
\citet{Scottetal2015} and by \citet{Grevesseetal2015} indicate very similar 
abundances. Indeed, the difference in dex ranges from +0.01 for Eu to 
$-$0.03 for Fe and Y, to $-$0.06 for La, and to $-$0.08 for Nd, while the
new Ce abundance is identical to the old one.

\subsubsection{Hyperfine structure and isotopic splitting}
\label{hfs}

Several lines used in our abundance analysis are affected by hyperfine 
structure (hereinafter HFS) and/or isotopic splitting in the line profile. 
We searched in the literature for atomic data required to compute the
fine-structure components that form these lines. We found that several
of them are already available and for those that are not available we 
adopted the same approach discussed in M13 to compute the HFS. The atomic 
data required to compute the HFS of La and Eu were taken from 
\citet{Lawleretal2001a} and \citet{Lawleretal2001b}, respectively. For the 
La line at 6774.27~\AA\ we adopted the HFS already computed by M13. The same 
outcome applies to Y: we adopted the HFS data given by M13, but they are 
only available for three (out of six) lines of this element
($\lambda$ = 5119.12, 5289.81, 5728.89~\AA).

No atomic data have been found in the literature for our lines of Ce and Nd. 
Only the odd isotopes $^{143}$Nd and $^{145}$Nd have HFS, but their effects 
can be safely ignored -- in the solar system these isotopes constitute only 
20.5\% of the total Nd abundance, their lines are very narrow, and recent 
laboratory transition probabilities by \citet{DenHartogetal2003} indicate no 
evident HFS structure for more than 700 lines of Nd~II.

\subsubsection{Abundances corrected from HFS}
\label{hfs_ab}

We derived the abundances of the current sample of 73 Cepheids by accounting 
for the HFS of the elements and lines mentioned in the previous section. 
In order to quantify the effects of the HFS on the derived abundances and 
their dependence on other parameters, we performed a comparison between our 
abundance results before and after performing the HFS analysis. The mean 
differences in abundance are summarized in Table~\ref{table_hfs_mean}. Note 
that the differences are much larger for some of the La lines, but they are 
very close to zero for both Y and Eu lines. Concerning possible dependences 
on other parameters, we found that these differences and their dispersions 
become smaller with increasing \teff\ (specially for \teff\ > 5500~K), with 
increasing surface gravity (specially for \logg\ > 0.5), and with 
decreasing pulsation period (\logP\ < 1.0). No clear correlation with 
metallicity is observed.

Finally, we note that the lines selected to measure n-capture elements 
are typically weak (typically smaller than 180~m\AA) and 
unsaturated. This means that the derived abundances for La, Ce, Nd and Eu 
are marginally affected by HFS sub-structure, while for Y they are small 
(see Table~\ref{table_hfs_mean}).

\begin{table}
\centering
\caption[]{Mean differences in the abundances derived before and after accounting for the HFS.}
\label{table_hfs_mean}
\begin{tabular}{lcc r@{ }r}
\hline\hline
 & & \\[-0.2cm]
specie &
$\lambda$ [\AA] &
\parbox[c]{3.3cm}{\centering ${\rm [X/H]}_{\rm HFS}-{\rm [X/H]}_{\rm no~HFS}$} &
\multicolumn{2}{c}{\parbox[c]{1.6cm}{\centering correction (min, max)}} \\[0.3cm]
\hline
 & & \\[-0.2cm]
\ion{Y}{II}  & 5119.12 &   +0.013 $\pm$ 0.006 &    0.00,&   +0.03 \\
\ion{Y}{II}  & 5289.81 &   +0.008 $\pm$ 0.005 &    0.00,&   +0.02 \\
\ion{Y}{II}  & 5728.89 &   +0.008 $\pm$ 0.005 &    0.00,&   +0.02 \\
\ion{La}{II} & 5114.56 & $-$0.172 $\pm$ 0.128 & $-$0.45,& $-$0.01 \\
\ion{La}{II} & 5290.82 & $-$0.010 $\pm$ 0.013 & $-$0.04,&   +0.01 \\
\ion{La}{II} & 5805.77 & $-$0.129 $\pm$ 0.098 & $-$0.53,&    0.00 \\
\ion{La}{II} & 6262.29 & $-$0.211 $\pm$ 0.178 & $-$0.90,& $-$0.01 \\
\ion{La}{II} & 6390.48 & $-$0.004 $\pm$ 0.009 & $-$0.03,&   +0.01 \\
\ion{La}{II} & 6774.27 &   +0.050 $\pm$ 0.050 & $-$0.17,&   +0.11 \\
\ion{Eu}{II} & 6437.64 & $-$0.006 $\pm$ 0.008 & $-$0.03,&   +0.01 \\
\ion{Eu}{II} & 6645.13 & $-$0.023 $\pm$ 0.021 & $-$0.10,&   +0.01 \\
\hline\noalign{\smallskip}
\end{tabular}
\tablefoot{The quoted errors in the third column represent the dispersion 
around the mean, and the forth column lists the minimum and maximum HFS
corrections applied to the abundances.}
\end{table}

Table~\ref{table_ab_heavy_spec} lists the abundances from
individual spectra. Column 3 shows the iron abundances derived by
G14, and column 4 the number of \ion{Fe}{i} and \ion{Fe}{ii} lines
used. The other columns show our results for the abundances of Y,
La, Ce, Nd, and Eu, corrected for HFS when possible, together with
the number of lines used. In Table~\ref{table_ab_heavy_mean} we
list the mean abundances computed for the stars with multiple
spectra. The HFS data that we adopted are listed in
Table~\ref{table_hfs} for Y, La, and Eu lines.

\subsection{Data available in the literature}
\label{liter}

We compared our abundance estimates, corrected for HFS, with the results 
provided by similar studies available in the literature:
\citet[][hereinafter LEM]{Lemasleetal2013}, \citet[][LII]{Lucketal2011}, 
\citet[][LIII]{LuckLambert2011}, and \citet[][YON]{Yongetal2006}. Note that 
none of the quoted investigations, but YON, take account of HFS corrections 
in their analysis of Cepheid spectra.

By comparing the stars in common among these different data sets, we evaluated the systematic difference among them. The mean differences between our measurements and those of LEM, LII, LIII, and YON range, in modulus, from 0.02~dex for Eu up to 0.31~dex for Fe. The details on these comparisons are listed in Table~\ref{table_diff_ab}, where we show the zero-point differences obtained by G14 for the iron abundances together with our determinations for the other elements. Each pair of data sets was chosen aiming to maximize the number of stars in common between them. To provide a homogeneous abundance scale for Galactic Cepheids, we applied these zero-point differences to the quoted data sets, putting them in the same scale of our current sample. The element abundances available in the literature and the rescaled values are listed in columns from 2 to 15 of Table~\ref{table_liter}.

The priority in using the abundances from the literature follows the same approach adopted by G14 and by G15: firstly, we adopt the abundances provided by our group, i.e., this study and the results from LEM, and finally those provided by the other studies, namely LIII, LII, and YON, in this order. We notice that the star HQ\,Car was also excluded from our analysis because it has been recently identified as a Type~II Cepheid by \citet{Lemasleetal2015}. The final sample has 435 Cepheids, with a homogeneous abundance scale for Fe, Y, La, Ce, Nd, and Eu.

%
%
\section{Neutron-capture element gradients}

\subsection{Neutron-capture gradients from Cepheids}
\label{grad_ceph}

In this section we investigate the radial gradients of Y, La, Ce, Nd, and Eu
across the Galactic disk using our sample of 73 classical Cepheids plus
a sample of 363 Cepheids available in literature. Homogeneous iron 
abundances and Galactocentric distances for the entire sample were provided 
by G14 (see their Table~1 and Table~4). The key advantage of the current 
approach when compared with similar investigations are the following:
{\it i)} the intrinsic parameters (\logg, \teff, \vt) were estimated using the same approach;
{\it ii)} elemental abundances are based on high-resolution and high signal-to-noise spectra and similar line lists;
{\it iii)} individual Cepheid distances were estimated using near-infrared Period-Wesenheit relations
that are reddening free and minimally affected by the metallicity \citep{Innoetal2013}.

In the following we discuss the radial gradients of four s-process
(Y, La, Ce, Nd) elements and a single r-process element (Eu). We
note that n-capture elements can be split according to solar
system abundances in pure s-process, pure r-process, and
mixed-parentage isotopes. 
Among the selected elements Eu is a pure r-process element, since 
the r-fraction abundance is 97\% \citep{Burrisetal2000,Simmereretal2004}. 
On the other hand, the selected s-process elements have s-fraction 
abundances ranging from roughly 50\% \citep[Nd, 58\%,][]{Snedenetal2008} to 
more than 70\% (Y, 72\%; La, 75\%; Ce, 81\%). Note that the quoted s- and
r-fraction abundances should be cautiously treated, since
\citet{Bisterzoetal2011}, using a different approach, found similar 
fractions for Eu (94\%), Nd (52\%), La (71\%) and Ce (81\%), but a 
significantly larger s-fraction for Y (92\%). 

\begin{figure}
\centering
\resizebox{\hsize}{!}{\includegraphics{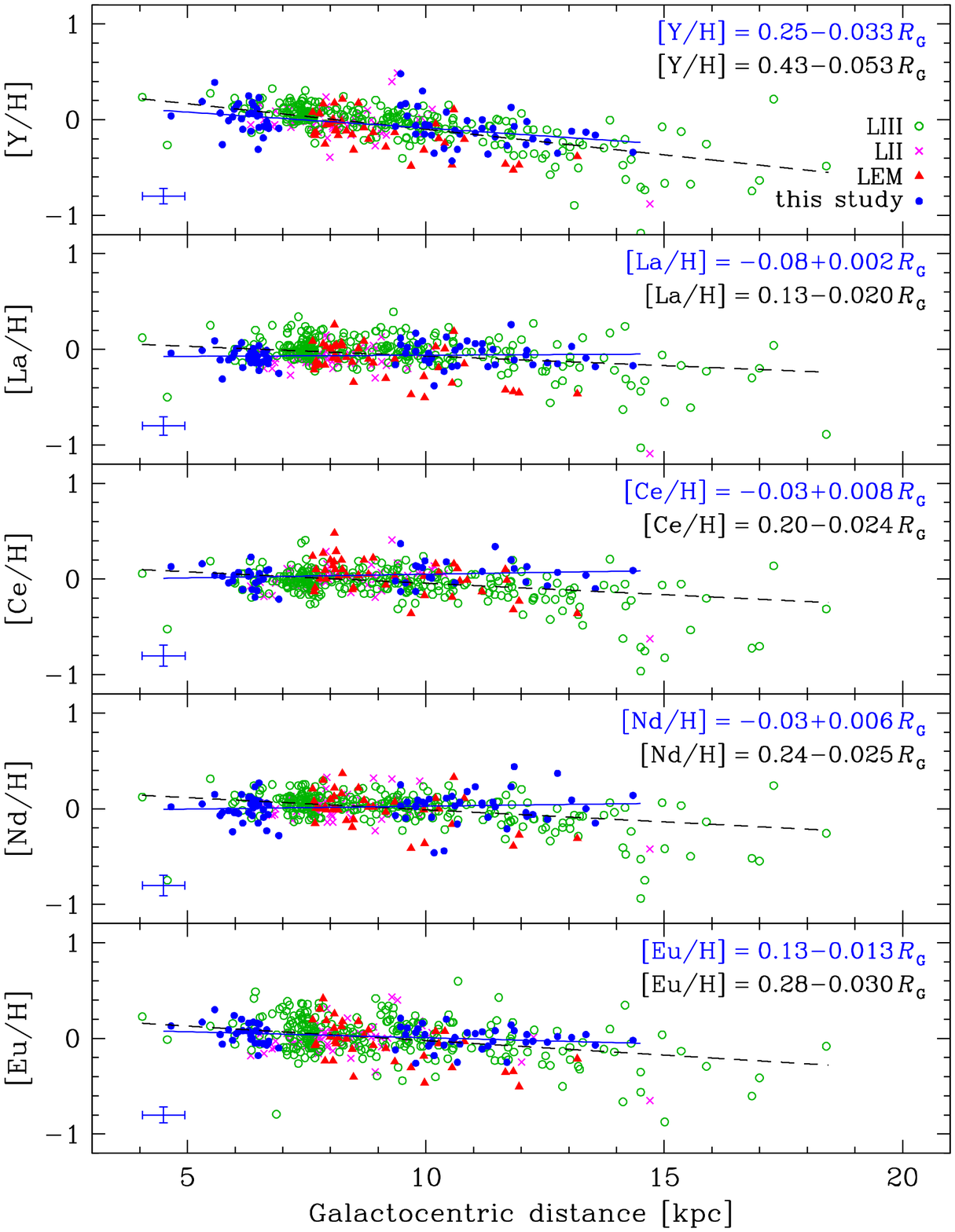}}
\caption{Abundances of neutron-capture elements as a function of \RG. Our results (filled blue circles) are compared with those of \citet[][LII, magenta crosses]{Lucketal2011}, \citet[][LIII open green circles]{LuckLambert2011}, and \citet[][LEM, red triangles]{Lemasleetal2013}. The blue solid line shows the linear regression of our Cepheid sample, while the black dashed line the linear regression of the entire Cepheid sample. The blue error bars display the mean spectroscopic error of the current sample. The abundances available in the literature have similar errors.}
\label{xh_Gdist_rs}
\end{figure}

Figure~\ref{xh_Gdist_rs} shows the abundances scaled to hydrogen of
Y, La, Ce, Nd, and Eu as a function of \RG\ for the final sample.
Stars plotted in this figure include the current 73 Cepheids plus
38 from LEM, 263 from LIII and 61 from LII. Note that for the Cepheid
XZ\,CMa the abundances of the above elements are not available, 
therefore, we ended up with a sample of 435 stars. The individual 
Cepheid Galactocentric distances were estimated by G14 and assume 
a solar Galactocentric distance of 7.94\,$\pm$\,0.37\,$\pm$\,0.26~kpc 
\citep[][]{Groenewegenetal2008,Matsunagaetal2013}. The individual 
\RG\ values are also listed in Table~\ref{table_ab_heavy_mean}. 
The typical uncertainty on the individual distances is $\sim$5\% 
and is mainly due to the accuracy of the zero-point in the adopted 
Period-Wesenheit relations \citep[for more details see][]{Innoetal2013}.

The Cepheid abundances from the literature plotted in Fig.~\ref{xh_Gdist_rs} 
were scaled adopting the zero-point differences listed in
Table~\ref{table_diff_ab}. A similar approach was adopted to scale both iron 
(G14) and $\alpha$-element abundances (G15). This figure also shows the 
linear Least Squares fits to the current sample of 73 Cepheids (blue solid 
line) and to the entire sample (435, black dashed line). To avoid thorny 
problems in the estimate of both the zero-point and the slope due to 
possible outliers, we applied a {\it biweight} procedure 
\citep{Beersetal1990}. The slopes and the zero-points of the two radial 
gradients are labeled. The slopes and the zero-points of the fits based on 
the entire sample together with their uncertainties and standard deviations 
are also listed in columns from 2 to 4 of Table~\ref{table_slopes}.

The empirical scenario emerging from the data plotted in this figures brings forward several interesting features.

{\it i)} -- Radial gradients -- The five investigated neutron-capture 
elements display well defined radial gradients. This evidence coupled with 
similar results concerning iron (see G14, and references therein), $\alpha$ 
(see G15, and references therein), and iron-group (LII, LIII) elements 
further indicates that young stellar tracers show radial gradients across 
the Galactic thin disk. A more quantitative discussion concerning the global 
behavior will be addressed in a forthcoming paper.
Finally, we note that the occurrence of well defined radial gradients for
light (Y) and heavy s-process elements (La, Ce, Nd) do not support the lack
of a radial gradient for Ba as recently suggested by
\citet{Andrievskyetal2014} and by \citet{Martinetal2015}. The quoted authors
used high-resolution spectra for a sizable sample of inner and outer disk
classical Cepheids and take account of NLTE effects.
However, Ba abundances in Classical Cepheids are affected by severe limits. In particular, \citet{Luck2014} noted that strong \ion{Ba}{ii} lines are affected by line-formation effects, while \citet{Andrievskyetal2013} discussed in detail physical and atomic (isotopic shifts) effects.
The reason for the lack of a Ba gradient remains still unclear.

{\it ii)} -- Slopes -- The slopes are quite similar and on average
of the order of $-$0.025\,$\pm$\,0.004~\dexkpc\ for La, Ce, Nd, and Eu.
The only exception is Y, for which the slope is more than a factor
of two steeper ($-$0.053\,$\pm$\,0.003~\dexkpc). The current
slopes agree quite well, within the errors, with similar estimates
available in the literature. We found that the slopes range from
$-$0.053\,$\pm$\,0.003~\dexkpc\ for [Y/H] to
$-$0.020\,$\pm$\,0.003 for [La/H]. The slopes estimated by
LII+LIII for the same elements range from $-$0.061\,$\pm$\,0.003
to $-$0.019\,$\pm$\,0.005~\dexkpc, while those estimated by LEM
range from $-$0.062\,$\pm$\,0.012 to
$-$0.045\,$\pm$\,0.012~\dexkpc. The latter is slightly
steeper and the difference might be due to the limited range in
Galactocentric distances covered by their Cepheid sample.
The main difference in the comparison with similar estimates available in
literature is for Nd. Indeed, LII found a flat distribution across the thin
disk. We performed several tests using different cuts in Galactocentric
distance and in sample size and we found that the slope is solid within
the current uncertainties (see labeled error bars). Moreover, the standard
deviation of the Nd gradient is the smallest among the investigated ones.
The reader interested in more details on the slopes of the available data 
sets is referred to columns from 6 to 9 of Table~\ref{table_slopes}.

{\it iii)} -- Spread -- The spread of the individual abundances attains
similar values across the thin disk. The outermost disk regions are
an exception, since the spread increases for \RG\ larger than 13~kpc.
The neutron capture elements display the same trend of iron and
$\alpha$-element abundances. Among the investigated elements, Y
seems to be once again an exception, since the spread is homogeneous
over the range of Galactocentric distances covered by the current sample.

{\it iv)} -- Comparison with theory -- Our results for La
($-$0.020 $\pm$ 0.003~\dexkpc) and for Eu ($-$0.030 $\pm$ 0.004~\dexkpc)
agree quite well with theoretical predictions by \citet{Cescuttietal2007}
for Galactocentric distances covering the entire thin disk
(4\,$\le$ \RG\ $\le$\,22~kpc). They found a slope of $-$0.021~\dexkpc\ for 
La and of $-$0.030~\dexkpc\ for Eu. The predicted slopes become steeper for
Galactocentric distances shorter than 14~kpc and shallower for distances
larger than 16~kpc (see their Table~5). Predictions for the other s-process 
elements are not available. In passing we note that the observed slope for
Y ($-$0.053\,$\pm$\,0.003~\dexkpc) is similar to the predicted slopes
for iron and iron-group elements in the Galactocentric range between 4 and
14~kpc.

\subsection{Comparisons with independent radial gradients}
\label{comp}

\begin{figure}
\centering
\resizebox{\hsize}{!}{\includegraphics{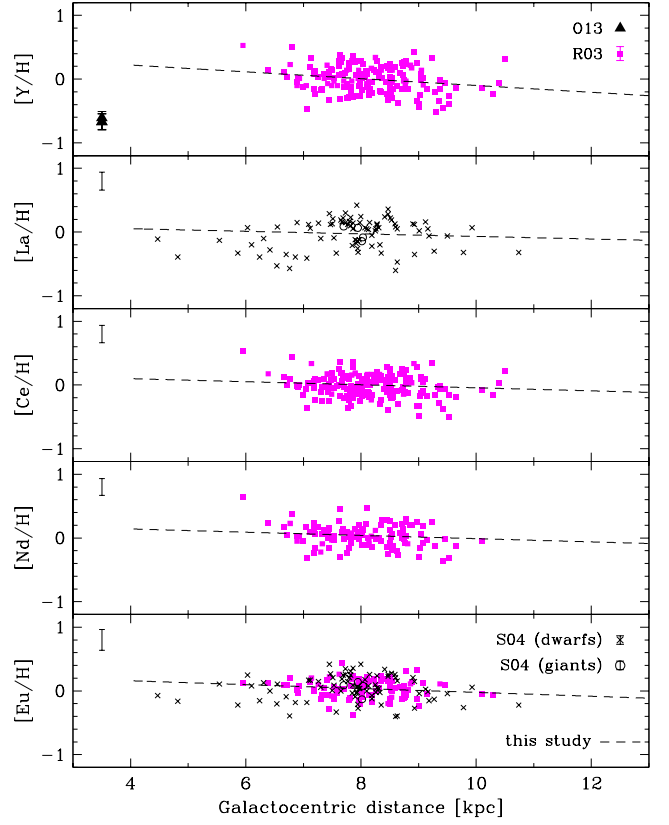}}
\caption{Abundances of neutron-capture elements as a function of \RG. The radial gradients we derived for Cepheid stars (dashed line) are compared with field dwarfs analyzed by \citet[][R03, magenta squares]{Reddyetal2003} and with field dwarfs (crosses) and giants (open circles) analyzed by \citet[][S04]{Simmereretal2004}. From the latter only stars with [Fe/H] > $-$1.0 are plotted. RSGs in the Scutum cluster analyzed by \citet[][O13, triangles]{Origliaetal2013} are also shown.}
\label{xh_Gdist_lit_rs}
\end{figure}

To further constrain the plausibility of the above radial gradients,
Fig.~\ref{xh_Gdist_lit_rs} shows the comparison between Cepheid gradients
and radial gradients of neutron-capture elements of Galactic field stars. 
The abundances of Y, Ce, Nd, and Eu for 181 F- and G-type dwarf stars 
provided by \citet[][hereinafter R03]{Reddyetal2003} are plotted.
The La and Eu abundances for 159 dwarf and giant stars were provided by 
\citet[][hereinafter S04]{Simmereretal2004}.
Their abundances were rescaled to the abundances of the solar mixture
adopted in the current investigation \citep{Grevesseetal1996}. Moreover,
to overcome possible differences between Cepheids and field stars concerning
either the different diagnostics adopted to determine distances or the use
of different spectral lines, in plotting their data we adopted the
zero-points of our gradients at the solar Galactocentric distance. Note that
in dealing with S04 data, we only selected the more metal-rich stars
([Fe/H] > $-$1.0) to be more consistent with the metallicity range of the
current Cepheids. The figure shows
that the radial gradient of the five neutron-capture elements based on
Cepheids agree quite well with the abundances for field dwarf stars in the
Galactic thin disk. The fact that the giants in the S04 sample covers only
a limited range of Galactocentric distances across the solar circle does not
allow us to constrain the radial gradient.

In Fig.~\ref{xh_Gdist_lit_rs} we also plot the Y abundances recently
provided by \citet[][hereinafter O13]{Origliaetal2013} for three red
supergiant (RSG) stars in the Scutum cluster. They used high-resolution
($R$ $\sim$ 50\,000) NIR (Y, J, H, K) spectra collected with GIANO at the
Telescopio Nazionale Galileo (TNG). The comparison further supports
previous results by \citet{Bonoetal2015} and G15 concerning the
underabundance of iron and $\alpha$-elements in blue and red supergiants
located either in the near end of the Galactic bar or in the Galactic
center. The Y abundances display the same underabundance
when compared with similar abundances of classical Cepheids located in
the inner edge of the Galactic thin disk.

\subsection{Age dependence of the [neutron-capture/H] ratios}
\label{dep_xh_logP}

The results concerning the abundance gradients discussed in the above
sections use the Galactocentric distance as independent variable. However,
classical Cepheids when compared with other stellar tracers have the key
advantage that their pulsation period is tightly anti-correlated with
their individual ages \citep[][G15]{Bonoetal2005}. The typical pulsation 
age of short period ($P$ $\sim$ 1.0-1.5~days) Cepheids is indeed of the 
order of 200~Myr, while for long period ($P$ $\sim$ 100~days) ones is of
the order of 10~Myr. The exact range in age does depend on the chemical 
composition and on the adopted evolutionary framework (see Table~4 and 5 in 
\citealt{Bonoetal2005}, and \citealt{Andersonetal2015}). This provides 
the unique opportunity to constrain the chemical enrichment history of the 
thin disk during the last $\sim$300~Myr (G14, G15).

\begin{figure}
\centering
\resizebox{\hsize}{!}{\includegraphics{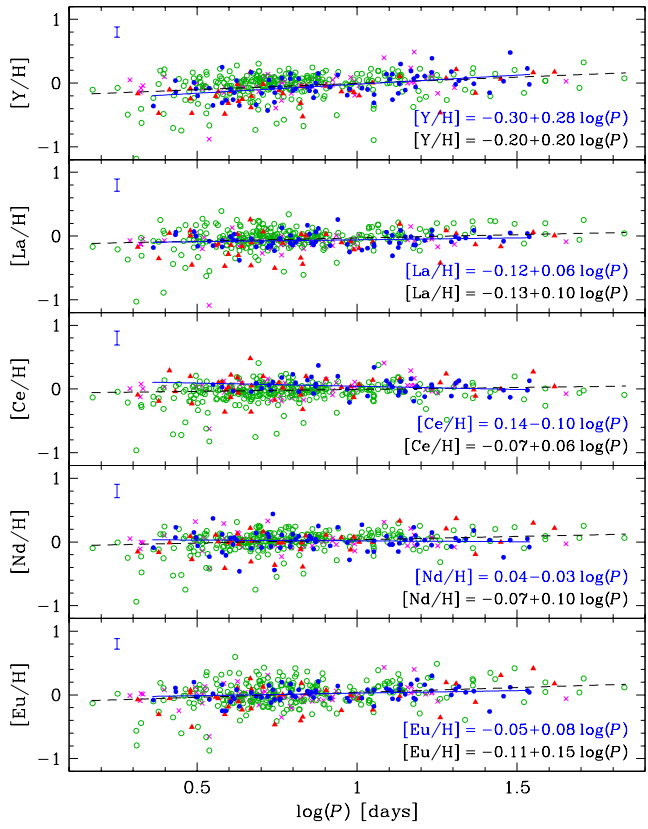}}
\caption{Abundances of neutron-capture elements as a function of the logarithmic pulsation period. Symbols and colors are the same as in Fig.~\ref{xh_Gdist_rs}.}
\label{xh_logP_rs}
\end{figure}

To constrain the age dependence of the metallicity gradients,
Fig.~\ref{xh_logP_rs} shows the same elemental abundances plotted in
Fig.~\ref{xh_Gdist_rs}, but as a function of the logarithmic period.
Data plotted in this figure show that the investigated neutron-capture
elements display well defined positive gradients as a function of the
pulsation period. The $\alpha$ elements (Mg, Si, Ca) and the light elements
(Na, Al) investigated by G15 show similar trends, but the current slopes
are on average shallower. The slopes of three out of the four s-process
elements (La, Ce, Nd) are equal or smaller than 0.10~dex per logarithmic 
day, however Y (s element) and Eu (r element) display steeper slopes
(0.20 and 0.15~dex per logarithmic day, respectively).

The above empirical evidence indicates that the elements that are more 
typically associated with explosive nucleosynthesis (Si, Ca, Eu) display age 
gradients ranging from 0.09 (Ca) to 0.15 (Si, Eu) dex per logarithmic day. 
On the other hand, Y shows a slope (0.20\,$\pm$\,0.03~dex per logarithmic 
day) that is at least a factor of two larger than the other s-process 
elements with similar s-fraction abundances (La, Ce). In this context it is 
worth mentioning that the 60-70\% of Y is produced in the main s-process, 
while 5-10\% comes from r-process and the remaining from the weak component. 
However, the significant difference in the Y slope when compared with the 
other s-process elements could suggest a larger contribution either from the 
r- and/or from the s-weak component. 

To take account for the above empirical evidence we could also use 
plain stellar evolutionary arguments. We start from the evidence that 
Cepheid stellar masses range, according to chemical composition, 
from 3.0--3.5 M$_\odot$ to 10--12 M$_\odot$ \citep{Bonoetal2010}. This 
means that a significant fraction of Cepheids evolve into the AGB 
phase. The difference in evolutionary time between the end of the 
so-called blue loop and the beginning of the AGB phase is negligible
when compared with H and He burning phases. This means that Cepheids and AGB 
stars with stellar masses ranging from $\sim$3 to $\sim$6 M$_\odot$ evolve 
with similar evolutionary lifetimes.
The current theoretical predictions indicate that intermediate-mass 
AGB stars in the quoted mass regime, mainly produce light s-process (ls) 
elements (such as Y), while the bulk of the heavy s-process (hs) elements 
(such as La) is mainly produced in low-mass (M $<$ 3 M$_\odot$) AGB stars. 
To further constrain this effect, we mention that an AGB star of 6~M$_\odot$ 
produces roughly 1/3 of the Y, but only 1/7 of the La produced by a
3~M$_\odot$ \citep{Cristalloetal2015a}. This would imply that Y is for 
younger Cepheids a good tracer of the recent chemical enrichment of 
intermediate-mass AGB stars. The same outcome applies for the slope of Eu, 
since this element is mainly produced in stellar structures that are either 
coeval or even younger than Cepheids. It goes without saying that the quoted 
scenario is qualitative and more detailed calculations based on chemo-
dynamical models are required to constrain the anti-correlation between s 
and r-process elements with age. 

\subsection{Radial gradient of [neutron-capture/Fe]}
\label{heavy_Fe}

\begin{figure}
\centering
\resizebox{\hsize}{!}{\includegraphics{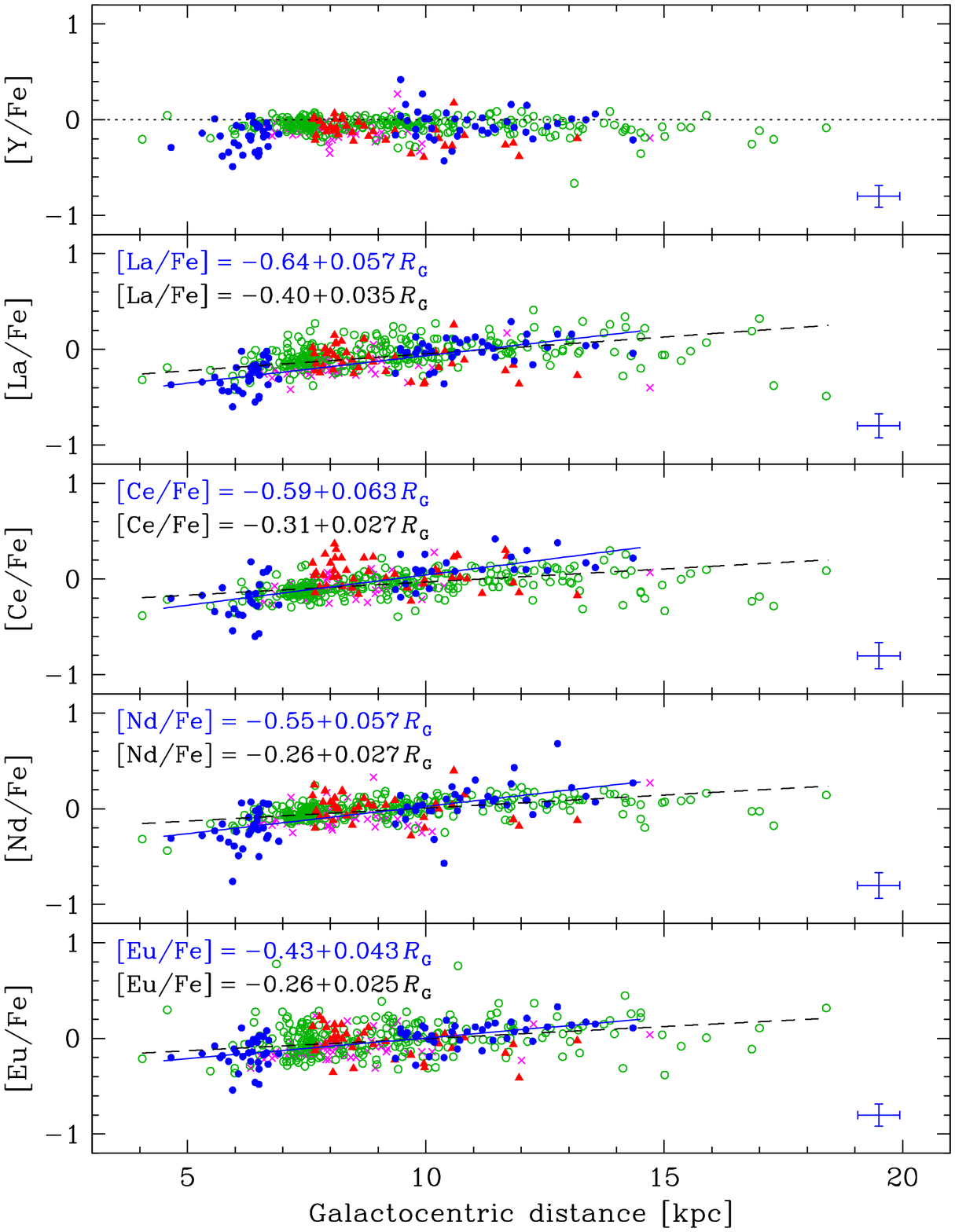}}
\caption{The same as Fig.~\ref{xh_Gdist_rs} but the abundances are scaled to iron. The dotted line displays the positions of solar abundance ratios.}
\label{xfe_Gdist_rs}
\end{figure}

Figure~\ref{xfe_Gdist_rs} shows the radial gradients of the abundance ratios
scaled to iron. Similar radial gradients for the $\alpha$ elements were
recently investigated by G15. The test was motivated by the similarity
in the slope of [Fe/H] and [$\alpha$/H] ratios. Indeed, they found that
the slopes of [$\alpha$/Fe] ratios as a function of the Galactocentric
distance are typically smaller than 0.018\,$\pm$\,0.002~\dexkpc.
The conclusion for the quoted elements was that they show, on average,
quite flat distribution across the entire thin disk.

Data plotted in Fig.~\ref{xfe_Gdist_rs} display a different empirical
scenario for neutron-capture elements. The s- (La, Ce, Nd) and the
r- (Eu) process elements display slopes that are on average a factor
of two larger when compared with [$\alpha$/Fe] ratios. The only element
to show a flat distribution over the entire disk is Y. The above evidence
is suggesting that the steady enhancement in four out of the five
neutron-capture elements investigated is mainly caused by the slopes of
La, Ce, Nd, and Eu radial gradients: they are at least a factor of two
smaller than the iron slope ($-$0.060\,$\pm$\,0.002~\dexkpc). The [Y/Fe]
ratio is flat because the slope of Y gradient
($-$0.053\,$\pm$\,0.003~\dexkpc) is quite similar to the iron one.

The above findings indicate that the chemical enrichment history of
La, Ce, Nd, and Eu across the Galactic thin disk is quite different when
compared with $\alpha$ elements and iron. Although Y is considered mainly
a s-process element, its abundance ratios appear to be more similar to
iron and to $\alpha$ elements than to the other neutron-capture elements.
It is worth mentioning that the spread in [element/Fe] of the five
investigated elements appears to be quite constant when moving from
the inner to the outer disk (see Fig.~\ref{xfe_Gdist_rs}). There is also
a mild evidence of a flattening in the above ratios towards the outer disk.
Indeed, the radial gradients based on the current sample are steeper than
the slopes based on the entire sample. The difference is mainly due to the
limited Galactocentric distance covered by our sample. However, the number
of Cepheids with Galactocentric distance larger than 13~kpc is limited,
and new identifications of classical Cepheids in the outer disk are
required to further constrain the quoted trends (see also G15).
In this context, it is worth mentioning that Gaia is going to play a crucial 
role, since detailed calculations indicate that the number of Galactic 
Cepheids will increase at least one order of magnitude 
\citep{Bono2003,Windmarketal2011}.

\subsection{Comparisons with independent radial gradients}
\label{comp_Fe}

\begin{figure}
\centering
\resizebox{\hsize}{!}{\includegraphics{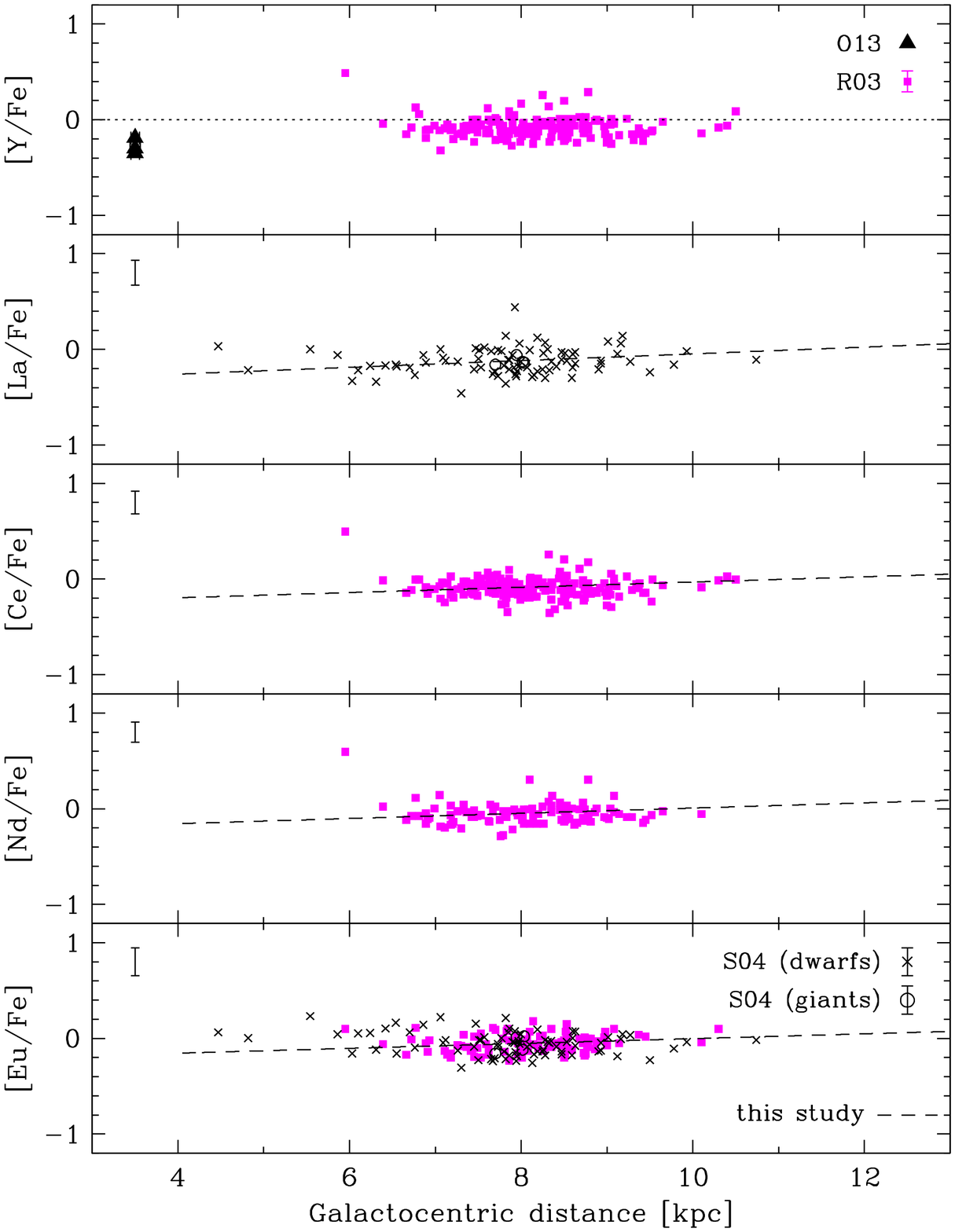}}
\caption{The same as Fig.~\ref{xh_Gdist_lit_rs}, but the abundances are scaled to iron. The dotted line displays the positions of solar abundance ratios.}
\label{xfe_Gdist_lit_rs}
\end{figure}

To validate the new slopes of the [neutron-capture/Fe] radial
gradients, Fig.~\ref{xfe_Gdist_lit_rs} shows the comparison with
similar data available in the literature. The colored symbols
denote the same field dwarf and giant stars plotted in
Fig.~\ref{xh_Gdist_lit_rs}. Note that the abundances ratios
plotted in this figure were scaled both in iron and in
neutron-capture element abundances. The flatness of the
[neutron-capture/Fe] ratios for Y and the increasing trends for
La, Ce, Nd, and Eu are quite similar to the results based on the
entire Cepheid sample. 

This evidence further supports our working hypothesis that 
neutron-capture elements -- but Y -- experienced during the last 
300~Myr a different chemical enrichment history from iron and 
$\alpha$ elements. The current predictions concerning the chemical 
enrichment of AGB stars indicate that ls elements (such as Y) are 
mainly synthesized in the more metal-rich ([Fe/H] > $-$0.6) regime, 
while the hs elements (such as La) are more favored in the 
metal-intermediate regime (see also \S~4). The quoted theoretical framework
supports the mild enhancement in hs elements when moving from the inner
(more metal-rich) to the outer (more metal-poor) Galactic thin disk. On the
other hand, the lack of a clear trend in the [Y/Fe] abundance ratio 
indicates a substantial balance across the entire disk. However, the 
most metal-rich ([Fe/H] $\sim$ 0.4-0.5) Cepheids in our sample that are 
located in the inner disk (5 $\le$ \RG\ $\le$ 7~kpc) show a downturn in 
[Y/Fe], suggesting an underabundance of Y at super-solar iron abundance.  
This finding further supports a similar trend in [Y/Fe] abundances provided 
by \citet[][hereinafter FG98]{FeltzingGustafsson1998} using high-resolution 
spectra for 47 dwarf stars with super-solar iron abundance 
(see their Fig.~22 and \S~6.13).
However, the presence of a mild enhancement of Eu in the outer disk is even
more compelling, since this is considered a solid r-process element mainly
produced by the same stellar masses producing $\alpha$ elements.

In passing we also note that the reduced spread of the above elements, 
at fixed \RG\ distance, is also suggesting a quite homogeneous spatial 
enrichment across the four Galactic quadrants. This is also an interesting 
evidence worth being investigated in more detail, since AGB stars can have 
both intermediate-age (1-9~Gyr) and old ($\sim$10~Gyr) progenitors.

\subsection{Age dependence of the [neutron-capture/Fe] ratios}
\label{dep_xfe_logP}

\begin{figure}
\centering
\resizebox{\hsize}{!}{\includegraphics{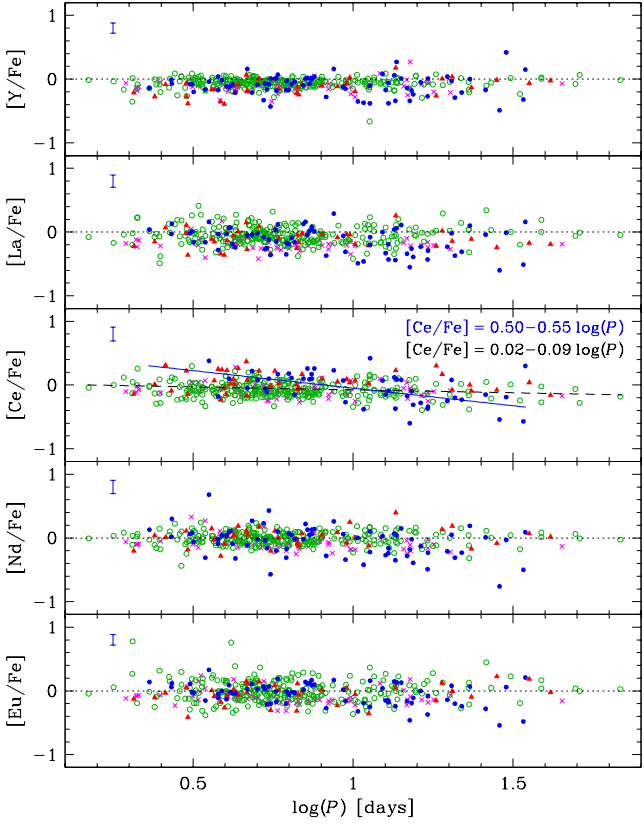}}
\caption{The same as Fig.~\ref{xh_logP_rs} but the abundances are scaled to iron. The dotted lines display the positions of solar abundance ratios. Symbols and colors are the same as in Fig.~\ref{xh_Gdist_rs}}
\label{xfe_logP_rs}
\end{figure}

To constrain the age dependence of the [neutron-capture/Fe]
abundance ratios, Fig.~\ref{xfe_logP_rs} shows the same elemental
abundances plotted in Fig.~\ref{xfe_Gdist_rs}, but as a function
of the logarithmic period. A glance at the data plotted in this
figure shows that the ratios are approximately constant over the
entire period range. The only exception is Ce, showing a mild
negative gradient ($-$0.09\,$\pm$\,0.02~dex per logarithmic day). Similar 
trends are also showed by light and $\alpha$ elements. Indeed, Ca showed
(see Fig.~5 in G15) a negative gradient ($-$0.11\,$\pm$\,0.02~dex per 
logarithmic day), while the others either a mild gradient (Al, Si) or
a flat distribution (Na, Mg). The flattening of the s-process elements is
once again an interesting finding, since it is suggesting that s
elements and iron enrichment across the Galactic thin disk have
been quite homogenous over a broad range in age. The zero-point,
the slope, their uncertainties, and the standard deviation of the
Ce gradient are listed in the bottom line of Table~\ref{table_slopes}.

%
%
\section{Neutron-capture element relative abundances}
\label{heavy_feh}

\subsection{Metallicity dependence of the [neutron-capture/Fe] ratios}

The comparison with abundances of neutron-capture elements
available in the literature discussed in the above sections were
limited to the data sets for which were also available individual
Galactocentric distances. In this section we perform the
comparison only using elemental abundances. In particular, we
selected:
{\it i)} Y, Ce, Nd, and Eu abundances of F- and G-type field dwarf 
stars provided by R03 (181 objects);
{\it ii)} Y and Eu abundances of F- and G-type field dwarf stars 
estimated by \citet[][hereinafter B05, 102 objects]{Bensbyetal2005} 
including both thin and thick disk stars;
{\it iii)} La and Eu abundances of field dwarf and giant stars 
provided by S04 (159 objects);
{\it iv)} Y abundances of field dwarfs analyzed by
\citet[][hereinafter E93, 157 objects]{Edvardssonetal1993};
{\it v)} Y, La, Nd, and Eu abundances of 47 super-metal-rich 
field dwarfs by FG98; and
{\it vi)} Y abundances of three RSG stars in the Scutum cluster
measured by O13.

\begin{figure}
\centering
\resizebox{\hsize}{!}{\includegraphics{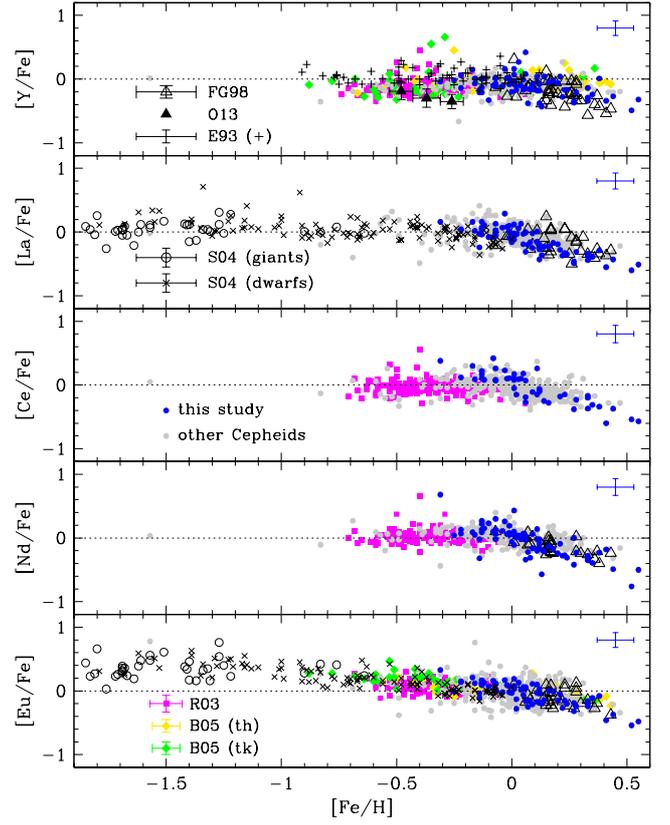}}
\caption{Abundances of neutron-capture elements as a function of the metallicity. Cepheid stars (filled circles) are compared with field dwarfs from the thin disk by R03 (magenta squares), from the thin (yellow diamonds) and thick (green diamonds) disks analyzed by \citet[][B05]{Bensbyetal2005}, with field dwarfs (crosses) and giants (open circles) by S04, and with field dwarfs analyzed by \citet[][E93, pluses]{Edvardssonetal1993} and by \citet[][FG98, open triangles]{FeltzingGustafsson1998}. RSG in the Scutum cluster by O13 (filled triangles) are also shown. The dotted lines display the positions of solar abundance ratios.}
\label{xfe_feh_lit_rs}
\end{figure}

Figure~\ref{xfe_feh_lit_rs} shows the comparison between Cepheid 
[neutron-capture/Fe] abundance ratios with the quoted data sets. 
Note that we applied a shift in the abundances by FG98, R03, and B05 in 
order to put them in the same scale of our data at solar metallicity. Data 
plotted in this figure show that the agreement between Cepheids and both 
field dwarfs and field giants in the Galactic disk is quite good over the 
entire metallicity range covered by the above samples. The trends are flat 
across solar iron abundances and display a modest abundance dispersion. 
Moreover, there is a clear decrease in the [element/Fe] ratios in the super 
metal-rich regime ([Fe/H] > 0.2). Thus suggesting a significant 
contribution in this iron regime from SNe type~Ia ejecta.  

The [Y/Fe] ratios has, once again, a different trend: it is underabundant 
and almost constant over the entire metallicity range (see the top panel 
of Fig.~\ref{xfe_feh_lit_rs}). This trend is supported by field dwarfs 
available in the literature, though for the super-metal-rich stars provided 
by FG98 the trend seems to be slightly steeper. The three RSGs observed by 
O13 also appear, within the errors, similar to the other field disk stars.

The [La/Fe] abundance ratio shows a steady enhancement when moving
from the metal-rich into the metal-poor regime. M13 suggested that
this trend is mainly caused by the metallicity dependence in the
production of the neutron-capture s-process elements \citep[see
also][]{Gallinoetal1998,Bussoetal1999}. The above ratio approaches
solar values for [Fe/H] $\sim$ $-$0.2 and attains a constant
value in the more metal-poor regime, thus suggesting no dependence
on iron in this metallicity range. The trend in the metal-rich
regime ([Fe/H] $\ge$ 0) is similar to those for Ce and Nd,
i.e., it is roughly 0.5~dex underabundant for [Fe/H] $\sim$
0.5. The quoted trend is quite evident for La, Ce, and Nd as
well as for Eu. The Y in the metal-rich regime shows a similar
trend, but the underabundance is milder when compared with the
above elements.

The [Eu/Fe] abundance ratio shows a different trend. The enhancement 
is steady over the entire metallicity range covered by Cepheids and 
by S04 sample. This evidence suggests a strong anticorrelation with 
iron concerning the Eu production. It is worth mentioning that data 
plotted in the bottom panel of this figure further support 
the contribution of SNe type~Ia to the iron abundance. Indeed, the 
steady decrease in [Eu/Fe] abundance ratio can be explained as a 
steady increase in iron abundance and a marginal, if any, production 
of Eu. This trend fully supports early results from FG98 for field 
super-metal-rich dwarfs.       

Finally, we also note that the spread in s-elements and in Eu is 
constant over the metallicity range covered by the current samples. 
There is solid empirical evidence that the spread in Eu increases 
in the most metal-poor regime, for [Fe/H] $\lesssim$ $-$2.0 
\citep{Cescuttietal2006}, but we still lack a detailed quantitative 
explanation of the observed trend.

\subsection{The [La/Eu] and [Y/Eu] abundance ratios}
\label{la_eu_ratios}

\begin{figure*}
\centering
\begin{minipage}[t]{0.33\textwidth}
\centering
\resizebox{\hsize}{!}{\includegraphics{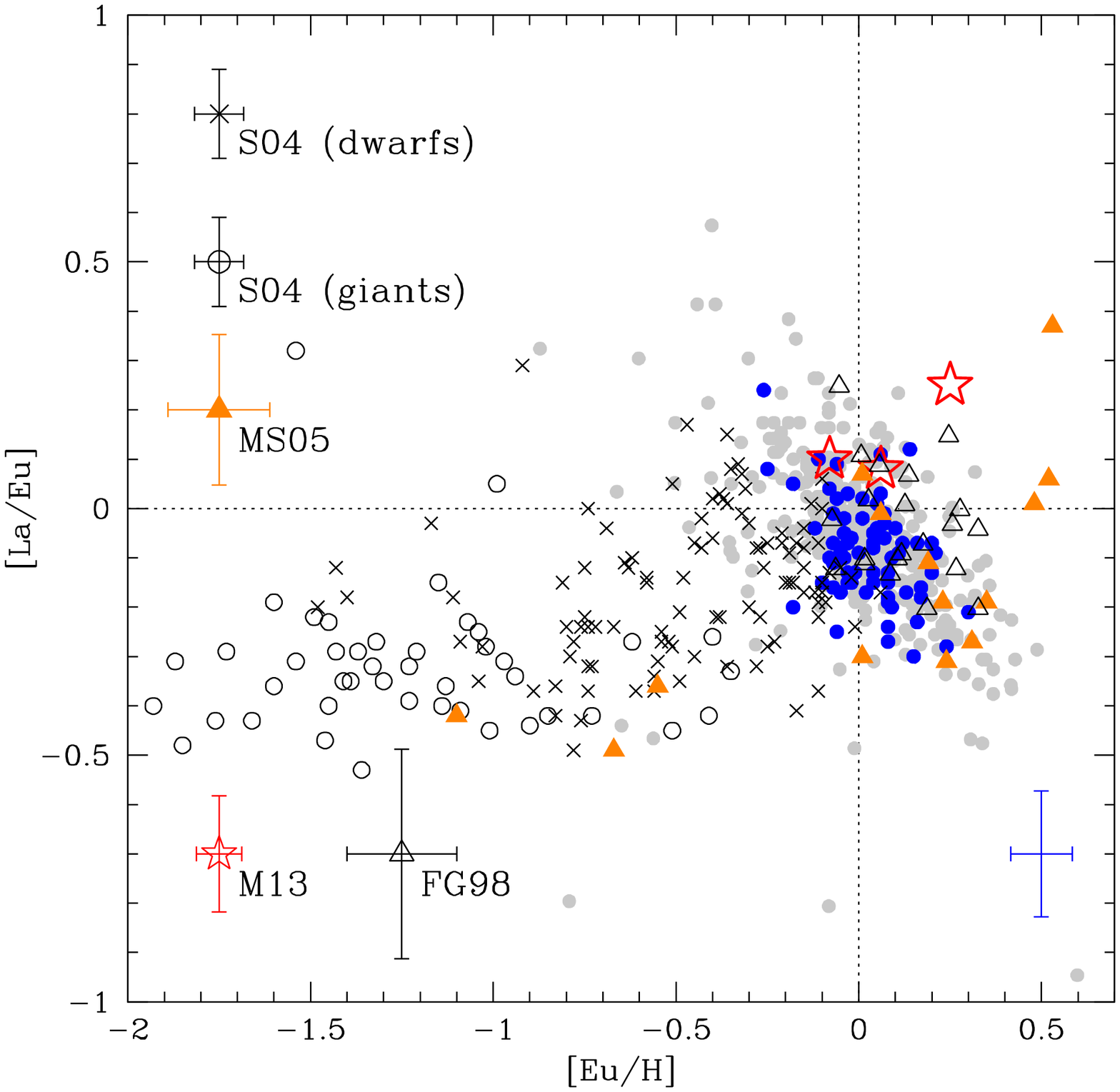}}
\end{minipage}
\begin{minipage}[t]{0.33\textwidth}
\centering
\resizebox{\hsize}{!}{\includegraphics{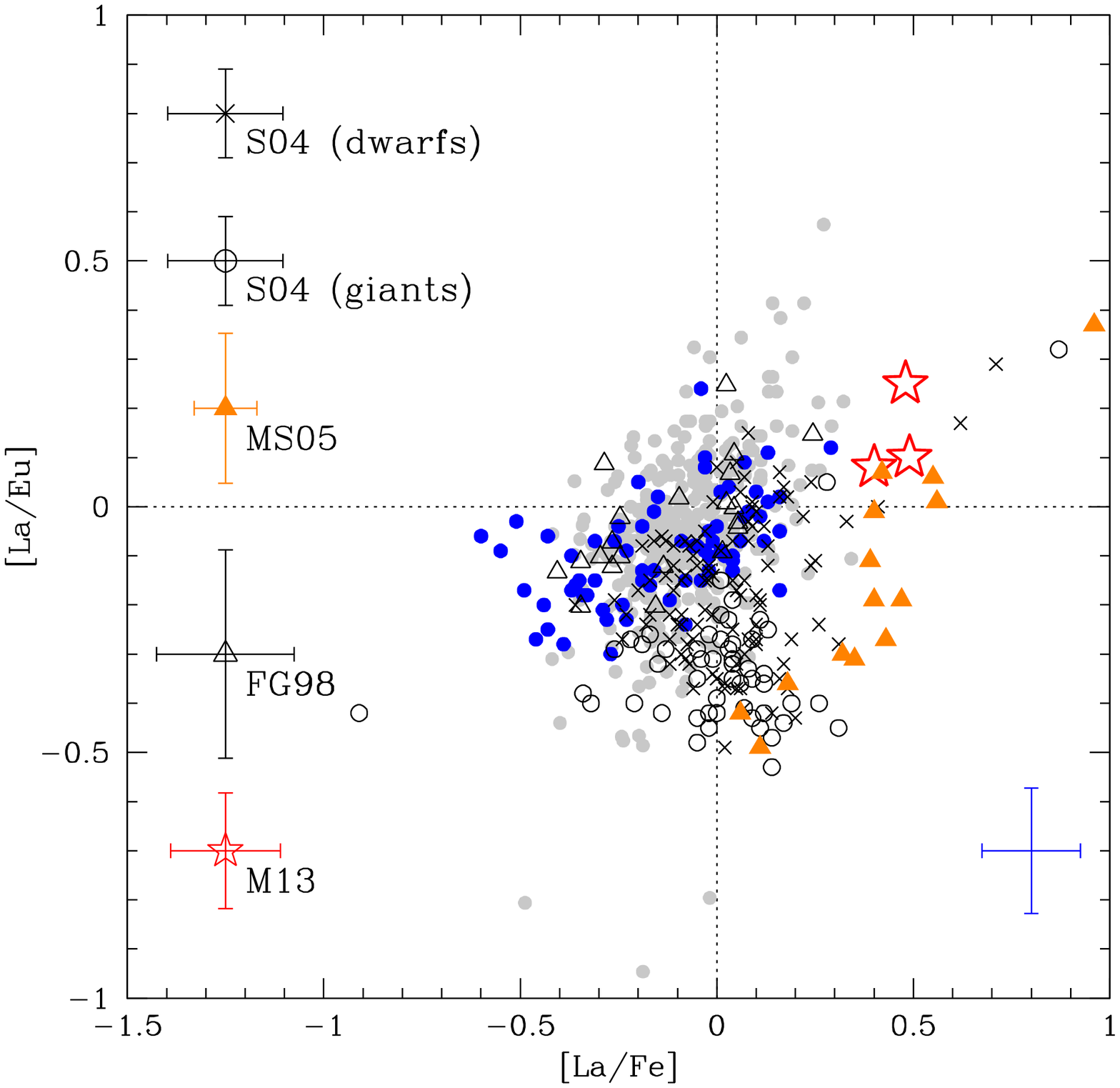}}
\end{minipage}
\begin{minipage}[t]{0.33\textwidth}
\centering
\resizebox{\hsize}{!}{\includegraphics{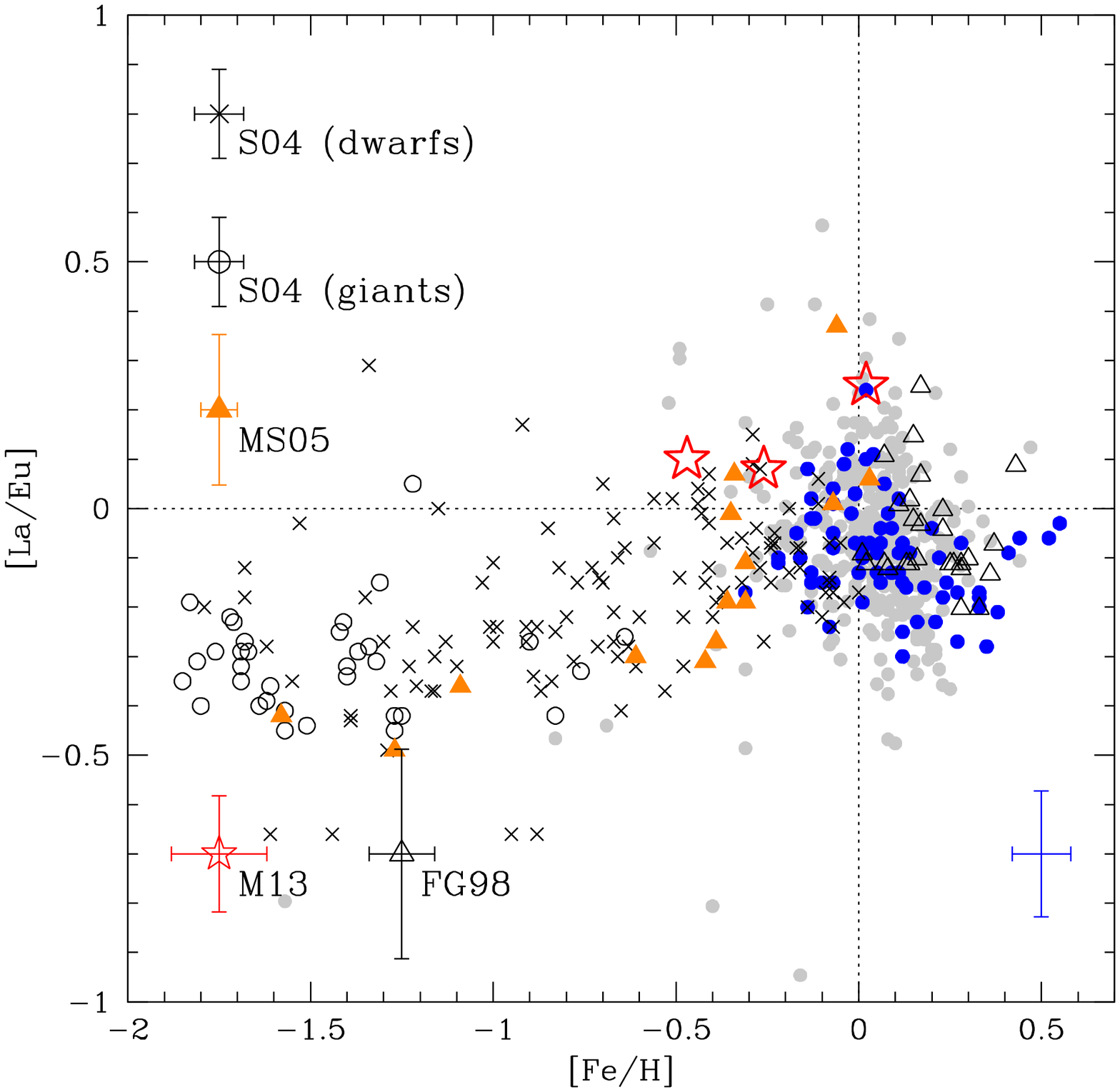}}
\end{minipage}
\caption{Abundance ratios between La and Eu as a function of [Eu/H]
({\it left panel}), [La/Fe] ({\it middle panel}), and [Fe/H] ({\it right
panel}). Cepheid stars (filled circles) are compared with Galactic field
dwarfs (crosses) and giants (open circles) by S04, with field dwarfs by FG98 
(open triangles) and with Sgr field stars studied by \citet[][MS05, filled 
triangles]{McWilliamSmeckerHane2005} and by
\citet[][M13, red stars]{McWilliametal2013}.}
\label{laeu}
\end{figure*}

The left panel of Fig.~\ref{laeu} shows the s to r abundance ratio
[La/Eu] vs. the [Eu/H] abundance. Note that the abundances derived by S04, 
\citet[][hereinafter MS05]{McWilliamSmeckerHane2005}, and M13 were only 
rescaled to take account of the solar mixture adopted in the current 
investigation \citep{Grevesseetal1996}. Note also that we are plotting the 
[La/Eu] ratio versus the [Eu/H] abundance to separate the role played by 
pure explosive nucleosynthesis of iron in SNe type~Ia and type~II from the 
neutron capture enrichments. The Cepheids in this plane show a well defined 
anticorrelation. The [La/Eu] ratio, when moving from the most Eu-rich to the 
most Eu-poor stars, increases by almost one dex. The Cepheid abundance ratio 
becomes even more compelling in the comparison with field giant and dwarf 
stars provided by S04. The latter sample shows an almost constant ratio over 
a broad range in Eu abundances and a mild increase in the approach to solar 
Eu abundances. The [La/Eu] abundance ratios provided by M13 and by MS05 for 
Sagittarius RGs show a similar distribution, but three stars display large 
Eu abundances. Two out of the three display a solar ratio, while the third 
one is 0.4~dex enhanced in La.

Data plotted in the middle panel of Fig.~\ref{laeu} shows the same abundance 
ratios, but versus the [La/Fe] ratio. The empirical scenario becomes more 
clear, and indeed we found that RGs in Sagittarius are systematically more 
enhanced in La when compared with Galactic thin disk stellar population. 
Indeed, only a few Sagittarius stars are located in the same region covered 
by thin disk stars. In passing we note that a similar enhancement in La has 
also been found in several dwarf spheroidal galaxies 
\citep{Shetroneetal2003,Geisleretal2005,Pompeiaetal2008,Letarteetal2010, Lemasleetal2014}.
Thus suggesting that the above plane is a good diagnostic to identify relic 
stars of dwarf galaxies that have been accreted by the Milky Way.

The right panel of Fig.~\ref{laeu} shows the same abundance ratio, but versus the iron abundance. The trend is quite similar to the left panel of the same figure. However, Cepheids and field stars display, at fixed [La/Eu], a larger spread in iron. The RGs in Sagittarius (MS05, M13) display a trend similar to the Galactic stars, and for [Fe/H] > $-$0.3 it is also similar to Galactic Cepheids. The above evidence indicates that s- and r-process elements in the Galactic thin disk have similar enrichment histories in the metal-poor regime ([Fe/H] $\le$ $-$0.3). The same ratio shows, in the more metal-rich regime, a well defined anticorrelation with iron and with Eu abundances.

\begin{figure*}
\centering
\begin{minipage}[t]{0.33\textwidth}
\centering
\resizebox{\hsize}{!}{\includegraphics{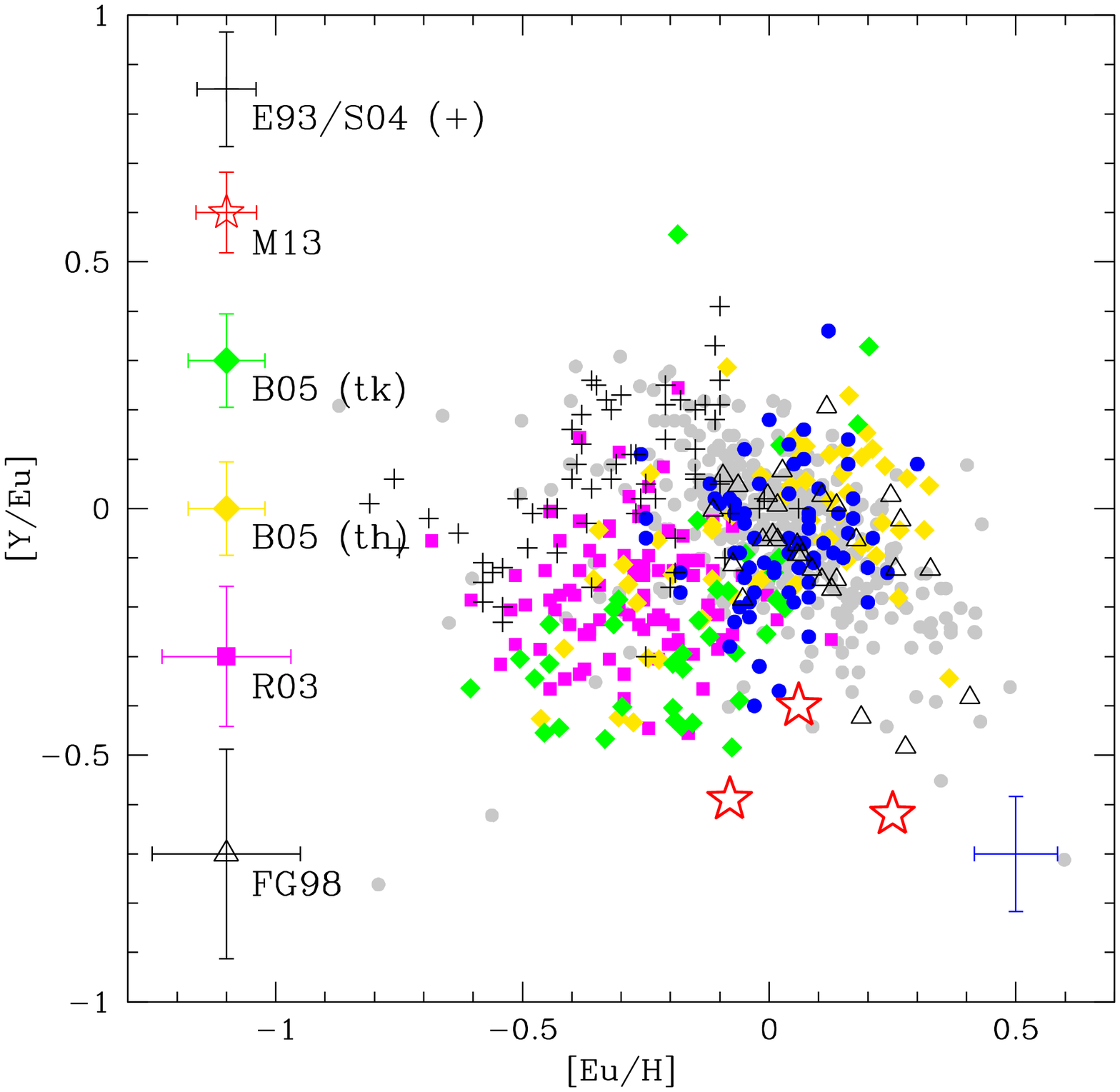}}
\end{minipage}
\begin{minipage}[t]{0.33\textwidth}
\centering
\resizebox{\hsize}{!}{\includegraphics{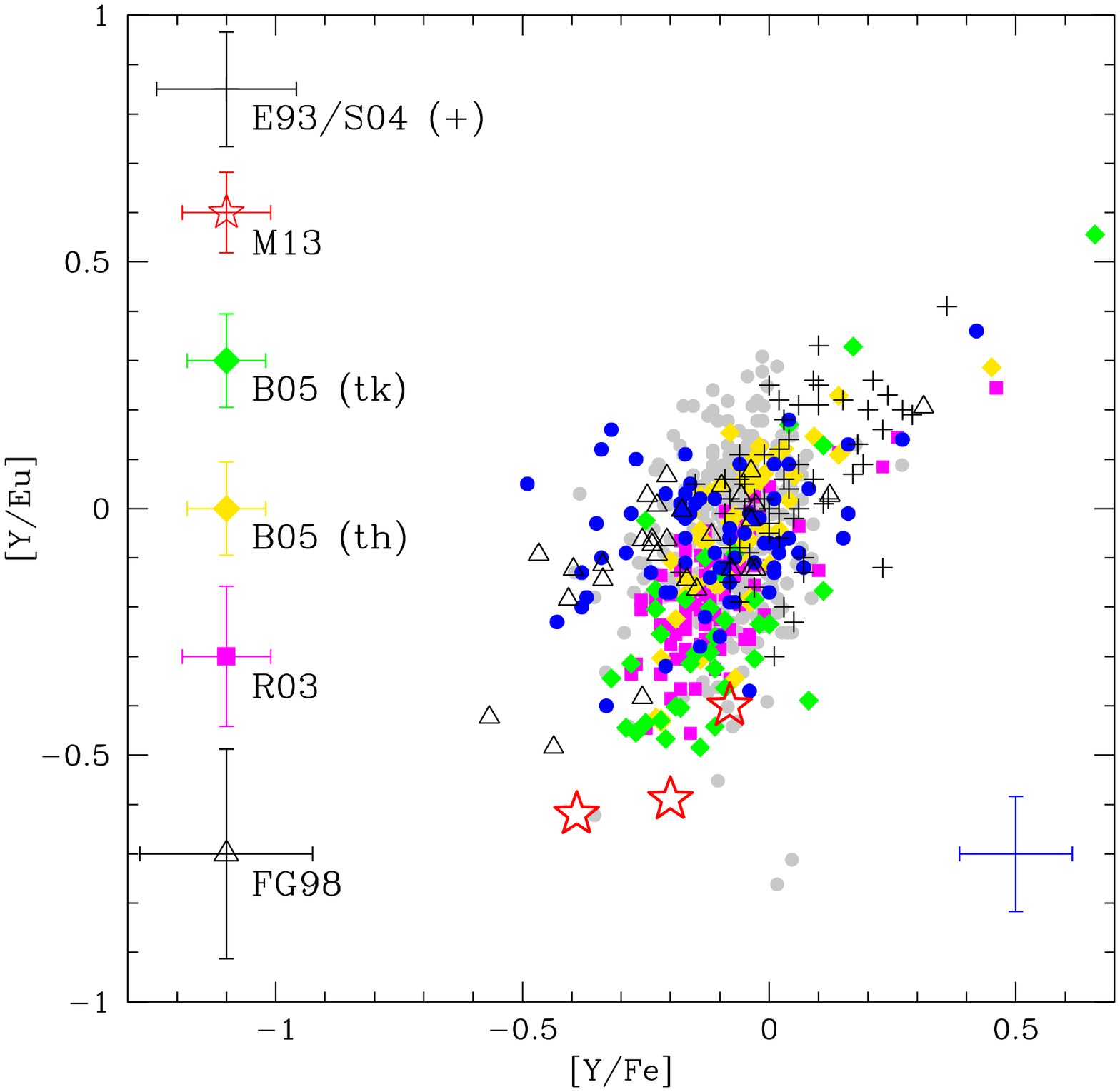}}
\end{minipage}
\begin{minipage}[t]{0.33\textwidth}
\centering
\resizebox{\hsize}{!}{\includegraphics{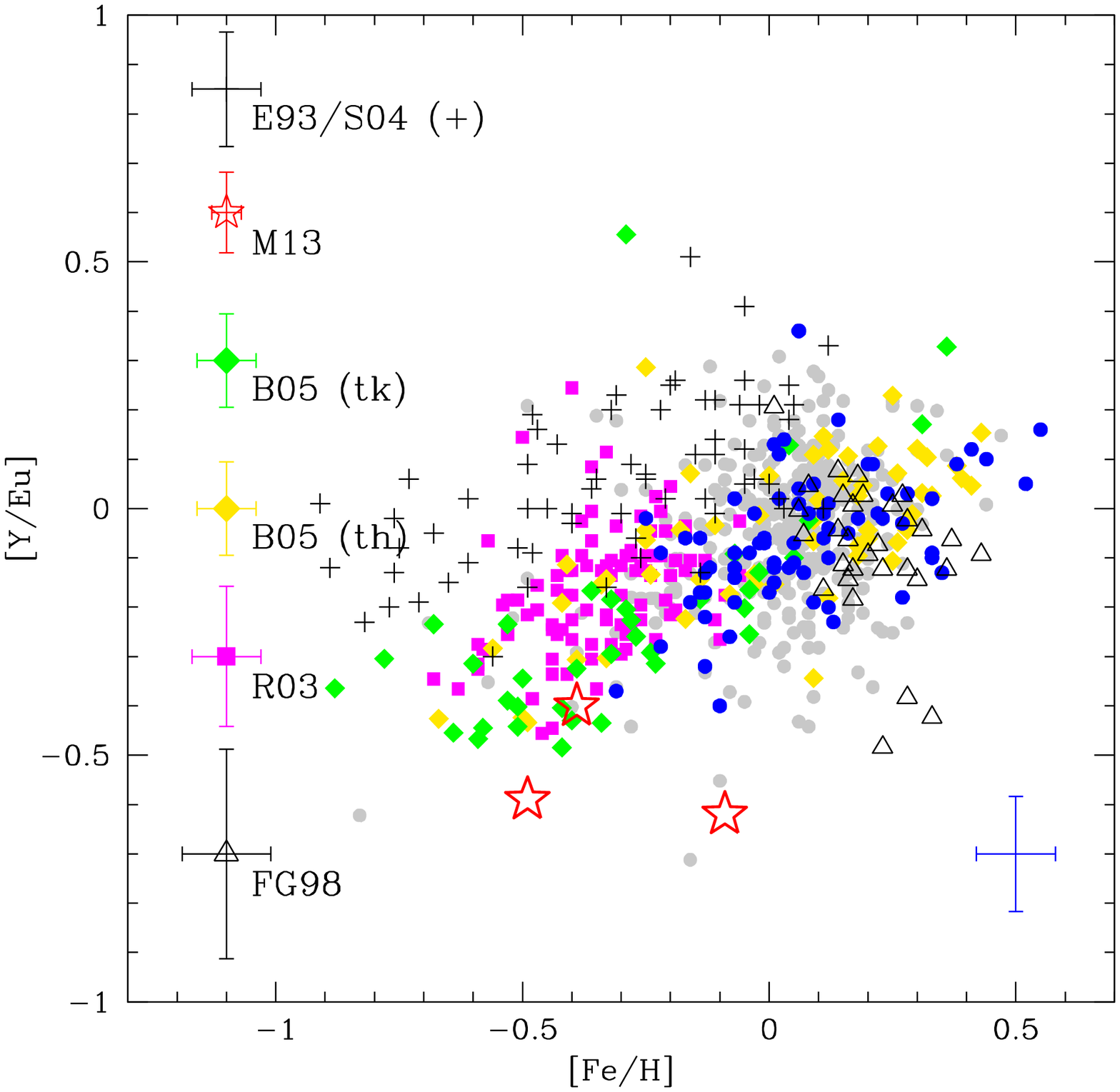}}
\end{minipage}
\caption{The same as in Fig.\ref{laeu}, but for Y and Eu ratios. Galactic
field dwarfs by E93/S04 (pluses), field dwarfs from the thin disk by R03
(magenta squares), and field dwarfs from the thin (yellow diamonds) and
thick (green diamonds) disks by B05 are also shown.}
\label{yeu}
\end{figure*}

To further constrain the ratio between s- and r- process elements we
also investigate the abundance ratio between a light s-element (Y) and Eu.
The left panel of Fig.~\ref{yeu} shows [Y/Eu] versus [Eu/H] plane. The
distribution of Cepheids in this plane is quite different than in the
[La/Eu] versus [Eu/H] plane. Indeed, Cepheids show a larger dispersion 
over the entire metallicity range they cover, and there is no clear
evidence of an anticorrelation with the europium content. On the other hand,
field stars show a mild evidence of a correlation with Eu abundance when
moving from the Eu-intermediate into the more Eu-rich regime.

Data plotted in the middle panel of Fig.~\ref{yeu} shows a well defined
correlation between [Y/Eu] and [Y/Fe]. This finding together with the
constant value of the [Y/Fe] as a function of both Cepheid ages and iron
abundance, is suggesting a different enrichment history between Y and Eu,
but also a difference between light (Y) and heavy (La) s-process elements.
Note that the three Sagittarius RGs attain in this plane the lowest values,
thus suggesting that they are quite Y poor when compared with field Galactic
stars.

The right panel of Fig.~\ref{yeu} shows the same data, but in the [Y/Eu]
versus iron abundance. The bulk of the data seems to suggest a correlation
between the s to r abundance ratio and iron content. In this context it is
worth mentioning that for iron abundances more metal-poor than the sun there
is a mild evidence of a possible dichotomous distribution. In particular,
field dwarf stars associated with the Galactic thick disk
provided by B05 display, at fixed iron content, lower [Y/Eu] abundance
ratios. The difference with similar abundances provided by E93 plus S04 and
by R03 is slightly larger than one sigma and needs to be further
investigated with larger homogeneous sample.

\subsection{The [La/Y] abundance ratio}
\label{la_y_ratios}

\begin{figure*}
\centering
\begin{minipage}[t]{0.33\textwidth}
\centering
\resizebox{\hsize}{!}{\includegraphics{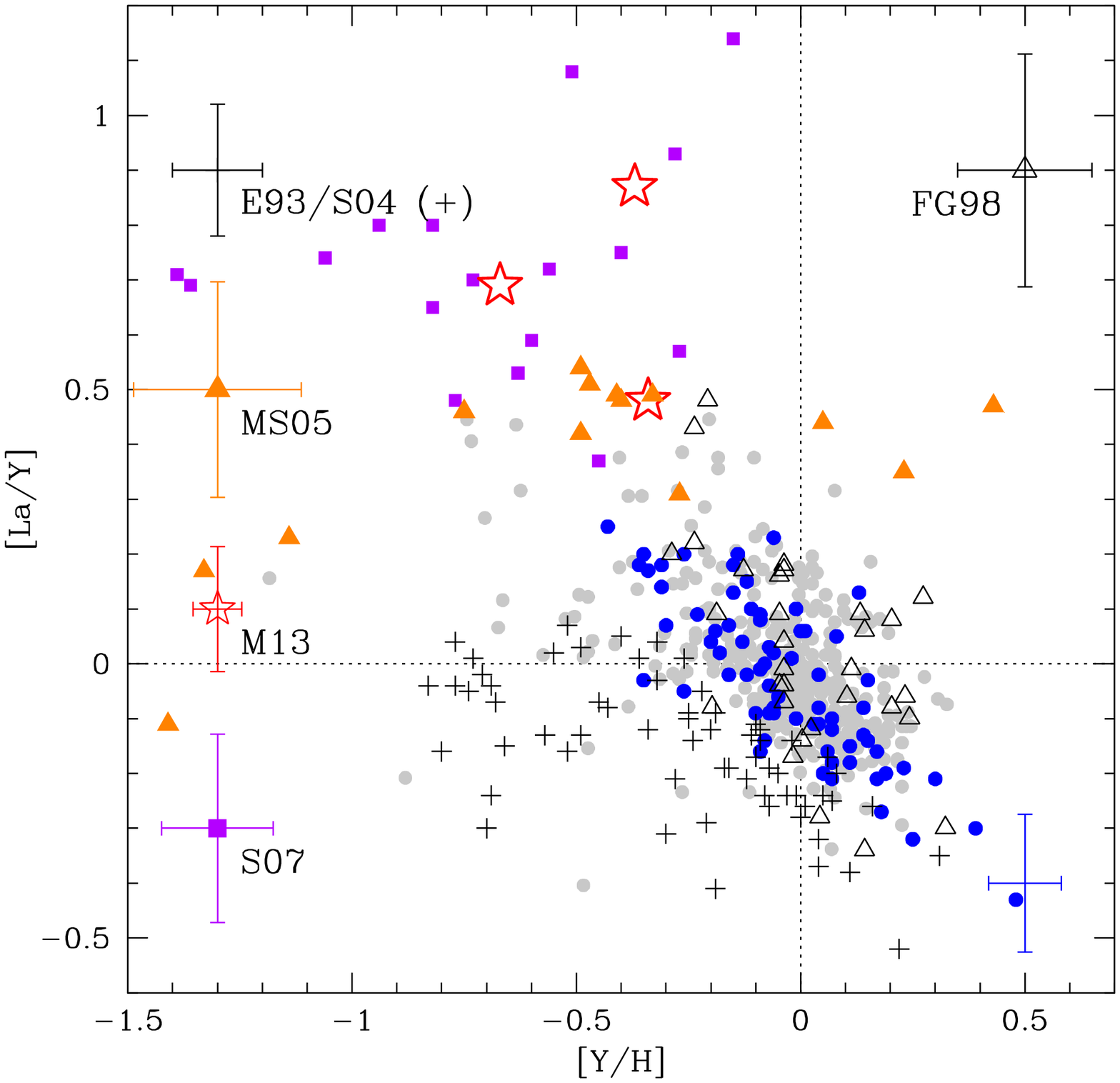}}
\end{minipage}
\begin{minipage}[t]{0.33\textwidth}
\centering
\resizebox{\hsize}{!}{\includegraphics{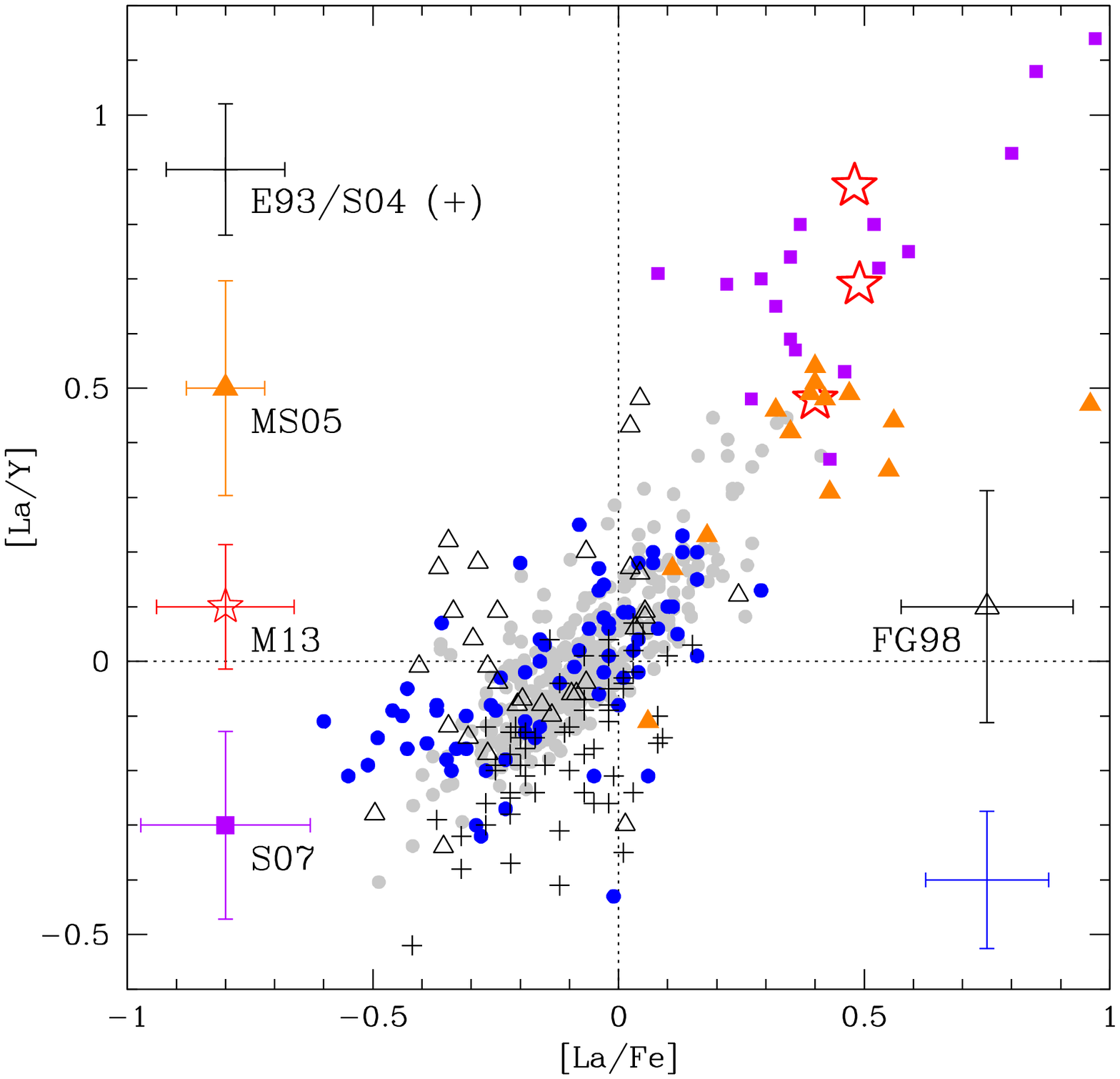}}
\end{minipage}
\begin{minipage}[t]{0.33\textwidth}
\centering
\resizebox{\hsize}{!}{\includegraphics{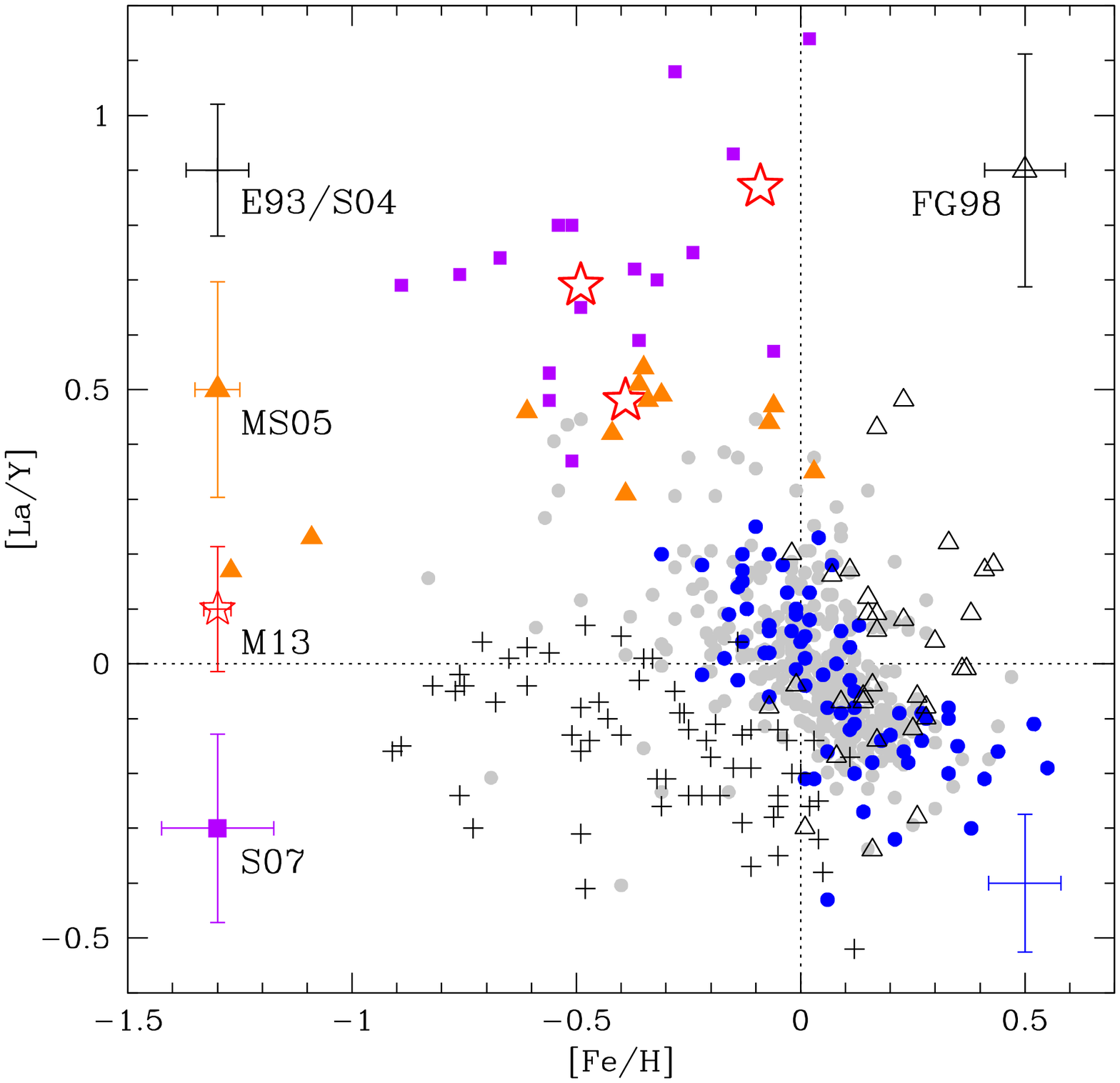}}
\end{minipage}
\caption{The same as in Fig.~\ref{laeu}, but for La and Y ratios.
Sgr stars studied by \citet[][S07, violet squares]{Sbordoneetal2007} are
also plotted.}
\label{lay}
\end{figure*}

Figure~\ref{lay} shows the ratio between a heavy (La) and
a light (Y) neutron-capture element. Such a ratio is a good 
diagnostic for the s-process index [hs/ls], i.e., the ratio 
between the heavy s-process elements and the light ones.  
The quoted ratio and its dependence on the metallicity are solid tracers 
of the role played by AGB stars in the chemical enrichment
\citep{Gallinoetal1998,Bussoetal1999,Bussoetal2001,Cristalloetal2009}.
The production of hs elements (such as La) is favored in the metal-
intermediate regime ([Fe/H] $\sim$ $-$0.6), while in the more metal-rich 
regime ls elements (such as Y) are mostly synthesized. Therefore, the ratio 
[hs/ls] is expected to be under-abundant in the metal-rich regime and 
enhanced in the more metal-poor regime. Our Cepheid data in the left panel 
of this figure display, in agreement with theoretical predictions
\citep{Cristalloetal2009,Cristalloetal2011,Cristalloetal2015b}, a
well defined anticorrelation between the [La/Y] ratio and the Y
abundance. The Galactic field stars measured by E93 and S04
display a large spread, at fixed Y abundance, but the trend is
similar. Interestingly enough, we found that Sagittarius RGs
-- provided by MS05, M13, and \citet[][hereinafter
S07]{Sbordoneetal2007} -- display two distinctive features:
{\it i)} a strong enhancement in La, with a marginal overlap with Galactic
stars; and
{\it ii)} the spread in [La/Y] abundance ratio is significantly
larger than Galactic stars. Thus suggesting that the enrichment 
of neutron-capture elements in Sagittarius is more complex than 
in the thin disk.

The trend of the data plotted in the left panel of Fig.~\ref{lay}
becomes even more clear in the [La/Y] versus [La/Fe] plane
(middle panel). Galactic stars and Sagittarius RGs display,
at fixed [La/Fe] abundance, a smaller spread in [La/Y] abundances.
Moreover, the separation between Sagittarius and Galactic stars
becomes even more solid. Indeed, only the three most metal-poor
objects in the Sagittarius sample overlap with Galactic thin disk
stars. This finding indicates a strong correlation between Y and
Fe over the entire metallicity range.

The right panel of Fig.~\ref{lay} shows the same data, but they are plotted 
as a function of the iron abundance. The distribution in this plane is
quite similar to the left panel, but with a larger spread in iron
abundances. In passing, we note that the separation between
Galactic and Sagittarius stars might be even more compelling than
suggested by current data. The 17 Cepheids with [La/Y] > 0.3
come from the LIII sample. The authors did not take account of the
hyperfine structure, moreover, 15 out of the 17 are located in the
1st quadrant and at Galactocentric distance larger than 9~kpc.
Cepheids in the outer disk will play a crucial role to further
constrain the use of the quoted chemical diagnostics to separate
Galactic and dwarf galaxy stars.

In this context it is worth mentioning that the star Sgr\,247 from the
M13 sample lies off the main trend in the left and in the right panel of
Fig.~\ref{lay}. The peculiar position of this object shows also up in the 
left and in the right panel of Fig.~\ref{yeu}. The current findings further 
support the results by M13 suggesting that this object was polluted by more 
metal-poor ([Fe/H] ranging from about $-$0.5 to about $-$1.0~dex) AGB 
ejecta. The AGB yields in this object were polluted but less than similar 
Sagittarius stars. This hypothesis is further supported by the evidence that 
the same object follow the main trend in the [hs/ls] and in the [s/r] 
abundance ratios (middle panels of Fig.~\ref{yeu} and Fig~\ref{lay}).  

\subsection{Comparison between predicted and observed [hs/ls] s-process index}

\begin{figure}
\centering
\resizebox{\hsize}{!}{\includegraphics{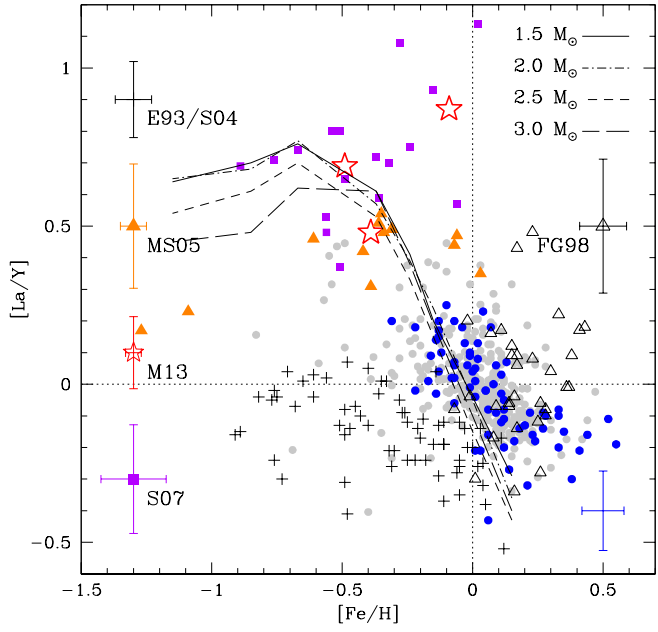}}
\caption{The same as in the {\it left panel} of Fig.~\ref{lay}, but
comparing the observational data with theoretical models 
available on the FRUITY database \citep{Cristalloetal2011,Cristalloetal2015b}.}
\label{lay_models}
\end{figure}

To further constrain the difference between the s-process index [hs/ls] in 
Cepheids with field Galactic stars and in nearby dwarf galaxies, 
Fig.~\ref{lay_models} shows the comparison between theoretical and observed 
[La/Y] as a function of iron content. The black lines display predicted 
final surface abundances for four low-mass (see labeled values) AGB models
available on the FRUITY database\footnote{fruity.oa-teramo.inaf.it}
\citet{Cristalloetal2011,Cristalloetal2015b}. The symbols and error bars for 
the data are the same as in the right panel of Fig.~\ref{lay}. The 
comparison brings forward several interesting new findings:

{\it i)} The agreement between theory and observations is quite good
over the entire metallicity range covered by Cepheids. In this context it
is worth mentioning that Cepheids offer a new opportunity to validate
the  [hs/ls] s-process index. Theoretical predictions are validated
using a broad range of s-enhanced stars -- O-rich and C-rich AGB stars,
post-AGB stars, Ba-rich stars and CH-rich stars -- for which the
evolutionary status is not well established. The advantage in using
Cepheids is that they do belong to the first stellar generation
formed after the recent enrichment of the interstellar medium. The
current comparison between theory and observations should be cautiously
treated because we are not accounting for dilution effects and for
detailed chemical evolution models.

However, it is worth mentioning that the above comparison was
performed overplotting predicted abundances on top of the observed
values. This means that once corrected for the adopted solar
abundances, we did not apply any shift in the predicted
abundances. Data plotted in this figure suggest that predicted
[La/Y] abundances display, in the metal-rich regime a spread that
is systematically smaller than observed. The reader interested in
a detailed discussion concerning the theoretical parameters
affecting the spread of the above s-process index is referred to
\citet{Cristalloetal2015a} (see also \citealt{Piersantietal2013}). In
this context we would like to stress the similarity in the slope
when moving from the metal-rich to the metal--poor regime of Galactic
Cepheids. The current empirical uncertainties  do not allow us to
constrain whether field dwarf stars provided by E93
and S04 do show a shallower slope when compared with Cepheids.

{\it ii)} Theory and observation display a steady increase in
[La/Y] when moving from the metal-rich into the metal-intermediate
regime, i.e., [Fe/H] $\sim$ $-$0.4/$-$0.7. The [La/Y]
abundances, as expected, decrease in the metal-poor regime
\citep{Cristalloetal2009}. There is a group of Sagittarius stars
showing [La/Y] abundances larger -- s-process enhanced -- than
predicted by AGB models and they have already been discussed by
M13. A similar discrepancy has also been found in CEMP stars at 
very low metallicities, which attain values of s-process
index of the order of +1.3~dex \citep{SpiteSpite2014,Beersetal2005}.
The lack of a sizable sample of Cepheids in the metal-intermediate
regime do not allow us to provide independent constraints on the
possible mismatch between predicted and observed [hs/ls] abundance
ratios. In passing we note that \citet{Misheninaetal2015}, in a 
recent investigation of more than two dozen of giant stars in 
five Galactic open clusters, found
solid evidence of [Ba/Fe] and [Ba/La] enhancement. They suggested
that the quoted empirical evidence might be explained assuming a
significant contribution from non standard s-process, i.e., the
intermediate neutron-capture process suggested by
\citet{CowanRose1977}.

To further validate the plausibility of the adopted theoretical framework 
for the production of s-process elements from AGB stars, we performed 
a plain test to constrain the slope of [Y/H] versus the Galactocentric distance. We performed a linear fit of the Cepheids plotted in
Fig.~\ref{lay_models} (i.e. [La/Y] vs. [Fe/H]). To overcome the increase in 
the spread in the more metal-poor and in the more metal-rich regime we 
selected the objects with iron abundances included between $-$0.3 and 
+0.3~dex. The current fit was combined with the analytical fits for [Fe/H] 
and [La/H] as a function of Galactocentric distance. We found that the 
expected slope for [Y/H] as a function of Galactocentric distance is quite 
similar to the observed slope ($-0.052$ vs. $-0.053$~\dexkpc, respectively). 
This evidence indicates that s-process elements predicted by AGB models take 
account for the observed slopes among the investigated elements.

%
%
\section{Summary and final remarks}
\label{concl}

This is the 10th of a series of papers focussed on the metallicity distribution of the Galactic thin disk using classical Cepheids as stellar tracers. The project (DIsk Optical Near-infrared Young Stellar Object Spectroscopy, DIONYSOS) is aimed at providing homogeneous and accurate elemental abundances and distances for a significant fraction of the known Galactic Cepheids.

In this investigation we present accurate and homogeneous measurements of
five neutron capture elements (Y, La, Ce, Nd, Eu) for 73 Galactic classical
Cepheids. The current abundances are based on high-spectral resolution
($R$ $\sim$ 38\,000) and high signal-to-noise ratio (S/N $\sim$ 50-300)
spectra collected with UVES at ESO VLT. They were derived by accounting for
the HFS of some lines of Y, La, and Eu, for which atomic data
are available in the literature.
The iron, $\alpha$ plus Na and Al abundances of the same Cepheids have
already been discussed in \citet{Genovalietal2013,Genovalietal2014,
Genovalietal2015}. Our Cepheids are representative of the Galactic sample,
and indeed they cover a broad range in pulsation periods
(0.36 $\leq$ \logP\ $\leq$ $\sim$ 1.54) and in Galactocentric distances
(4.6 $\leq$ \RG\ $\leq$ 14.3~kpc).

We also selected similar abundances for Galactic Cepheids available in the literature and we ended up with homogenous measurements for 435 Galactic Cepheids. Roughly one third of the entire sample have measurements provided by our group (current plus LEM), while the others come from LII, LIII, and YON. The different samples have from one to 4 dozen of Cepheids in common, which allowed us to provide homogeneous abundance scales for the quoted five elements plus iron and $\alpha$ elements (G14, G15).

The individual distances for the entire Cepheid sample are based on homogeneous NIR photometry, transformed into the 2MASS photometric system, and on the Period-Wesenheit relations provided by \citet{Innoetal2013}. The main findings of the current analysis are the following:

{\it i) [element/H] radial gradients:} The investigated neutron
capture elements display well defined radial gradients. The slopes
for four (La, Ce, Nd, Eu) out of the five elements are quite
average ($-$0.025\,$\pm$\,0.004~\dexkpc). The Y slope is more than
a factor of two steeper and more similar to the slopes of iron and
$\alpha$ elements. The current estimates agree quite well with
similar radial gradients available in the literature. However, we
provide firm constraints concerning the Nd gradient for which it
was suggested a flat distribution when moving from the inner to
the outer disk. Moreover, the difference in the slope between Y
and the other three s-process elements (La, Ce, Nd) brings forward
a more complex enrichment history for this element.

{\it ii) Comparison with theory:}
The comparison with radial gradients predicted by chemical evolution models provided by \citet{Cescuttietal2006,Cescuttietal2007}, indicates a very good agreement for the slopes of both La and Eu.

{\it iii) Comparison with observations:}
The comparison with similar abundances for field thin and thick dwarf and 
giant stars provided by E93, FG98, R03, S04, B05, and O13 indicates a 
very good agreement over the Galactocentric distances covered 
by the quoted samples.

{\it iv) Age dependence:}
We took advantage of the tight anticorrelation between pulsation period 
and age to constrain the age dependence of the investigated elements. 
We found that the slopes are positive, i.e., they are more abundant 
in young (a few tens of Myrs) than in old ($\sim$300~Myr) Cepheids. 
However, the slopes of La, Ce, and Nd are shallower than for iron, 
$\alpha$ elements, and light elements, while for Y and Eu are more similar.

{\it v) [element/Fe] radial gradients:} We found that three
s-process dominated elements (La, Ce, Nd) and one
r-process dominated element (Eu) display slopes that are
on average a factor of two larger than similar slopes of the
$\alpha$ and light elements investigated by G15. The slope of Y is
once again an exception, and indeed this element shows a flat
distribution across the entire disk. The quoted trends are the
consequence of the difference/similarity with the iron radial
gradient. 

{\it vi) [element/Fe] abundance ratios in the super-metal-rich-regime:}
We found that s- and r-process abundance ratios display a steady decrease 
for iron abundances larger than solar. The change in the slope indicates 
a clear contribution from SNe type~Ia ejects. The trend in the [Eu/Fe] 
abundance ratio as a function of iron abundance further supports the 
above hypothesis with a steady decrease in the slope when moving from 
[Fe/H] $\sim$ $-$1.2 to [Fe/H] $\sim$ $-$0.5. The current findings support 
previous results for super-metal-rich field dwarfs by FG98.     

{\it vii) Spatial and temporal homogeneity:}
The reduced scatter in the above radial gradients at fixed 
Galactocentric distance and the lack of well defined slopes 
for [element/Fe] as a function of the pulsation period (but Ce) 
is indicating that the chemical enrichment across the Galactic 
thin disk is characterized by firm spatial and temporal homogeneity.

{\it viii) s to r abundance ratio:}
We found that Cepheid [La/Eu] abundance ratios show a well defined
anticorrelation when plotted as a function of Eu and Fe abundances. Field
stars display a different trend. Indeed, they attain an almost constant 
ratio in the metal-poor regime and only for [Eu/H] and [Fe/H] larger than
$\sim$$-$0.5~dex show a mild enhancement in La. The light s- to r-process
element abundance ratio ([Y/Eu]) shows a different trend. The Cepheids do
not show a clear anticorrelation with [Y/H] and with [Y/Fe]. On the other
hand, field stars display a correlation with both Y and iron. Moreover,
[Y/Eu] shows a well defined correlation with iron abundance. This trend 
appears as the consequence of the strong correlation between Y and iron 
abundances.

{\it ix) Heavy to light s element abundance ratio:} We found
that Cepheid [La/Y] abundance ratios show a strong anticorrelation
when plotted as a function of Y and Fe abundances. Field Galactic
stars display the same trend, thus supporting the metallicity
dependence of heavy (La, Ce, Nd) and light (Y) s-process elements
on the metal content
\citep{Cristalloetal2009,Cristalloetal2011,Cristalloetal2015b}.
Moreover, we also found that the dispersion in [La/Y] as a
function of [La/Fe] is small among Galactic and Sagittarius stars,
further supporting similarity in the origin of Fe and Y.
Interestingly enough, we also found that in the quoted planes, in
particular, in the [La/Y] vs. [La/Fe] one, the Sagittarius RGs are
well separated by Galactic stars, due to their La enhancement.
Thus suggesting that they can be adopted as solid diagnostics to
identify relic stars of dwarf galaxies accreted  by our galaxies.

{\it x) Comparison between predicted and observed s-process index:}
We performed a detailed comparison between predicted and observed
s-process index [La/Y]. We found that final surface abundances of
low-mass (1.5 $\le$ $M/M_\odot$ $\le$ 3.0) AGB stars agree quite well over
the entire metallicity range covered by the current sample of
classical Cepheids.

\begin{acknowledgements}
This work was partially supported by PRIN-MIUR (2010LY5N2T)
``Chemical and dynamical evolution of the Milky Way and Local
Group galaxies'' (P.I.: F. Matteucci) and PRIN-MIUR (20128PCN59)
``Nucleosynthesis in AGB stars: an integrated approach'' project
(P.I.: L. Gialanella). We also acknowledge an anonymous referee for the 
positive opinions concerning this experiment and for the very pertinent 
suggestions that improved the content and the readability of the paper.
\end{acknowledgements}

\bibliographystyle{aa}
\bibliography{daSilvaetal2015_aph}

\clearpage
\clearpage
\clearpage
\clearpage
\clearpage

\include{daSilvaetal2015_table}

\end{document}

%% file: daSilvaetal2015_table.tex
\clearpage
\begin{table*}[p]
\centering
\caption[]{Abundances of heavy elements for our sample of classical
Cepheids derived based on individual spectra.}
\label{table_ab_heavy_spec}
{\scriptsize 
\begin{tabular}{rc r@{ }l c r@{ }l c r@{ }l c r@{ }l c r@{ }l c r@{ }l c}
\noalign{\smallskip}\hline\hline\noalign{\smallskip}
Name & MJD &
\multicolumn{2}{c}{[Fe/H]} & \parbox[c]{1.0cm}{\centering $N_{\rm L}$ (\ion{Fe}{i},\ion{Fe}{ii})} &
\multicolumn{2}{c}{[Y/H]} & $N_{\rm L}$ &
\multicolumn{2}{c}{[La/H]} & $N_{\rm L}$ &
\multicolumn{2}{c}{[Ce/H]} & $N_{\rm L}$ &
\multicolumn{2}{c}{[Nd/H]} & $N_{\rm L}$ &
\multicolumn{2}{c}{[Eu/H]} & $N_{\rm L}$ \\
\noalign{\smallskip}\hline\noalign{\smallskip}
 V340\,Ara & 56137.137 &    0.27 & $\pm$ 0.10 &  (23, 2) & $-$0.24 & $\pm$ 0.16 &      5 & $-$0.19 & $\pm$ 0.07 &      5 & $-$0.09 & $\pm$ 0.16 &      2 & $-$0.25 & $\pm$ 0.16 &      5 & $-$0.01 & $\pm$ 0.10 &      2 \\
 V340\,Ara & 54708.065 &    0.53 & $\pm$ 0.09 &  (53, 4) & $-$0.02 & $\pm$ 0.13 &      3 & $-$0.07 & $\pm$ 0.11 &      4 &    0.06 & $\pm$ 0.18 &      2 & $-$0.15 & $\pm$ 0.15 &      2 &    0.00 & $\pm$ 0.03 &      2 \\
 V340\,Ara & 54709.079 &    0.53 & $\pm$ 0.16 &  (26, 3) & $-$0.05 & $\pm$ 0.10 &      3 &    0.01 & $\pm$ 0.11 &      4 &    0.05 & $\pm$ 0.02 &      2 & $-$0.04 & $\pm$ 0.19 &      2 &    0.17 & $\pm$ 0.03 &      2 \\
 V340\,Ara & 56138.094 &    0.32 & $\pm$ 0.09 &  (41, 2) & $-$0.08 & $\pm$ 0.16 &      5 & $-$0.08 & $\pm$ 0.04 &      5 & $-$0.06 & $\pm$ 0.15 &      2 & $-$0.02 & $\pm$ 0.15 &      5 &    0.04 & $\pm$ 0.01 &      2 \\
 V340\,Ara & 56139.185 &    0.22 & $\pm$ 0.01 &  (51, 2) & $-$0.17 & $\pm$ 0.10 &      6 & $-$0.11 & $\pm$ 0.06 &      4 & $-$0.23 & $\pm$ 0.14 &      2 & $-$0.15 & $\pm$ 0.19 &      4 &    0.05 & $\pm$ 0.04 &      2 \\
 V340\,Ara & 56152.054 &    0.24 & $\pm$ 0.18 &  (15, 2) &    0.04 & $\pm$ 0.12 &      3 &    0.18 & $\pm$ 0.07 &      4 &    0.52 & $\pm$ 0.15 &      2 & $-$0.04 & $\pm$ 0.12 &      2 &    0.27 & $\pm$ 0.04 &      2 \\
   AS\,Aur & 54845.136 &    0.00 & $\pm$ 0.08 &  (74, 8) & $-$0.20 & $\pm$ 0.13 &      2 & $-$0.16 & $\pm$ 0.19 &      3 &	   & \ldots	& \ldots & $-$0.06 & $\pm$ 0.10 &      2 & $-$0.03 & $\pm$ 0.02 &      2 \\
   KN\,Cen & 54862.355 &    0.55 & $\pm$ 0.12 &  (14, 3) &    0.23 & $\pm$ 0.08 &      1 &    0.04 & $\pm$ 0.15 &      4 & $-$0.02 & $\pm$ 0.11 &      1 &    0.05 & $\pm$ 0.21 &      2 &    0.07 & $\pm$ 0.16 &      2 \\
   MZ\,Cen & 54584.280 &    0.27 & $\pm$ 0.10 &  (45, 4) & $-$0.08 & $\pm$ 0.14 &      2 & $-$0.22 & $\pm$ 0.15 &      4 &	   & \ldots	& \ldots &    0.27 & $\pm$ 0.11 &      1 & $-$0.05 & $\pm$ 0.08 &      2 \\
   OO\,Cen & 54585.060 &    0.20 & $\pm$ 0.06 &  (30, 4) &    0.14 & $\pm$ 0.07 &      2 &    0.01 & $\pm$ 0.11 &      3 &	   & \ldots	& \ldots & $-$0.04 & $\pm$ 0.05 &      2 &    0.05 & $\pm$ 0.07 &      2 \\
   TX\,Cen & 54862.363 &    0.44 & $\pm$ 0.12 &  (78, 7) &    0.17 & $\pm$ 0.06 &      2 &    0.01 & $\pm$ 0.01 &      2 &    0.07 & $\pm$ 0.11 &      1 & $-$0.05 & $\pm$ 0.48 &      2 &    0.07 & $\pm$ 0.09 &      2 \\
 V339\,Cen & 54584.304 &    0.06 & $\pm$ 0.03 &  (39, 3) & $-$0.09 & $\pm$ 0.03 &      2 & $-$0.25 & $\pm$ 0.13 &      4 & $-$0.21 & $\pm$ 0.11 &      1 & $-$0.28 & $\pm$ 0.37 &      2 & $-$0.10 & $\pm$ 0.03 &      2 \\
   VW\,Cen & 54862.359 &    0.41 & $\pm$ 0.08 &  (43, 2) &    0.07 & $\pm$ 0.08 &      1 & $-$0.14 & $\pm$ 0.07 &      4 & $-$0.19 & $\pm$ 0.11 &      1 &    0.23 & $\pm$ 0.11 &      1 & $-$0.05 & $\pm$ 0.08 &      2 \\
   AO\,CMa & 54839.053 &    0.01 & $\pm$ 0.06 &  (75, 5) &    0.08 & $\pm$ 0.10 &      2 &    0.13 & $\pm$ 0.08 &      4 &    0.18 & $\pm$ 0.11 &      1 &    0.11 & $\pm$ 0.08 &      2 &    0.20 & $\pm$ 0.08 &      2 \\
   RW\,CMa & 54839.138 & $-$0.07 & $\pm$ 0.08 &  (83, 5) & $-$0.06 & $\pm$ 0.01 &      2 & $-$0.04 & $\pm$ 0.05 &      4 &	   & \ldots	& \ldots &	   & \ldots	& \ldots & $-$0.08 & $\pm$ 0.08 &      1 \\
   SS\,CMa & 54839.066 &    0.06 & $\pm$ 0.04 &  (57, 5) &    0.14 & $\pm$ 0.13 &      2 &    0.06 & $\pm$ 0.08 &      4 &    0.13 & $\pm$ 0.11 &      1 &    0.09 & $\pm$ 0.15 &      2 &    0.10 & $\pm$ 0.01 &      2 \\
   TV\,CMa & 54847.246 &    0.01 & $\pm$ 0.07 &  (89, 6) &    0.17 & $\pm$ 0.14 &      2 & $-$0.04 & $\pm$ 0.03 &      4 & $-$0.04 & $\pm$ 0.11 &      1 & $-$0.10 & $\pm$ 0.19 &      2 &    0.04 & $\pm$ 0.04 &      2 \\
   TW\,CMa & 54839.077 &    0.04 & $\pm$ 0.09 &  (38, 4) & $-$0.06 & $\pm$ 0.27 &      2 &    0.17 & $\pm$ 0.29 &      3 &    0.14 & $\pm$ 0.11 &      1 &    0.02 & $\pm$ 0.11 &      1 &    0.06 & $\pm$ 0.08 &      1 \\
   AA\,Gem & 54846.149 & $-$0.08 & $\pm$ 0.05 &  (74, 5) & $-$0.18 & $\pm$ 0.06 &      2 & $-$0.16 & $\pm$ 0.08 &      4 &    0.34 & $\pm$ 0.45 &      2 &    0.02 & $\pm$ 0.21 &      2 &    0.08 & $\pm$ 0.03 &      2 \\
   AD\,Gem & 54846.221 & $-$0.14 & $\pm$ 0.06 &  (70, 7) & $-$0.31 & $\pm$ 0.11 &      2 & $-$0.17 & $\pm$ 0.17 &      2 &	   & \ldots	& \ldots & $-$0.16 & $\pm$ 0.11 &      1 & $-$0.25 & $\pm$ 0.08 &      1 \\
   BW\,Gem & 54845.122 & $-$0.22 & $\pm$ 0.09 &  (99, 6) & $-$0.36 & $\pm$ 0.06 &      2 & $-$0.18 & $\pm$ 0.03 &      4 &	   & \ldots	& \ldots & $-$0.09 & $\pm$ 0.11 &      1 & $-$0.08 & $\pm$ 0.13 &      2 \\
   DX\,Gem & 54846.196 & $-$0.01 & $\pm$ 0.09 &  (72, 6) & $-$0.09 & $\pm$ 0.01 &      2 &    0.00 & $\pm$ 0.09 &      3 &	   & \ldots	& \ldots &    0.07 & $\pm$ 0.11 &      1 & $-$0.03 & $\pm$ 0.08 &      1 \\
   RZ\,Gem & 54845.094 & $-$0.16 & $\pm$ 0.03 &  (44, 5) & $-$0.15 & $\pm$ 0.08 &      1 &	   & \ldots	& \ldots &	   & \ldots	& \ldots &	   & \ldots	& \ldots &	   & \ldots	& \ldots \\
   BE\,Mon & 54846.201 &    0.05 & $\pm$ 0.09 &  (78, 5) &    0.04 & $\pm$ 0.04 &      2 &    0.02 & $\pm$ 0.05 &      4 &	   & \ldots	& \ldots &    0.07 & $\pm$ 0.10 &      2 &    0.11 & $\pm$ 0.04 &      2 \\
   CV\,Mon & 54846.182 &    0.09 & $\pm$ 0.09 &  (52, 2) & $-$0.07 & $\pm$ 0.16 &      2 & $-$0.16 & $\pm$ 0.10 &      4 & $-$0.02 & $\pm$ 0.11 &      1 & $-$0.07 & $\pm$ 0.49 &      2 & $-$0.12 & $\pm$ 0.06 &      2 \\
   FT\,Mon & 54845.104 & $-$0.13 & $\pm$ 0.08 &  (61, 8) & $-$0.34 & $\pm$ 0.13 &      2 & $-$0.17 & $\pm$ 0.08 &      4 &    0.09 & $\pm$ 0.11 &      1 &    0.14 & $\pm$ 0.21 &      2 & $-$0.02 & $\pm$ 0.05 &      2 \\
   SV\,Mon & 54845.119 &    0.12 & $\pm$ 0.08 &  (54, 8) & $-$0.06 & $\pm$ 0.01 &      2 & $-$0.14 & $\pm$ 0.16 &      4 &    0.02 & $\pm$ 0.11 &      1 &    0.10 & $\pm$ 0.24 &      2 & $-$0.07 & $\pm$ 0.05 &      2 \\
   TW\,Mon & 54796.347 & $-$0.13 & $\pm$ 0.07 &  (75, 6) & $-$0.12 & $\pm$ 0.02 &      2 &    0.03 & $\pm$ 0.06 &      4 &	   & \ldots	& \ldots &    0.09 & $\pm$ 0.04 &      2 &    0.01 & $\pm$ 0.04 &      2 \\
   TX\,Mon & 54798.345 & $-$0.03 & $\pm$ 0.05 &  (76, 4) &    0.13 & $\pm$ 0.03 &      2 &    0.26 & $\pm$ 0.19 &      4 &    0.20 & $\pm$ 0.11 &      1 &    0.23 & $\pm$ 0.16 &      2 &    0.14 & $\pm$ 0.09 &      2 \\
   TY\,Mon & 54846.139 &    0.02 & $\pm$ 0.08 &  (85, 6) & $-$0.09 & $\pm$ 0.03 &      2 & $-$0.01 & $\pm$ 0.07 &      2 &	   & \ldots	& \ldots &    0.06 & $\pm$ 0.11 &      1 & $-$0.11 & $\pm$ 0.08 &      1 \\
   TZ\,Mon & 54847.237 & $-$0.02 & $\pm$ 0.07 &  (94, 6) &    0.00 & $\pm$ 0.04 &      2 &    0.06 & $\pm$ 0.07 &      4 &    0.08 & $\pm$ 0.11 &      1 &    0.04 & $\pm$ 0.02 &      2 &    0.07 & $\pm$ 0.01 &      2 \\
 V465\,Mon & 54847.241 & $-$0.07 & $\pm$ 0.07 & (107, 6) & $-$0.14 & $\pm$ 0.15 &      2 &    0.06 & $\pm$ 0.10 &      4 &	   & \ldots	& \ldots &    0.23 & $\pm$ 0.11 &      1 &    0.05 & $\pm$ 0.08 &      1 \\
 V495\,Mon & 54846.167 & $-$0.13 & $\pm$ 0.07 &  (73, 4) & $-$0.26 & $\pm$ 0.04 &      2 & $-$0.06 & $\pm$ 0.07 &      3 & $-$0.03 & $\pm$ 0.11 &      1 & $-$0.04 & $\pm$ 0.09 &      2 & $-$0.04 & $\pm$ 0.01 &      2 \\
 V508\,Mon & 54847.232 & $-$0.04 & $\pm$ 0.10 & (118, 7) & $-$0.15 & $\pm$ 0.01 &      2 &    0.03 & $\pm$ 0.08 &      4 &	   & \ldots	& \ldots &    0.07 & $\pm$ 0.11 &      1 & $-$0.06 & $\pm$ 0.08 &      1 \\
 V510\,Mon & 54846.153 & $-$0.16 & $\pm$ 0.06 &  (80, 3) & $-$0.23 & $\pm$ 0.08 &      2 & $-$0.14 & $\pm$ 0.10 &      4 & $-$0.07 & $\pm$ 0.11 &      1 & $-$0.11 & $\pm$ 0.06 &      2 & $-$0.04 & $\pm$ 0.05 &      2 \\
   XX\,Mon & 54798.335 &    0.01 & $\pm$ 0.08 &  (55, 2) & $-$0.07 & $\pm$ 0.04 &      2 & $-$0.11 & $\pm$ 0.18 &      4 &    0.11 & $\pm$ 0.11 &      1 &    0.44 & $\pm$ 0.20 &      2 &    0.08 & $\pm$ 0.01 &      2 \\
   GU\,Nor & 54667.205 &    0.08 & $\pm$ 0.06 &  (80, 7) & $-$0.08 & $\pm$ 0.04 &      2 & $-$0.08 & $\pm$ 0.19 &      2 &	   & \ldots	& \ldots & $-$0.23 & $\pm$ 0.11 &      1 & $-$0.07 & $\pm$ 0.08 &      1 \\
   IQ\,Nor & 54584.299 &    0.22 & $\pm$ 0.07 &  (63, 7) & $-$0.06 & $\pm$ 0.15 &      2 & $-$0.15 & $\pm$ 0.09 &      3 &	   & \ldots	& \ldots & $-$0.06 & $\pm$ 0.11 &      1 & $-$0.05 & $\pm$ 0.06 &      2 \\
   QZ\,Nor & 54863.366 &    0.18 & $\pm$ 0.08 &  (81, 3) & $-$0.08 & $\pm$ 0.11 &      2 &    0.00 & $\pm$ 0.02 &      4 &    0.09 & $\pm$ 0.13 &      3 & $-$0.19 & $\pm$ 0.03 &      2 &    0.15 & $\pm$ 0.06 &      2 \\
   QZ\,Nor & 54923.345 &    0.23 & $\pm$ 0.07 &  (86, 2) & $-$0.08 & $\pm$ 0.10 &      2 & $-$0.07 & $\pm$ 0.02 &      4 &    0.03 & $\pm$ 0.06 &      3 &    0.21 & $\pm$ 0.11 &      1 &    0.15 & $\pm$ 0.04 &      2 \\
   RS\,Nor & 54863.361 &    0.18 & $\pm$ 0.08 &  (82, 5) &    0.15 & $\pm$ 0.09 &      2 &    0.01 & $\pm$ 0.04 &      3 &	   & \ldots	& \ldots &    0.01 & $\pm$ 0.11 &      1 &    0.17 & $\pm$ 0.02 &      2 \\
   SY\,Nor & 54708.061 &    0.27 & $\pm$ 0.10 &  (46, 5) & $-$0.06 & $\pm$ 0.05 &      3 & $-$9.99 & $\pm$ 0.10 &      0 &    0.14 & $\pm$ 0.03 &      3 & $-$0.39 & $\pm$ 0.27 &      2 &    0.02 & $\pm$ 0.08 &      2 \\
   SY\,Nor & 54709.075 &    0.20 & $\pm$ 0.09 &  (58, 4) & $-$0.04 & $\pm$ 0.15 &      3 & $-$0.03 & $\pm$ 0.04 &      3 &    0.01 & $\pm$ 0.10 &      3 & $-$0.16 & $\pm$ 0.10 &      2 &    0.09 & $\pm$ 0.02 &      2 \\
   TW\,Nor & 54666.127 &    0.27 & $\pm$ 0.10 &  (69, 7) & $-$0.10 & $\pm$ 0.18 &      2 & $-$0.19 & $\pm$ 0.15 &      4 & $-$0.11 & $\pm$ 0.11 &      1 & $-$0.15 & $\pm$ 0.50 &      2 &    0.08 & $\pm$ 0.08 &      2 \\
 V340\,Nor & 54873.376 &    0.07 & $\pm$ 0.07 &  (47, 4) & $-$0.31 & $\pm$ 0.04 &      2 & $-$0.13 & $\pm$ 0.26 &      4 &	   & \ldots	& \ldots & $-$0.15 & $\pm$ 0.11 &      1 & $-$0.18 & $\pm$ 0.02 &      2 \\
   CS\,Ori & 54845.085 & $-$0.25 & $\pm$ 0.06 &  (68, 6) & $-$0.27 & $\pm$ 0.09 &      2 &	   & \ldots	& \ldots &	   & \ldots	& \ldots & $-$0.21 & $\pm$ 0.11 &      1 & $-$0.25 & $\pm$ 0.08 &      1 \\
   RS\,Ori & 54845.100 &    0.11 & $\pm$ 0.09 &  (71, 5) &    0.15 & $\pm$ 0.01 &      2 &    0.12 & $\pm$ 0.11 &      3 &    0.37 & $\pm$ 0.01 &      2 &    0.25 & $\pm$ 0.11 &      1 &    0.21 & $\pm$ 0.08 &      2 \\
   AQ\,Pup & 54839.075 &    0.06 & $\pm$ 0.05 &  (14, 2) &    0.48 & $\pm$ 0.08 &      1 &    0.05 & $\pm$ 0.14 &      3 & $-$0.13 & $\pm$ 0.11 &      1 &    0.03 & $\pm$ 0.11 &      1 &    0.12 & $\pm$ 0.15 &      2 \\
   BC\,Pup & 54839.147 & $-$0.31 & $\pm$ 0.07 &  (57, 3) & $-$0.35 & $\pm$ 0.16 &      2 & $-$0.15 & $\pm$ 0.10 &      4 &    0.07 & $\pm$ 0.11 &      1 &    0.37 & $\pm$ 0.11 &      1 &    0.02 & $\pm$ 0.10 &      2 \\
   BM\,Pup & 54839.086 & $-$0.07 & $\pm$ 0.08 &  (61, 7) & $-$0.05 & $\pm$ 0.01 &      2 & $-$0.11 & $\pm$ 0.16 &      4 &    0.19 & $\pm$ 0.11 &      1 &    0.06 & $\pm$ 0.23 &      2 &    0.04 & $\pm$ 0.08 &      2 \\
   BN\,Pup & 54839.109 &    0.03 & $\pm$ 0.05 &  (69, 4) &    0.30 & $\pm$ 0.18 &      2 &    0.09 & $\pm$ 0.14 &      4 &    0.12 & $\pm$ 0.11 &      1 &    0.07 & $\pm$ 0.13 &      2 &    0.16 & $\pm$ 0.16 &      2 \\
   CK\,Pup & 54839.113 & $-$0.15 & $\pm$ 0.08 &  (72, 4) & $-$0.15 & $\pm$ 0.12 &      3 &    0.01 & $\pm$ 0.04 &      3 &    0.22 & $\pm$ 0.11 &      3 & $-$0.16 & $\pm$ 0.11 &      2 &    0.08 & $\pm$ 0.11 &      2 \\
   CK\,Pup & 54839.173 & $-$0.12 & $\pm$ 0.08 &  (78,11) & $-$0.26 & $\pm$ 0.02 &      2 & $-$0.11 & $\pm$ 0.04 &      4 &    0.11 & $\pm$ 0.09 &      3 & $-$0.17 & $\pm$ 0.02 &      2 & $-$0.01 & $\pm$ 0.12 &      2 \\
   HW\,Pup & 54792.249 & $-$0.22 & $\pm$ 0.09 &  (70, 3) & $-$0.16 & $\pm$ 0.07 &      2 & $-$0.18 & $\pm$ 0.12 &      4 & $-$0.10 & $\pm$ 0.11 &      1 & $-$0.15 & $\pm$ 0.16 &      2 & $-$0.07 & $\pm$ 0.06 &      2 \\
   LS\,Pup & 54839.081 & $-$0.12 & $\pm$ 0.11 &  (18, 1) & $-$0.11 & $\pm$ 0.06 &      2 & $-$0.01 & $\pm$ 0.13 &      4 & $-$0.04 & $\pm$ 0.11 &      1 &    0.03 & $\pm$ 0.33 &      2 &    0.01 & $\pm$ 0.01 &      2 \\
   VW\,Pup & 54832.331 & $-$0.14 & $\pm$ 0.06 &  (50, 4) & $-$0.35 & $\pm$ 0.16 &      2 & $-$0.38 & $\pm$ 0.05 &      4 &	   & \ldots	& \ldots & $-$0.46 & $\pm$ 0.11 &      1 & $-$0.18 & $\pm$ 0.08 &      2 \\
   VZ\,Pup & 54839.096 & $-$0.01 & $\pm$ 0.04 &  (27, 2) & $-$0.01 & $\pm$ 0.05 &      2 &    0.09 & $\pm$ 0.11 &      4 &	   & \ldots	& \ldots &    0.18 & $\pm$ 0.11 &      1 &    0.06 & $\pm$ 0.08 &      1 \\
   WW\,Pup & 54839.091 &    0.13 & $\pm$ 0.16 &  (18, 1) & $-$0.30 & $\pm$ 0.06 &      2 & $-$0.23 & $\pm$ 0.09 &      4 &	   & \ldots	& \ldots & $-$0.44 & $\pm$ 0.11 &      1 & $-$0.07 & $\pm$ 0.07 &      2 \\
   WY\,Pup & 54839.100 & $-$0.10 & $\pm$ 0.08 &  (49, 6) & $-$0.43 & $\pm$ 0.08 &      1 & $-$0.18 & $\pm$ 0.04 &      3 &	   & \ldots	& \ldots &    0.13 & $\pm$ 0.03 &      2 & $-$0.03 & $\pm$ 0.08 &      1 \\
   WZ\,Pup & 54839.104 & $-$0.07 & $\pm$ 0.06 &  (72, 7) & $-$0.16 & $\pm$ 0.08 &      2 & $-$0.09 & $\pm$ 0.05 &      3 &    0.01 & $\pm$ 0.11 &      1 & $-$0.09 & $\pm$ 0.30 &      2 & $-$0.04 & $\pm$ 0.02 &      2 \\
\hline\noalign{\smallskip}
\multicolumn{20}{r}{\it {\footnotesize continued on next page}} \\
\end{tabular}}
\tablefoot{Column 3 lists the weighted mean and standard deviation of the
\ion{Fe}{i} and \ion{Fe}{ii} abundances derived by \citet{Genovalietal2014}.
Column 4 lists the respective number ($N_{\rm L}$) of iron lines used. The
other $N_{\rm L}$ values indicate the number of lines used for the other
elements to derive their abundances. For these elements, the quoted errors
represent either the dispersion around the mean if two or more lines were
measured, or the mean dispersion computed for the eleven calibrating stars
if only one line was available.}
\end{table*}
\addtocounter{table}{-1}
\begin{table*}[p]
\centering
\caption[]{continued.}
{\scriptsize 
\begin{tabular}{rc r@{ }l c r@{ }l c r@{ }l c r@{ }l c r@{ }l c r@{ }l c}
\noalign{\smallskip}\hline\hline\noalign{\smallskip}
Name & MJD &
\multicolumn{2}{c}{[Fe/H]} & \parbox[c]{1.0cm}{\centering $N_{\rm L}$ (\ion{Fe}{i},\ion{Fe}{ii})} &
\multicolumn{2}{c}{[Y/H]} & $N_{\rm L}$ &
\multicolumn{2}{c}{[La/H]} & $N_{\rm L}$ &
\multicolumn{2}{c}{[Ce/H]} & $N_{\rm L}$ &
\multicolumn{2}{c}{[Nd/H]} & $N_{\rm L}$ &
\multicolumn{2}{c}{[Eu/H]} & $N_{\rm L}$ \\
\noalign{\smallskip}\hline\noalign{\smallskip}
    X\,Pup & 54839.070 &    0.02 & $\pm$ 0.08 &  (15, 2) & $-$0.15 & $\pm$ 0.02 &      2 & $-$0.02 & $\pm$ 0.10 &      3 & $-$0.13 & $\pm$ 0.11 &      1 &    0.03 & $\pm$ 0.11 &      1 & $-$0.26 & $\pm$ 0.01 &      2 \\
   KQ\,Sco & 56139.021 &    0.26 & $\pm$ 0.15 &  (32, 0) & $-$0.22 & $\pm$ 0.16 &      5 & $-$0.24 & $\pm$ 0.18 &      5 & $-$0.21 & $\pm$ 0.24 &      2 & $-$0.24 & $\pm$ 0.09 &      5 & $-$0.12 & $\pm$ 0.04 &      2 \\
   KQ\,Sco & 54873.379 &    0.52 & $\pm$ 0.08 &  (51, 4) & $-$0.14 & $\pm$ 0.37 &      2 & $-$0.01 & $\pm$ 0.02 &      4 &    0.03 & $\pm$ 0.27 &      2 & $-$0.75 & $\pm$ 0.11 &      1 & $-$0.13 & $\pm$ 0.08 &      2 \\
   KQ\,Sco & 56152.097 &    0.30 & $\pm$ 0.21 &  (16, 1) &    0.07 & $\pm$ 0.15 &      5 &    0.04 & $\pm$ 0.12 &      3 &    0.27 & $\pm$ 0.23 &      2 & $-$0.13 & $\pm$ 0.08 &      4 &    0.06 & $\pm$ 0.08 &      2 \\
   KQ\,Sco & 56163.004 &    0.22 & $\pm$ 0.27 &  (20, 1) &    0.00 & $\pm$ 0.19 &      3 & $-$0.15 & $\pm$ 0.09 &      4 & $-$0.29 & $\pm$ 0.28 &      2 & $-$0.21 & $\pm$ 0.08 &      4 & $-$0.09 & $\pm$ 0.02 &      2 \\
   KQ\,Sco & 56166.004 &    0.21 & $\pm$ 0.28 &  (15, 1) & $-$0.12 & $\pm$ 0.08 &      2 & $-$0.18 & $\pm$ 0.09 &      5 &    0.02 & $\pm$ 0.15 &      2 & $-$0.15 & $\pm$ 0.04 &      3 & $-$0.06 & $\pm$ 0.01 &      2 \\
   RY\,Sco & 56140.187 &    0.06 & $\pm$ 0.01 &  (74, 2) &    0.04 & $\pm$ 0.12 &      5 &    0.07 & $\pm$ 0.09 &      5 &    0.12 & $\pm$ 0.11 &      2 &    0.08 & $\pm$ 0.18 &      6 &    0.10 & $\pm$ 0.02 &      2 \\
   RY\,Sco & 54599.412 &    0.06 & $\pm$ 0.02 &  (34, 5) & $-$0.07 & $\pm$ 0.02 &      3 &    0.01 & $\pm$ 0.10 &      4 &    0.03 & $\pm$ 0.04 &      2 & $-$0.22 & $\pm$ 0.02 &      2 &    0.07 & $\pm$ 0.02 &      2 \\
   RY\,Sco & 56152.143 &    0.01 & $\pm$ 0.03 &  (75, 3) & $-$0.16 & $\pm$ 0.07 &      6 & $-$0.06 & $\pm$ 0.09 &      5 &    0.01 & $\pm$ 0.08 &      2 & $-$0.08 & $\pm$ 0.18 &      6 & $-$0.01 & $\pm$ 0.01 &      2 \\
   RY\,Sco & 56162.170 & $-$0.03 & $\pm$ 0.05 &  (66, 2) & $-$0.13 & $\pm$ 0.12 &      6 & $-$0.03 & $\pm$ 0.08 &      5 &    0.05 & $\pm$ 0.12 &      2 & $-$0.02 & $\pm$ 0.20 &      6 &    0.02 & $\pm$ 0.02 &      2 \\
   RY\,Sco & 56167.085 & $-$0.04 & $\pm$ 0.08 &  (45, 2) & $-$0.01 & $\pm$ 0.01 &      4 &    0.01 & $\pm$ 0.13 &      5 &    0.09 & $\pm$ 0.09 &      2 & $-$0.11 & $\pm$ 0.18 &      6 &    0.10 & $\pm$ 0.01 &      2 \\
 V470\,Sco & 54708.073 &    0.16 & $\pm$ 0.06 &  (66, 4) &    0.11 & $\pm$ 0.09 &      2 & $-$0.07 & $\pm$ 0.09 &      4 & $-$0.12 & $\pm$ 0.11 &      1 &    0.03 & $\pm$ 0.11 &      1 &    0.16 & $\pm$ 0.10 &      2 \\
 V500\,Sco & 56140.191 & $-$0.01 & $\pm$ 0.05 &  (86, 3) & $-$0.17 & $\pm$ 0.07 &      6 &    0.01 & $\pm$ 0.07 &      5 &    0.08 & $\pm$ 0.07 &      2 &    0.10 & $\pm$ 0.15 &      5 &    0.05 & $\pm$ 0.04 &      2 \\
 V500\,Sco & 56152.092 &    0.00 & $\pm$ 0.10 &  (67, 3) & $-$0.18 & $\pm$ 0.17 &      6 & $-$0.11 & $\pm$ 0.07 &      5 & $-$0.01 & $\pm$ 0.09 &      2 & $-$0.18 & $\pm$ 0.18 &      6 & $-$0.07 & $\pm$ 0.02 &      2 \\
 V500\,Sco & 56162.998 & $-$0.03 & $\pm$ 0.12 &  (53, 3) & $-$0.32 & $\pm$ 0.05 &      4 & $-$0.26 & $\pm$ 0.08 &      5 & $-$0.12 & $\pm$ 0.08 &      2 & $-$0.25 & $\pm$ 0.18 &      5 & $-$0.19 & $\pm$ 0.02 &      2 \\
 V500\,Sco & 56167.077 & $-$0.11 & $\pm$ 0.07 &  (97, 5) & $-$0.19 & $\pm$ 0.07 &      6 & $-$0.03 & $\pm$ 0.07 &      5 &    0.03 & $\pm$ 0.04 &      2 & $-$0.08 & $\pm$ 0.20 &      6 &    0.00 & $\pm$ 0.04 &      2 \\
   EV\,Sct & 54708.086 &    0.09 & $\pm$ 0.07 &  (57, 3) &    0.01 & $\pm$ 0.04 &      2 &    0.07 & $\pm$ 0.09 &      3 &	   & \ldots	& \ldots &    0.15 & $\pm$ 0.11 &      2 &    0.20 & $\pm$ 0.09 &      2 \\
   RU\,Sct & 54906.414 &    0.16 & $\pm$ 0.05 &  (95, 7) & $-$0.03 & $\pm$ 0.16 &      3 & $-$0.13 & $\pm$ 0.09 &      4 & $-$0.05 & $\pm$ 0.21 &      2 & $-$0.25 & $\pm$ 0.21 &      2 & $-$0.03 & $\pm$ 0.04 &      2 \\
   RU\,Sct & 54923.375 &    0.09 & $\pm$ 0.07 &  (45, 5) &    0.06 & $\pm$ 0.15 &      3 & $-$0.06 & $\pm$ 0.08 &      4 &    0.13 & $\pm$ 0.40 &      2 & $-$0.19 & $\pm$ 0.17 &      2 & $-$0.05 & $\pm$ 0.10 &      2 \\
   UZ\,Sct & 56137.160 &    0.28 & $\pm$ 0.12 &  (17, 4) &    0.15 & $\pm$ 0.08 &      4 & $-$0.08 & $\pm$ 0.11 &      4 &    0.23 & $\pm$ 0.24 &      2 & $-$0.36 & $\pm$ 0.17 &      4 &    0.29 & $\pm$ 0.02 &      2 \\
   UZ\,Sct & 54906.400 &    0.36 & $\pm$ 0.10 &  (34, 5) &    0.19 & $\pm$ 0.05 &      2 &    0.00 & $\pm$ 0.08 &      4 &    0.03 & $\pm$ 0.02 &      2 & $-$0.05 & $\pm$ 0.17 &      2 &    0.03 & $\pm$ 0.09 &      2 \\
   UZ\,Sct & 54923.366 &    0.45 & $\pm$ 0.07 &  (63, 7) &    0.09 & $\pm$ 0.02 &      2 & $-$0.08 & $\pm$ 0.14 &      4 & $-$0.16 & $\pm$ 0.06 &      2 & $-$0.22 & $\pm$ 0.13 &      2 &    0.09 & $\pm$ 0.04 &      2 \\
   UZ\,Sct & 56152.064 &    0.25 & $\pm$ 0.28 &  ( 8, 0) &    0.11 & $\pm$ 0.13 &      4 &    0.00 & $\pm$ 0.11 &      4 &    0.45 & $\pm$ 0.04 &      2 & $-$0.23 & $\pm$ 0.08 &      2 &    0.32 & $\pm$ 0.10 &      2 \\
   UZ\,Sct & 56160.167 &    0.36 & $\pm$ 0.10 &  (47, 2) &    0.18 & $\pm$ 0.13 &      4 &    0.01 & $\pm$ 0.08 &      5 &    0.08 & $\pm$ 0.08 &      2 &    0.03 & $\pm$ 0.17 &      5 &    0.00 & $\pm$ 0.03 &      2 \\
   UZ\,Sct & 56175.049 &    0.31 & $\pm$ 0.21 &  (36, 2) & $-$0.01 & $\pm$ 0.17 &      5 & $-$0.06 & $\pm$ 0.08 &      5 &    0.04 & $\pm$ 0.15 &      2 & $-$0.05 & $\pm$ 0.18 &      4 & $-$0.05 & $\pm$ 0.19 &      2 \\
 V367\,Sct & 56137.147 &    0.13 & $\pm$ 0.07 &  (72, 3) & $-$0.13 & $\pm$ 0.09 &      6 & $-$0.08 & $\pm$ 0.07 &      5 &    0.02 & $\pm$ 0.20 &      2 & $-$0.03 & $\pm$ 0.17 &      5 &    0.03 & $\pm$ 0.09 &      2 \\
 V367\,Sct & 54709.128 & $-$0.04 & $\pm$ 0.04 &  (56, 4) & $-$0.05 & $\pm$ 0.05 &      2 & $-$0.08 & $\pm$ 0.11 &      4 & $-$0.09 & $\pm$ 0.06 &      3 & $-$0.28 & $\pm$ 0.31 &      2 &    0.07 & $\pm$ 0.04 &      2 \\
 V367\,Sct & 56175.105 &    0.14 & $\pm$ 0.06 &  (50, 3) & $-$0.13 & $\pm$ 0.07 &      5 & $-$0.08 & $\pm$ 0.09 &      5 & $-$0.02 & $\pm$ 0.12 &      2 &    0.14 & $\pm$ 0.21 &      3 &    0.04 & $\pm$ 0.09 &      2 \\
 V367\,Sct & 56184.000 &    0.03 & $\pm$ 0.07 &  (84, 4) & $-$0.32 & $\pm$ 0.07 &      6 & $-$0.22 & $\pm$ 0.13 &      5 & $-$0.17 & $\pm$ 0.04 &      2 & $-$0.15 & $\pm$ 0.19 &      4 & $-$0.14 & $\pm$ 0.05 &      2 \\
    X\,Sct & 54709.122 &    0.12 & $\pm$ 0.09 &  (72, 9) &    0.04 & $\pm$ 0.16 &      2 & $-$0.07 & $\pm$ 0.10 &      4 &	   & \ldots	& \ldots &    0.03 & $\pm$ 0.11 &      2 &    0.08 & $\pm$ 0.07 &      2 \\
    Z\,Sct & 56137.123 &    0.10 & $\pm$ 0.16 &  (20, 0) & $-$0.40 & $\pm$ 0.09 &      3 & $-$0.38 & $\pm$ 0.05 &      3 & $-$0.43 & $\pm$ 0.03 &      2 & $-$0.51 & $\pm$ 0.17 &      4 & $-$0.27 & $\pm$ 0.08 &      1 \\
    Z\,Sct & 54678.090 &    0.18 & $\pm$ 0.09 &  (41, 3) & $-$0.43 & $\pm$ 0.11 &      3 & $-$0.36 & $\pm$ 0.08 &      4 & $-$0.57 & $\pm$ 0.37 &      2 & $-$0.60 & $\pm$ 0.48 &      2 & $-$0.21 & $\pm$ 0.07 &      2 \\
    Z\,Sct & 56152.073 &    0.11 & $\pm$ 0.02 &  (49, 2) & $-$0.22 & $\pm$ 0.05 &      4 & $-$0.18 & $\pm$ 0.13 &      5 & $-$0.22 & $\pm$ 0.13 &      2 & $-$0.38 & $\pm$ 0.18 &      4 &    0.04 & $\pm$ 0.08 &      2 \\
    Z\,Sct & 56159.186 &    0.26 & $\pm$ 0.08 &  (45, 2) & $-$0.17 & $\pm$ 0.12 &      5 & $-$0.19 & $\pm$ 0.06 &      5 & $-$0.08 & $\pm$ 0.13 &      2 & $-$0.17 & $\pm$ 0.15 &      3 &    0.01 & $\pm$ 0.01 &      2 \\
    Z\,Sct & 56175.038 &    0.00 & $\pm$ 0.30 &  (25, 0) & $-$0.42 & $\pm$ 0.18 &      5 & $-$0.42 & $\pm$ 0.16 &      5 & $-$0.31 & $\pm$ 0.07 &      2 & $-$0.36 & $\pm$ 0.09 &      5 & $-$0.03 & $\pm$ 0.11 &      2 \\
   AA\,Ser & 54708.040 &    0.38 & $\pm$ 0.20 &  (24, 1) &    0.39 & $\pm$ 0.06 &      2 &    0.09 & $\pm$ 0.15 &      4 &    0.04 & $\pm$ 0.11 &      1 &    0.15 & $\pm$ 0.71 &      2 &    0.30 & $\pm$ 0.18 &      2 \\
   CR\,Ser & 54709.116 &    0.12 & $\pm$ 0.08 &  (53, 5) &    0.05 & $\pm$ 0.04 &      2 & $-$0.15 & $\pm$ 0.16 &      4 &    0.06 & $\pm$ 0.11 &      1 &    0.08 & $\pm$ 0.49 &      2 &    0.15 & $\pm$ 0.18 &      2 \\
   AV\,Sgr & 56136.169 &    0.40 & $\pm$ 0.15 &  (29, 2) &    0.09 & $\pm$ 0.09 &      4 &    0.03 & $\pm$ 0.12 &      4 &    0.11 & $\pm$ 0.19 &      2 & $-$0.23 & $\pm$ 0.10 &      4 &    0.26 & $\pm$ 0.08 &      2 \\
   AV\,Sgr & 56136.192 &    0.44 & $\pm$ 0.15 &  (31, 2) & $-$0.01 & $\pm$ 0.05 &      4 & $-$0.06 & $\pm$ 0.08 &      5 &    0.01 & $\pm$ 0.17 &      2 & $-$0.22 & $\pm$ 0.09 &      4 &    0.24 & $\pm$ 0.04 &      2 \\
   AV\,Sgr & 54923.348 &    0.53 & $\pm$ 0.17 &  (16, 2) &    0.06 & $\pm$ 0.09 &      2 & $-$0.06 & $\pm$ 0.06 &      3 & $-$0.18 & $\pm$ 0.27 &      2 & $-$0.61 & $\pm$ 0.40 &      2 &    0.03 & $\pm$ 0.05 &      2 \\
   AV\,Sgr & 56152.082 &    0.42 & $\pm$ 0.17 &  (24, 2) & $-$0.07 & $\pm$ 0.04 &      3 &    0.07 & $\pm$ 0.03 &      4 &    0.06 & $\pm$ 0.21 &      2 & $-$0.32 & $\pm$ 0.19 &      4 &    0.18 & $\pm$ 0.01 &      2 \\
   AV\,Sgr & 56168.049 &    0.30 & $\pm$ 0.22 &  (19, 1) &    0.08 & $\pm$ 0.16 &      4 &    0.06 & $\pm$ 0.10 &      4 &    0.15 & $\pm$ 0.29 &      2 & $-$0.33 & $\pm$ 0.17 &      4 &    0.27 & $\pm$ 0.07 &      2 \\
   AY\,Sgr & 54599.398 &    0.11 & $\pm$ 0.06 &  (58, 5) &    0.07 & $\pm$ 0.16 &      2 & $-$0.05 & $\pm$ 0.04 &      4 & $-$0.04 & $\pm$ 0.11 &      1 &    0.09 & $\pm$ 0.11 &      1 &    0.08 & $\pm$ 0.08 &      2 \\
V1954\,Sgr & 54599.389 &    0.24 & $\pm$ 0.10 &  (61, 4) &    0.07 & $\pm$ 0.08 &      2 & $-$0.11 & $\pm$ 0.12 &      4 &	   & \ldots	& \ldots & $-$0.07 & $\pm$ 0.11 &      1 &    0.04 & $\pm$ 0.06 &      2 \\
 V773\,Sgr & 54669.207 &    0.11 & $\pm$ 0.06 &  (58, 8) & $-$0.07 & $\pm$ 0.03 &      2 & $-$0.04 & $\pm$ 0.09 &      4 &	   & \ldots	& \ldots & $-$0.09 & $\pm$ 0.11 &      1 & $-$0.06 & $\pm$ 0.18 &      2 \\
   VY\,Sgr & 56160.179 &    0.27 & $\pm$ 0.25 &  (14, 1) & $-$0.01 & $\pm$ 0.02 &      2 & $-$0.17 & $\pm$ 0.12 &      5 &    0.00 & $\pm$ 0.14 &      2 & $-$0.44 & $\pm$ 0.18 &      4 &    0.05 & $\pm$ 0.02 &      2 \\
   VY\,Sgr & 54923.356 &    0.42 & $\pm$ 0.14 &  (30, 6) &    0.05 & $\pm$ 0.04 &      2 &    0.04 & $\pm$ 0.15 &      4 &    0.13 & $\pm$ 0.31 &      2 & $-$0.08 & $\pm$ 0.25 &      2 &    0.06 & $\pm$ 0.05 &      2 \\
   VY\,Sgr & 56162.162 &    0.32 & $\pm$ 0.27 &  (17, 1) & $-$0.24 & $\pm$ 0.15 &      4 & $-$0.32 & $\pm$ 0.07 &      5 & $-$0.32 & $\pm$ 0.06 &      2 & $-$0.34 & $\pm$ 0.16 &      5 & $-$0.13 & $\pm$ 0.05 &      2 \\
   VY\,Sgr & 56168.062 &    0.31 & $\pm$ 0.05 &  (51, 2) & $-$0.02 & $\pm$ 0.18 &      4 &    0.00 & $\pm$ 0.07 &      5 &    0.09 & $\pm$ 0.19 &      2 &    0.10 & $\pm$ 0.19 &      5 &    0.17 & $\pm$ 0.13 &      2 \\
   WZ\,Sgr & 56132.190 &    0.18 & $\pm$ 0.08 &  (56, 2) & $-$0.08 & $\pm$ 0.09 &      4 &    0.04 & $\pm$ 0.12 &      5 & $-$0.02 & $\pm$ 0.07 &      2 & $-$0.02 & $\pm$ 0.15 &      6 &    0.17 & $\pm$ 0.02 &      2 \\
   WZ\,Sgr & 54599.395 &    0.35 & $\pm$ 0.08 &  (42, 2) &    0.17 & $\pm$ 0.06 &      2 & $-$0.15 & $\pm$ 0.14 &      4 & $-$0.21 & $\pm$ 0.07 &      2 & $-$9.99 & $\pm$ 0.11 &      0 & $-$0.01 & $\pm$ 0.05 &      2 \\
   WZ\,Sgr & 56136.213 &    0.24 & $\pm$ 0.01 &  (48, 2) & $-$0.02 & $\pm$ 0.18 &      5 &    0.05 & $\pm$ 0.10 &      5 &    0.17 & $\pm$ 0.11 &      2 &    0.06 & $\pm$ 0.14 &      5 &    0.09 & $\pm$ 0.14 &      2 \\
   WZ\,Sgr & 56152.044 &    0.28 & $\pm$ 0.12 &  (28, 2) & $-$0.23 & $\pm$ 0.16 &      5 & $-$0.14 & $\pm$ 0.08 &      5 & $-$0.09 & $\pm$ 0.12 &      2 & $-$0.30 & $\pm$ 0.16 &      5 & $-$0.07 & $\pm$ 0.12 &      2 \\
   WZ\,Sgr & 56159.125 &    0.37 & $\pm$ 0.06 &  (44, 2) &    0.08 & $\pm$ 0.14 &      5 &    0.04 & $\pm$ 0.09 &      5 &    0.19 & $\pm$ 0.12 &      2 &    0.21 & $\pm$ 0.19 &      4 & $-$0.10 & $\pm$ 0.07 &      2 \\
   XX\,Sgr & 56054.234 & $-$0.01 & $\pm$ 0.10 & (100, 5) & $-$0.14 & $\pm$ 0.08 &      5 &    0.03 & $\pm$ 0.09 &      5 &    0.12 & $\pm$ 0.05 &      2 & $-$0.02 & $\pm$ 0.17 &      6 &    0.06 & $\pm$ 0.03 &      2 \\
   XX\,Sgr & 54599.404 & $-$0.07 & $\pm$ 0.07 &  (59, 4) & $-$0.13 & $\pm$ 0.11 &      2 & $-$0.04 & $\pm$ 0.05 &      3 &    0.08 & $\pm$ 0.12 &      3 & $-$0.19 & $\pm$ 0.20 &      2 & $-$0.01 & $\pm$ 0.07 &      2 \\
   XX\,Sgr & 56136.223 & $-$0.05 & $\pm$ 0.05 & (101, 6) & $-$0.27 & $\pm$ 0.07 &      5 & $-$0.09 & $\pm$ 0.10 &      5 & $-$0.06 & $\pm$ 0.08 &      2 & $-$0.15 & $\pm$ 0.15 &      5 & $-$0.10 & $\pm$ 0.09 &      2 \\
   XX\,Sgr & 56152.047 &    0.05 & $\pm$ 0.03 &  (43, 3) & $-$0.15 & $\pm$ 0.04 &      4 & $-$0.12 & $\pm$ 0.12 &      5 & $-$0.02 & $\pm$ 0.06 &      2 & $-$0.07 & $\pm$ 0.19 &      6 & $-$0.06 & $\pm$ 0.03 &      2 \\
   XX\,Sgr & 56159.128 & $-$0.02 & $\pm$ 0.12 &  (64, 4) & $-$0.14 & $\pm$ 0.12 &      5 & $-$0.12 & $\pm$ 0.15 &      5 &    0.07 & $\pm$ 0.06 &      2 & $-$0.12 & $\pm$ 0.18 &      5 & $-$0.04 & $\pm$ 0.03 &      2 \\
   EZ\,Vel & 54759.348 & $-$0.17 & $\pm$ 0.15 &  (23, 1) & $-$0.02 & $\pm$ 0.08 &      2 & $-$0.01 & $\pm$ 0.23 &      3 &    0.13 & $\pm$ 0.11 &      1 & $-$0.08 & $\pm$ 0.11 &      1 &    0.04 & $\pm$ 0.10 &      2 \\
\hline
\end{tabular}}
\end{table*}

\clearpage
\begin{table*}[p]
\centering
\caption[]{Mean abundances of heavy elements for our sample of classical
Cepheids.}
\label{table_ab_heavy_mean}
{\scriptsize 
\begin{tabular}{rc r@{ }l r@{ }l r@{ }l r@{ }l r@{ }l r@{ }l r@{ }l r}
\noalign{\smallskip}\hline\hline\noalign{\smallskip}
Name &
\parbox[c]{0.6cm}{\centering \logP\ [days]} &
\multicolumn{2}{c}{\parbox[c]{0.4cm}{\centering \RG\ [pc]}} &
\multicolumn{2}{c}{[Fe/H]} &
\multicolumn{2}{c}{[Y/H]} &
\multicolumn{2}{c}{[La/H]} &
\multicolumn{2}{c}{[Ce/H]} &
\multicolumn{2}{c}{[Nd/H]} &
\multicolumn{2}{c}{[Eu/H]} & $N_{\rm S}$ \\
\noalign{\smallskip}\hline\noalign{\smallskip}
 V340\,Ara & 1.3183 &  4657 & $\pm$ 427 &    0.33 & $\pm$ 0.09 &    0.04 & $\pm$ 0.14 & $-$0.04 & $\pm$ 0.09 &    0.13 & $\pm$ 0.13 &	 0.02 & $\pm$ 0.25 &	0.13 & $\pm$ 0.07 &  6     \\
   QZ\,Nor & 0.5782 &  6283 & $\pm$ 447 &    0.21 & $\pm$ 0.06 &    0.25 & $\pm$ 0.32 & $-$0.07 & $\pm$ 0.02 &    0.03 & $\pm$ 0.11 &	 0.12 & $\pm$ 0.01 &	0.16 & $\pm$ 0.07 &  2     \\
   SY\,Nor & 1.1019 &  6286 & $\pm$ 446 &    0.23 & $\pm$ 0.07 &    0.06 & $\pm$ 0.01 & $-$0.10 & $\pm$ 0.08 &    0.08 & $\pm$ 0.11 & $-$0.04 & $\pm$ 0.06 &	0.08 & $\pm$ 0.01 &  2     \\
   CK\,Pup & 0.8703 & 13357 & $\pm$ 423 & $-$0.13 & $\pm$ 0.06 & $-$0.13 & $\pm$ 0.04 & $-$0.09 & $\pm$ 0.04 &    0.04 & $\pm$ 0.11 &	 0.00 & $\pm$ 0.11 &	0.04 & $\pm$ 0.16 &  2     \\
   KQ\,Sco & 1.4577 &  5948 & $\pm$ 451 &    0.52 & $\pm$ 0.08 &    0.03 & $\pm$ 0.15 & $-$0.08 & $\pm$ 0.12 & $-$0.02 & $\pm$ 0.25 & $-$0.24 & $\pm$ 0.25 & $-$0.02 & $\pm$ 0.07 &  5     \\
   RY\,Sco & 1.3078 &  6663 & $\pm$ 453 &    0.01 & $\pm$ 0.06 & $-$0.02 & $\pm$ 0.08 & $-$0.01 & $\pm$ 0.11 &    0.09 & $\pm$ 0.11 &	 0.06 & $\pm$ 0.16 &	0.09 & $\pm$ 0.02 &  5     \\
 V500\,Sco & 0.9693 &  6590 & $\pm$ 453 & $-$0.07 & $\pm$ 0.08 & $-$0.19 & $\pm$ 0.10 & $-$0.13 & $\pm$ 0.09 &    0.00 & $\pm$ 0.12 & $-$0.01 & $\pm$ 0.20 & $-$0.05 & $\pm$ 0.04 &  4     \\
   RU\,Sct & 1.2945 &  6361 & $\pm$ 449 &    0.14 & $\pm$ 0.04 &    0.18 & $\pm$ 0.21 & $-$0.09 & $\pm$ 0.10 & $-$0.11 & $\pm$ 0.11 & $-$0.07 & $\pm$ 0.29 &	0.00 & $\pm$ 0.13 &  2     \\
   UZ\,Sct & 1.1686 &  5309 & $\pm$ 448 &    0.33 & $\pm$ 0.08 &    0.19 & $\pm$ 0.09 & $-$0.01 & $\pm$ 0.12 &    0.16 & $\pm$ 0.12 &	 0.05 & $\pm$ 0.32 &	0.17 & $\pm$ 0.12 &  6     \\
 V367\,Sct & 0.7989 &  6332 & $\pm$ 451 &    0.05 & $\pm$ 0.08 & $-$0.12 & $\pm$ 0.10 & $-$0.14 & $\pm$ 0.10 &    0.23 & $\pm$ 0.11 &	 0.12 & $\pm$ 0.21 & $-$0.01 & $\pm$ 0.08 &  4     \\
    Z\,Sct & 1.1106 &  5733 & $\pm$ 445 &    0.12 & $\pm$ 0.09 & $-$0.26 & $\pm$ 0.12 & $-$0.31 & $\pm$ 0.11 &    0.03 & $\pm$ 0.12 & $-$0.04 & $\pm$ 0.25 & $-$0.06 & $\pm$ 0.08 &  5     \\
   AV\,Sgr & 1.1879 &  5980 & $\pm$ 455 &    0.35 & $\pm$ 0.17 &    0.11 & $\pm$ 0.15 & $-$0.04 & $\pm$ 0.20 &    0.04 & $\pm$ 0.25 & $-$0.04 & $\pm$ 0.26 &	0.24 & $\pm$ 0.10 &  5     \\
   VY\,Sgr & 1.1322 &  5862 & $\pm$ 453 &    0.33 & $\pm$ 0.12 & $-$0.01 & $\pm$ 0.08 & $-$0.11 & $\pm$ 0.12 & $-$0.04 & $\pm$ 0.13 & $-$0.02 & $\pm$ 0.28 &	0.09 & $\pm$ 0.08 &  4     \\
   WZ\,Sgr & 1.3394 &  6326 & $\pm$ 453 &    0.28 & $\pm$ 0.08 &    0.07 & $\pm$ 0.08 & $-$0.03 & $\pm$ 0.11 &    0.05 & $\pm$ 0.12 &	 0.05 & $\pm$ 0.23 &	0.04 & $\pm$ 0.11 &  5     \\
   XX\,Sgr & 0.8078 &  6706 & $\pm$ 453 & $-$0.01 & $\pm$ 0.06 & $-$0.09 & $\pm$ 0.06 & $-$0.10 & $\pm$ 0.12 &    0.10 & $\pm$ 0.07 &	 0.04 & $\pm$ 0.15 & $-$0.03 & $\pm$ 0.02 &  5     \\
\noalign{\smallskip}\hline\noalign{\smallskip}
   AS\,Aur & 0.5017 & 12244 & $\pm$ 469 &    0.00 & $\pm$ 0.08 & $-$0.20 & $\pm$ 0.13 & $-$0.16 & $\pm$ 0.19 &         & \ldots     & $-$0.06 & $\pm$ 0.10 & $-$0.03 & $\pm$ 0.02 &  1     \\
   KN\,Cen & 1.5321 &  6498 & $\pm$ 417 &    0.55 & $\pm$ 0.12 &    0.23 & $\pm$ 0.08 &    0.04 & $\pm$ 0.15 & $-$0.02 & $\pm$ 0.11 &	 0.05 & $\pm$ 0.21 &	0.07 & $\pm$ 0.16 &  1     \\
   MZ\,Cen & 1.0151 &  6501 & $\pm$ 391 &    0.27 & $\pm$ 0.10 & $-$0.08 & $\pm$ 0.14 & $-$0.22 & $\pm$ 0.15 &         & \ldots     &	 0.27 & $\pm$ 0.11 & $-$0.05 & $\pm$ 0.08 &  1     \\
   OO\,Cen & 1.1099 &  6025 & $\pm$ 389 &    0.20 & $\pm$ 0.06 &    0.14 & $\pm$ 0.07 &    0.01 & $\pm$ 0.11 &         & \ldots     & $-$0.04 & $\pm$ 0.05 &	0.05 & $\pm$ 0.07 &  1     \\
   TX\,Cen & 1.2328 &  6070 & $\pm$ 419 &    0.44 & $\pm$ 0.12 &    0.17 & $\pm$ 0.06 &    0.01 & $\pm$ 0.01 &    0.07 & $\pm$ 0.11 & $-$0.05 & $\pm$ 0.48 &	0.07 & $\pm$ 0.09 &  1     \\
 V339\,Cen & 0.9762 &  6917 & $\pm$ 446 &    0.06 & $\pm$ 0.03 & $-$0.09 & $\pm$ 0.03 & $-$0.25 & $\pm$ 0.13 & $-$0.21 & $\pm$ 0.11 & $-$0.28 & $\pm$ 0.37 & $-$0.10 & $\pm$ 0.03 &  1     \\
   VW\,Cen & 1.1771 &  6417 & $\pm$ 405 &    0.41 & $\pm$ 0.08 &    0.07 & $\pm$ 0.08 & $-$0.14 & $\pm$ 0.07 & $-$0.19 & $\pm$ 0.11 &	 0.23 & $\pm$ 0.11 & $-$0.05 & $\pm$ 0.08 &  1     \\
   AO\,CMa & 0.7646 & 10430 & $\pm$ 433 &    0.01 & $\pm$ 0.06 &    0.08 & $\pm$ 0.10 &    0.13 & $\pm$ 0.08 &    0.18 & $\pm$ 0.11 &	 0.11 & $\pm$ 0.08 &	0.20 & $\pm$ 0.08 &  1     \\
   RW\,CMa & 0.7581 & 10057 & $\pm$ 445 & $-$0.07 & $\pm$ 0.08 & $-$0.06 & $\pm$ 0.01 & $-$0.04 & $\pm$ 0.05 &         & \ldots     &	      & \ldots     & $-$0.08 & $\pm$ 0.08 &  1     \\
   SS\,CMa & 1.0921 &  9829 & $\pm$ 439 &    0.06 & $\pm$ 0.04 &    0.14 & $\pm$ 0.13 &    0.06 & $\pm$ 0.08 &    0.13 & $\pm$ 0.11 &	 0.09 & $\pm$ 0.15 &	0.10 & $\pm$ 0.01 &  1     \\
   TV\,CMa & 0.6693 &  9575 & $\pm$ 447 &    0.01 & $\pm$ 0.07 &    0.17 & $\pm$ 0.14 & $-$0.04 & $\pm$ 0.03 & $-$0.04 & $\pm$ 0.11 & $-$0.10 & $\pm$ 0.19 &	0.04 & $\pm$ 0.04 &  1     \\
   TW\,CMa & 0.8448 &  9788 & $\pm$ 445 &    0.04 & $\pm$ 0.09 & $-$0.06 & $\pm$ 0.27 &    0.17 & $\pm$ 0.29 &    0.14 & $\pm$ 0.11 &	 0.02 & $\pm$ 0.11 &	0.06 & $\pm$ 0.08 &  1     \\
   AA\,Gem & 1.0532 & 11454 & $\pm$ 459 & $-$0.08 & $\pm$ 0.05 & $-$0.18 & $\pm$ 0.06 & $-$0.16 & $\pm$ 0.08 &    0.34 & $\pm$ 0.45 &	 0.02 & $\pm$ 0.21 &	0.08 & $\pm$ 0.03 &  1     \\
   AD\,Gem & 0.5784 & 10662 & $\pm$ 455 & $-$0.14 & $\pm$ 0.06 & $-$0.31 & $\pm$ 0.11 & $-$0.17 & $\pm$ 0.17 &         & \ldots     & $-$0.16 & $\pm$ 0.11 & $-$0.25 & $\pm$ 0.08 &  1     \\
   BW\,Gem & 0.3633 & 11302 & $\pm$ 463 & $-$0.22 & $\pm$ 0.09 & $-$0.36 & $\pm$ 0.06 & $-$0.18 & $\pm$ 0.03 &         & \ldots     & $-$0.09 & $\pm$ 0.11 & $-$0.08 & $\pm$ 0.13 &  1     \\
   DX\,Gem & 0.4966 & 11407 & $\pm$ 473 & $-$0.01 & $\pm$ 0.09 & $-$0.09 & $\pm$ 0.01 &    0.00 & $\pm$ 0.09 &         & \ldots     &	 0.07 & $\pm$ 0.11 & $-$0.03 & $\pm$ 0.08 &  1     \\
   RZ\,Gem & 0.7427 &  9973 & $\pm$ 454 & $-$0.16 & $\pm$ 0.03 & $-$0.15 & $\pm$ 0.08 & 	& \ldots     &         & \ldots     &	      & \ldots     &	     & \ldots	  &  1     \\
   BE\,Mon & 0.4322 &  9609 & $\pm$ 452 &    0.05 & $\pm$ 0.09 &    0.04 & $\pm$ 0.04 &    0.02 & $\pm$ 0.05 &         & \ldots     &	 0.07 & $\pm$ 0.10 &	0.11 & $\pm$ 0.04 &  1     \\
   CV\,Mon & 0.7307 &  9362 & $\pm$ 452 &    0.09 & $\pm$ 0.09 & $-$0.07 & $\pm$ 0.16 & $-$0.16 & $\pm$ 0.10 & $-$0.02 & $\pm$ 0.11 & $-$0.07 & $\pm$ 0.49 & $-$0.12 & $\pm$ 0.06 &  1     \\
   FT\,Mon & 0.6843 & 14344 & $\pm$ 468 & $-$0.13 & $\pm$ 0.08 & $-$0.34 & $\pm$ 0.13 & $-$0.17 & $\pm$ 0.08 &    0.09 & $\pm$ 0.11 &	 0.14 & $\pm$ 0.21 & $-$0.02 & $\pm$ 0.05 &  1     \\
   SV\,Mon & 1.1828 & 10070 & $\pm$ 453 &    0.12 & $\pm$ 0.08 & $-$0.06 & $\pm$ 0.01 & $-$0.14 & $\pm$ 0.16 &    0.02 & $\pm$ 0.11 &	 0.10 & $\pm$ 0.24 & $-$0.07 & $\pm$ 0.05 &  1     \\
   TW\,Mon & 0.8511 & 13059 & $\pm$ 457 & $-$0.13 & $\pm$ 0.07 & $-$0.12 & $\pm$ 0.02 &    0.03 & $\pm$ 0.06 &         & \ldots     &	 0.09 & $\pm$ 0.04 &	0.01 & $\pm$ 0.04 &  1     \\
   TX\,Mon & 0.9396 & 11790 & $\pm$ 452 & $-$0.03 & $\pm$ 0.05 &    0.13 & $\pm$ 0.03 &    0.26 & $\pm$ 0.19 &    0.20 & $\pm$ 0.11 &	 0.23 & $\pm$ 0.16 &	0.14 & $\pm$ 0.09 &  1     \\
   TY\,Mon & 0.6045 & 11180 & $\pm$ 451 &    0.02 & $\pm$ 0.08 & $-$0.09 & $\pm$ 0.03 & $-$0.01 & $\pm$ 0.07 &         & \ldots     &	 0.06 & $\pm$ 0.11 & $-$0.11 & $\pm$ 0.08 &  1     \\
   TZ\,Mon & 0.8709 & 11183 & $\pm$ 451 & $-$0.02 & $\pm$ 0.07 &    0.00 & $\pm$ 0.04 &    0.06 & $\pm$ 0.07 &    0.08 & $\pm$ 0.11 &	 0.04 & $\pm$ 0.02 &	0.07 & $\pm$ 0.01 &  1     \\
 V465\,Mon & 0.4335 & 11037 & $\pm$ 450 & $-$0.07 & $\pm$ 0.07 & $-$0.14 & $\pm$ 0.15 &    0.06 & $\pm$ 0.10 &         & \ldots     &	 0.23 & $\pm$ 0.11 &	0.05 & $\pm$ 0.08 &  1     \\
 V495\,Mon & 0.6124 & 12098 & $\pm$ 453 & $-$0.13 & $\pm$ 0.07 & $-$0.26 & $\pm$ 0.04 & $-$0.06 & $\pm$ 0.07 & $-$0.03 & $\pm$ 0.11 & $-$0.04 & $\pm$ 0.09 & $-$0.04 & $\pm$ 0.01 &  1     \\
 V508\,Mon & 0.6163 & 10714 & $\pm$ 452 & $-$0.04 & $\pm$ 0.10 & $-$0.15 & $\pm$ 0.01 &    0.03 & $\pm$ 0.08 &         & \ldots     &	 0.07 & $\pm$ 0.11 & $-$0.06 & $\pm$ 0.08 &  1     \\
 V510\,Mon & 0.8637 & 12550 & $\pm$ 456 & $-$0.16 & $\pm$ 0.06 & $-$0.23 & $\pm$ 0.08 & $-$0.14 & $\pm$ 0.10 & $-$0.07 & $\pm$ 0.11 & $-$0.11 & $\pm$ 0.06 & $-$0.04 & $\pm$ 0.05 &  1     \\
   XX\,Mon & 0.7369 & 11854 & $\pm$ 451 &    0.01 & $\pm$ 0.08 & $-$0.07 & $\pm$ 0.04 & $-$0.11 & $\pm$ 0.18 &    0.11 & $\pm$ 0.11 &	 0.44 & $\pm$ 0.20 &	0.08 & $\pm$ 0.01 &  1     \\
   GU\,Nor & 0.5382 &  6663 & $\pm$ 450 &    0.08 & $\pm$ 0.06 & $-$0.08 & $\pm$ 0.04 & $-$0.08 & $\pm$ 0.19 &         & \ldots     & $-$0.23 & $\pm$ 0.11 & $-$0.07 & $\pm$ 0.08 &  1     \\
   IQ\,Nor & 0.9159 &  6691 & $\pm$ 448 &    0.22 & $\pm$ 0.07 & $-$0.06 & $\pm$ 0.15 & $-$0.15 & $\pm$ 0.09 &         & \ldots     & $-$0.06 & $\pm$ 0.11 & $-$0.05 & $\pm$ 0.06 &  1     \\
   RS\,Nor & 0.7923 &  6385 & $\pm$ 449 &    0.18 & $\pm$ 0.08 &    0.15 & $\pm$ 0.09 &    0.01 & $\pm$ 0.04 &         & \ldots     &	 0.01 & $\pm$ 0.11 &	0.17 & $\pm$ 0.02 &  1     \\
   TW\,Nor & 1.0329 &  6160 & $\pm$ 447 &    0.27 & $\pm$ 0.10 & $-$0.10 & $\pm$ 0.18 & $-$0.19 & $\pm$ 0.15 & $-$0.11 & $\pm$ 0.11 & $-$0.15 & $\pm$ 0.50 &	0.08 & $\pm$ 0.08 &  1     \\
 V340\,Nor & 1.0526 &  6483 & $\pm$ 449 &    0.07 & $\pm$ 0.07 & $-$0.31 & $\pm$ 0.04 & $-$0.13 & $\pm$ 0.26 &         & \ldots     & $-$0.15 & $\pm$ 0.11 & $-$0.18 & $\pm$ 0.02 &  1     \\
   CS\,Ori & 0.5899 & 11701 & $\pm$ 458 & $-$0.25 & $\pm$ 0.06 & $-$0.27 & $\pm$ 0.09 & 	& \ldots     &         & \ldots     & $-$0.21 & $\pm$ 0.11 & $-$0.25 & $\pm$ 0.08 &  1     \\
   RS\,Ori & 0.8789 &  9470 & $\pm$ 453 &    0.11 & $\pm$ 0.09 &    0.15 & $\pm$ 0.01 &    0.12 & $\pm$ 0.11 &    0.37 & $\pm$ 0.01 &	 0.25 & $\pm$ 0.11 &	0.21 & $\pm$ 0.08 &  1     \\
   AQ\,Pup & 1.4786 &  9472 & $\pm$ 436 &    0.06 & $\pm$ 0.05 &    0.48 & $\pm$ 0.08 &    0.05 & $\pm$ 0.14 & $-$0.13 & $\pm$ 0.11 &	 0.03 & $\pm$ 0.11 &	0.12 & $\pm$ 0.15 &  1     \\
   BC\,Pup & 0.5495 & 12763 & $\pm$ 426 & $-$0.31 & $\pm$ 0.07 & $-$0.35 & $\pm$ 0.16 & $-$0.15 & $\pm$ 0.10 &    0.07 & $\pm$ 0.11 &	 0.37 & $\pm$ 0.11 &	0.02 & $\pm$ 0.10 &  1     \\
   BM\,Pup & 0.8572 &  9981 & $\pm$ 435 & $-$0.07 & $\pm$ 0.08 & $-$0.05 & $\pm$ 0.01 & $-$0.11 & $\pm$ 0.16 &    0.19 & $\pm$ 0.11 &	 0.06 & $\pm$ 0.23 &	0.04 & $\pm$ 0.08 &  1     \\
   BN\,Pup & 1.1359 &  9930 & $\pm$ 428 &    0.03 & $\pm$ 0.05 &    0.30 & $\pm$ 0.18 &    0.09 & $\pm$ 0.14 &    0.12 & $\pm$ 0.11 &	 0.07 & $\pm$ 0.13 &	0.16 & $\pm$ 0.16 &  1     \\
   HW\,Pup & 1.1289 & 13554 & $\pm$ 436 & $-$0.22 & $\pm$ 0.09 & $-$0.16 & $\pm$ 0.07 & $-$0.18 & $\pm$ 0.12 & $-$0.10 & $\pm$ 0.11 & $-$0.15 & $\pm$ 0.16 & $-$0.07 & $\pm$ 0.06 &  1     \\
   LS\,Pup & 1.1506 & 10610 & $\pm$ 423 & $-$0.12 & $\pm$ 0.11 & $-$0.11 & $\pm$ 0.06 & $-$0.01 & $\pm$ 0.13 & $-$0.04 & $\pm$ 0.11 &	 0.03 & $\pm$ 0.33 &	0.01 & $\pm$ 0.01 &  1     \\
   VW\,Pup & 0.6320 & 10175 & $\pm$ 443 & $-$0.14 & $\pm$ 0.06 & $-$0.35 & $\pm$ 0.16 & $-$0.38 & $\pm$ 0.05 &         & \ldots     & $-$0.46 & $\pm$ 0.11 & $-$0.18 & $\pm$ 0.08 &  1     \\
   VZ\,Pup & 1.3649 & 10867 & $\pm$ 425 & $-$0.01 & $\pm$ 0.04 & $-$0.01 & $\pm$ 0.05 &    0.09 & $\pm$ 0.11 &         & \ldots     &	 0.18 & $\pm$ 0.11 &	0.06 & $\pm$ 0.08 &  1     \\
   WW\,Pup & 0.7417 & 10382 & $\pm$ 436 &    0.13 & $\pm$ 0.16 & $-$0.30 & $\pm$ 0.06 & $-$0.23 & $\pm$ 0.09 &         & \ldots     & $-$0.44 & $\pm$ 0.11 & $-$0.07 & $\pm$ 0.07 &  1     \\
   WY\,Pup & 0.7202 & 10549 & $\pm$ 430 & $-$0.10 & $\pm$ 0.08 & $-$0.43 & $\pm$ 0.08 & $-$0.18 & $\pm$ 0.04 &         & \ldots     &	 0.13 & $\pm$ 0.03 & $-$0.03 & $\pm$ 0.08 &  1     \\
   WZ\,Pup & 0.7013 & 10123 & $\pm$ 437 & $-$0.07 & $\pm$ 0.06 & $-$0.16 & $\pm$ 0.08 & $-$0.09 & $\pm$ 0.05 &    0.01 & $\pm$ 0.11 & $-$0.09 & $\pm$ 0.30 & $-$0.04 & $\pm$ 0.02 &  1     \\
    X\,Pup & 1.4143 &  9788 & $\pm$ 441 &    0.02 & $\pm$ 0.08 & $-$0.15 & $\pm$ 0.02 & $-$0.02 & $\pm$ 0.10 & $-$0.13 & $\pm$ 0.11 &	 0.03 & $\pm$ 0.11 & $-$0.26 & $\pm$ 0.01 &  1     \\
 V470\,Sco & 1.2112 &  6461 & $\pm$ 454 &    0.16 & $\pm$ 0.06 &    0.11 & $\pm$ 0.09 & $-$0.07 & $\pm$ 0.09 & $-$0.12 & $\pm$ 0.11 &	 0.03 & $\pm$ 0.11 &	0.16 & $\pm$ 0.10 &  1     \\
   EV\,Sct & 0.4901 &  6135 & $\pm$ 449 &    0.09 & $\pm$ 0.07 &    0.01 & $\pm$ 0.04 &    0.07 & $\pm$ 0.09 &         & \ldots     &	 0.15 & $\pm$ 0.11 &	0.20 & $\pm$ 0.09 &  1     \\
    X\,Sct & 0.6230 &  6464 & $\pm$ 452 &    0.12 & $\pm$ 0.09 &    0.04 & $\pm$ 0.16 & $-$0.07 & $\pm$ 0.10 &         & \ldots     &	 0.03 & $\pm$ 0.11 &	0.08 & $\pm$ 0.07 &  1     \\
   AA\,Ser & 1.2340 &  5572 & $\pm$ 437 &    0.38 & $\pm$ 0.20 &    0.39 & $\pm$ 0.06 &    0.09 & $\pm$ 0.15 &    0.04 & $\pm$ 0.11 &	 0.15 & $\pm$ 0.71 &	0.30 & $\pm$ 0.18 &  1     \\
   CR\,Ser & 0.7244 &  6510 & $\pm$ 452 &    0.12 & $\pm$ 0.08 &    0.05 & $\pm$ 0.04 & $-$0.15 & $\pm$ 0.16 &    0.06 & $\pm$ 0.11 &	 0.08 & $\pm$ 0.49 &	0.15 & $\pm$ 0.18 &  1     \\
   AY\,Sgr & 0.8175 &  6429 & $\pm$ 452 &    0.11 & $\pm$ 0.06 &    0.07 & $\pm$ 0.16 & $-$0.05 & $\pm$ 0.04 & $-$0.04 & $\pm$ 0.11 &	 0.09 & $\pm$ 0.11 &	0.08 & $\pm$ 0.08 &  1     \\
V1954\,Sgr & 0.7909 &  5687 & $\pm$ 456 &    0.24 & $\pm$ 0.10 &    0.07 & $\pm$ 0.08 & $-$0.11 & $\pm$ 0.12 &         & \ldots     & $-$0.07 & $\pm$ 0.11 &	0.04 & $\pm$ 0.06 &  1     \\
 V773\,Sgr & 0.7596 &  6595 & $\pm$ 454 &    0.11 & $\pm$ 0.06 & $-$0.07 & $\pm$ 0.03 & $-$0.04 & $\pm$ 0.09 &         & \ldots     & $-$0.09 & $\pm$ 0.11 & $-$0.06 & $\pm$ 0.18 &  1     \\
   EZ\,Vel & 1.5383 & 12119 & $\pm$ 358 & $-$0.17 & $\pm$ 0.15 & $-$0.02 & $\pm$ 0.08 & $-$0.01 & $\pm$ 0.23 &    0.13 & $\pm$ 0.11 & $-$0.08 & $\pm$ 0.11 &	0.04 & $\pm$ 0.10 &  1     \\
\hline
\end{tabular}}
\tablefoot{The weighted (in the case of iron) or the arithmetic (for the
other elements) mean abundances of the stars with multiple spectra
(Table~\ref{table_ab_heavy_spec}) are listed first. Columns 2 and 3 shows
the logarithmic of the pulsation period and the Galactocentric distance
(\RG), respectively. The $N_{\rm S}$ values indicate the number of spectra
available for each star.}
\end{table*}

\clearpage
\begin{table*}[p]
\centering
\caption[]{Hyperfine structure list for some of the lines in our linelist.}
\label{table_hfs}
\begin{tabular}{lcc | lcc | lcc}
\hline\hline
 & & & & & \\[-0.2cm]
specie & $\lambda$ [\AA] & \loggf\ &
specie & $\lambda$ [\AA] & \loggf\ &
specie & $\lambda$ [\AA] & \loggf\ \\[0.1cm]
\hline
 & & & & & \\[-0.2cm]
\ion{Y}{II}  & 5119.113 & --1.758 & \ion{La}{II} & 6262.166 & --2.471 & \ion{Eu}{II} & 6437.630 & --2.281 \\
\ion{Y}{II}  & 5119.109 & --2.904 & \ion{La}{II} & 6262.169 & --2.596 & \ion{Eu}{II} & 6437.634 & --2.281 \\
\ion{Y}{II}  & 5119.111 & --1.603 & \ion{La}{II} & 6262.171 & --3.269 & \ion{Eu}{II} & 6437.637 & --1.391 \\
\ion{Y}{II}  & 5289.814 & --2.248 & \ion{La}{II} & 6262.211 & --2.286 & \ion{Eu}{II} & 6437.637 & --1.463 \\
\ion{Y}{II}  & 5289.816 & --3.394 & \ion{La}{II} & 6262.215 & --2.535 & \ion{Eu}{II} & 6437.640 & --2.238 \\
\ion{Y}{II}  & 5289.815 & --2.093 & \ion{La}{II} & 6262.218 & --3.290 & \ion{Eu}{II} & 6437.640 & --2.238 \\
\ion{Y}{II}  & 5728.889 & --1.564 & \ion{La}{II} & 6262.269 & --2.130 & \ion{Eu}{II} & 6437.642 & --2.292 \\
\ion{Y}{II}  & 5728.891 & --2.518 & \ion{La}{II} & 6262.274 & --2.531 & \ion{Eu}{II} & 6437.642 & --2.489 \\
\ion{Y}{II}  & 5728.888 & --2.518 & \ion{La}{II} & 6262.277 & --3.400 & \ion{Eu}{II} & 6437.642 & --1.569 \\
\ion{Y}{II}  & 5728.890 & --1.372 & \ion{La}{II} & 6262.340 & --1.994 & \ion{Eu}{II} & 6437.643 & --2.319 \\
\ion{La}{II} & 5114.512 & --1.624 & \ion{La}{II} & 6262.346 & --2.597 & \ion{Eu}{II} & 6437.644 & --1.711 \\
\ion{La}{II} & 5114.529 & --1.820 & \ion{La}{II} & 6262.350 & --3.612 & \ion{Eu}{II} & 6437.644 & --1.659 \\
\ion{La}{II} & 5114.556 & --1.820 & \ion{La}{II} & 6262.425 & --1.873 & \ion{Eu}{II} & 6437.644 & --2.292 \\
\ion{La}{II} & 5114.573 & --3.005 & \ion{La}{II} & 6262.431 & --2.802 & \ion{Eu}{II} & 6437.645 & --2.489 \\
\ion{La}{II} & 5114.586 & --1.824 & \ion{La}{II} & 6262.437 & --4.015 & \ion{Eu}{II} & 6437.650 & --2.319 \\
\ion{La}{II} & 5114.608 & --1.824 & \ion{La}{II} & 6390.460 & --2.012 & \ion{Eu}{II} & 6437.656 & --1.502 \\
\ion{La}{II} & 5114.621 & --2.079 & \ion{La}{II} & 6390.472 & --2.753 & \ion{Eu}{II} & 6437.662 & --2.276 \\
\ion{La}{II} & 5290.787 & --3.048 & \ion{La}{II} & 6390.472 & --2.183 & \ion{Eu}{II} & 6437.667 & --2.276 \\
\ion{La}{II} & 5290.788 & --2.872 & \ion{La}{II} & 6390.482 & --3.753 & \ion{Eu}{II} & 6437.672 & --1.607 \\
\ion{La}{II} & 5290.796 & --2.872 & \ion{La}{II} & 6390.482 & --2.570 & \ion{Eu}{II} & 6437.677 & --2.330 \\
\ion{La}{II} & 5290.797 & --4.796 & \ion{La}{II} & 6390.483 & --2.390 & \ion{Eu}{II} & 6437.679 & --2.330 \\
\ion{La}{II} & 5290.796 & --2.699 & \ion{La}{II} & 6390.491 & --3.335 & \ion{Eu}{II} & 6437.684 & --1.697 \\
\ion{La}{II} & 5290.808 & --2.699 & \ion{La}{II} & 6390.491 & --2.536 & \ion{Eu}{II} & 6437.687 & --2.527 \\
\ion{La}{II} & 5290.809 & --3.370 & \ion{La}{II} & 6390.492 & --2.661 & \ion{Eu}{II} & 6437.689 & --2.527 \\
\ion{La}{II} & 5290.810 & --2.688 & \ion{La}{II} & 6390.498 & --3.101 & \ion{Eu}{II} & 6437.692 & --1.749 \\
\ion{La}{II} & 5290.825 & --2.688 & \ion{La}{II} & 6390.499 & --2.595 & \ion{Eu}{II} & 6645.098 & --0.837 \\
\ion{La}{II} & 5290.826 & --2.674 & \ion{La}{II} & 6390.499 & --3.079 & \ion{Eu}{II} & 6645.099 & --2.106 \\
\ion{La}{II} & 5290.827 & --2.846 & \ion{La}{II} & 6390.503 & --2.955 & \ion{Eu}{II} & 6645.100 & --0.799 \\
\ion{La}{II} & 5290.846 & --2.846 & \ion{La}{II} & 6390.504 & --2.778 & \ion{Eu}{II} & 6645.101 & --3.749 \\
\ion{La}{II} & 5290.847 & --2.276 & \ion{La}{II} & 6390.507 & --2.858 & \ion{Eu}{II} & 6645.104 & --2.144 \\
\ion{La}{II} & 5805.620 & --3.387 & \ion{La}{II} & 6774.157 & --3.869 & \ion{Eu}{II} & 6645.112 & --3.787 \\
\ion{La}{II} & 5805.621 & --3.241 & \ion{La}{II} & 6774.168 & --3.392 & \ion{Eu}{II} & 6645.114 & --0.875 \\
\ion{La}{II} & 5805.636 & --3.045 & \ion{La}{II} & 6774.159 & --3.392 & \ion{Eu}{II} & 6645.116 & --1.911 \\
\ion{La}{II} & 5805.638 & --3.058 & \ion{La}{II} & 6774.187 & --3.170 & \ion{Eu}{II} & 6645.121 & --3.432 \\
\ion{La}{II} & 5805.639 & --3.718 & \ion{La}{II} & 6774.173 & --3.170 & \ion{Eu}{II} & 6645.123 & --0.954 \\
\ion{La}{II} & 5805.663 & --2.811 & \ion{La}{II} & 6774.190 & --4.015 & \ion{Eu}{II} & 6645.124 & --0.913 \\
\ion{La}{II} & 5805.666 & --2.936 & \ion{La}{II} & 6774.214 & --3.071 & \ion{Eu}{II} & 6645.128 & --1.866 \\
\ion{La}{II} & 5805.668 & --3.609 & \ion{La}{II} & 6774.196 & --3.071 & \ion{Eu}{II} & 6645.131 & --1.037 \\
\ion{La}{II} & 5805.702 & --2.626 & \ion{La}{II} & 6774.219 & --3.385 & \ion{Eu}{II} & 6645.131 & --1.949 \\
\ion{La}{II} & 5805.705 & --2.875 & \ion{La}{II} & 6774.250 & --3.045 & \ion{Eu}{II} & 6645.134 & --3.359 \\
\ion{La}{II} & 5805.708 & --3.630 & \ion{La}{II} & 6774.226 & --3.045 & \ion{Eu}{II} & 6645.136 & --1.121 \\
\ion{La}{II} & 5805.752 & --2.470 & \ion{La}{II} & 6774.256 & --2.999 & \ion{Eu}{II} & 6645.137 & --1.917 \\
\ion{La}{II} & 5805.756 & --2.871 & \ion{La}{II} & 6774.293 & --3.098 & \ion{Eu}{II} & 6645.139 & --3.470 \\
\ion{La}{II} & 5805.759 & --3.740 & \ion{La}{II} & 6774.265 & --3.098 & \ion{Eu}{II} & 6645.140 & --1.204 \\
\ion{La}{II} & 5805.813 & --2.334 & \ion{La}{II} & 6774.301 & --2.713 & \ion{Eu}{II} & 6645.142 & --2.112 \\
\ion{La}{II} & 5805.818 & --2.937 & \ion{La}{II} & 6774.343 & --3.293 & \ion{Eu}{II} & 6645.143 & --3.527 \\
\ion{La}{II} & 5805.822 & --3.952 & \ion{La}{II} & 6774.311 & --3.293 & \ion{Eu}{II} & 6645.145 & --0.993 \\
\ion{La}{II} & 5805.886 & --2.213 & \ion{La}{II} & 6774.353 & --2.485 & \ion{Eu}{II} & 6645.153 & --1.904 \\
\ion{La}{II} & 5805.891 & --3.142 & \ion{Eu}{II} & 6437.610 & --1.243 & \ion{Eu}{II} & 6645.160 & --3.397 \\
\ion{La}{II} & 5805.896 & --4.355 & \ion{Eu}{II} & 6437.611 & --1.281 & \ion{Eu}{II} & 6645.164 & --1.075 \\
\ion{La}{II} & 6262.116 & --3.047 & \ion{Eu}{II} & 6437.613 & --2.473 & \ion{Eu}{II} & 6645.171 & --1.956 \\
\ion{La}{II} & 6262.117 & --2.901 & \ion{Eu}{II} & 6437.619 & --2.511 & \ion{Eu}{II} & 6645.177 & --3.565 \\
\ion{La}{II} & 6262.134 & --2.705 & \ion{Eu}{II} & 6437.624 & --2.473 & \ion{Eu}{II} & 6645.179 & --1.159 \\
\ion{La}{II} & 6262.136 & --2.718 & \ion{Eu}{II} & 6437.627 & --1.353 & \ion{Eu}{II} & 6645.185 & --2.150 \\
\ion{La}{II} & 6262.138 & --3.378 & \ion{Eu}{II} & 6437.629 & --2.511 & \ion{Eu}{II} & 6645.190 & --1.242 \\
\hline\noalign{\smallskip}
\end{tabular}
\tablefoot{The second column displays the wavelength of each component
that forms the line profile.}
\end{table*}

\clearpage
\begin{table*}[p]
\centering
\caption[]{Abundance difference of stars in common among the current sample
           and other data sets.}
\label{table_diff_ab}
\begin{tabular}{cc r@{ }l c}
\noalign{\smallskip}\hline\hline\noalign{\smallskip}
\parbox[c]{1.6cm}{\centering Abundance ratio} &
Data sets$^1$ &
\multicolumn{2}{c}{\parbox[c]{1.6cm}{\centering Zero-point difference}} &
$N_{\rm Common}$ \\
\noalign{\smallskip}\hline\noalign{\smallskip}
 {[Fe/H]} & LII--G14  & $-$0.05 & $\pm$ 0.11 & 45 \\
 {[Fe/H]} & LIII--G14 &    0.03 & $\pm$ 0.08 & 33 \\
 {[Fe/H]} & LII--LEM  &    0.08 & $\pm$ 0.12 & 51 \\
 {[Fe/H]} & LIII--YON &    0.34 & $\pm$ 0.20 & 20 \\[0.1cm]
 {[Y/H]}  & LII--TS   &    0.21 & $\pm$ 0.20 & 37 \\
 {[Y/H]}  & LIII--TS  &    0.15 & $\pm$ 0.18 & 34 \\
 {[Y/H]}  & LII--LEM  &    0.09 & $\pm$ 0.15 & 46 \\[0.1cm]
 {[La/H]} & LII--TS   &    0.29 & $\pm$ 0.18 & 40 \\
 {[La/H]} & LIII--TS  &    0.31 & $\pm$ 0.16 & 34 \\
 {[La/H]} & LII--LEM  &    0.06 & $\pm$ 0.20 & 47 \\
 {[La/H]} & LIII--YON &    0.27 & $\pm$ 0.33 & 16 \\[0.1cm]
 {[Ce/H]} & LII--TS   & $-$0.06 & $\pm$ 0.16 & 24 \\
 {[Ce/H]} & LIII--TS  &    0.17 & $\pm$ 0.17 & 19 \\
 {[Ce/H]} & LII--LEM  & $-$0.21 & $\pm$ 0.21 & 50 \\[0.1cm]
 {[Nd/H]} & LII--TS   &    0.09 & $\pm$ 0.24 & 42 \\
 {[Nd/H]} & LIII--TS  &    0.15 & $\pm$ 0.18 & 33 \\
 {[Nd/H]} & LII--LEM  & $-$0.08 & $\pm$ 0.29 & 50 \\[0.1cm]
 {[Eu/H]} & LII--TS   &    0.10 & $\pm$ 0.17 & 41 \\
 {[Eu/H]} & LIII--TS  &    0.11 & $\pm$ 0.21 & 34 \\
 {[Eu/H]} & LII--LEM  & $-$0.09 & $\pm$ 0.20 & 53 \\
 {[Eu/H]} & LIII--YON &    0.12 & $\pm$ 0.25 & 15 \\
\hline\noalign{\smallskip}
\end{tabular}
\tablefoot{\tablefoottext{1}{G14: \citet{Genovalietal2014}; TS: this study;
LII: \citet{Lucketal2011}; LIII: \citet{LuckLambert2011};
LEM: \citet{Lemasleetal2013}; YON: \citet{Yongetal2006}. For Y, La, and Eu,
the differences were computed after accounting for the HFS affecting some
lines of these elements in the currect sample. The quoted errors represent
the dispersion around the mean.}}
\end{table*}

\clearpage
\begin{table*}[p]
\centering
\caption[]{Slopes and zero-points of the abundance gradients as a function
           of the Galactocentric distance and of the pulsation period.}
\label{table_slopes}
{\footnotesize 
\begin{tabular}{c r@{ }l r@{ }l cc r@{ }l r@{ }l r@{ }l r@{ }l}
\noalign{\smallskip}\hline\hline\noalign{\smallskip}
\parbox[c]{1.6cm}{\centering Abundance ratio} &
\multicolumn{2}{c}{\parbox[c]{1.4cm}{\centering Slope$^a$}} &
\multicolumn{2}{c}{\parbox[c]{1.6cm}{\centering Zero-point [dex]}} &
\parbox[c]{0.8cm}{\centering $\sigma$ [dex]} &
$N$ &
\multicolumn{2}{c}{\parbox[c]{1.4cm}{\centering Slope$^a$ (TS)}} &
\multicolumn{2}{c}{\parbox[c]{1.7cm}{\centering Slope$^a$ (LEM)}} &
\multicolumn{2}{c}{\parbox[c]{1.5cm}{\centering Slope$^a$ (LII)}} &
\multicolumn{2}{c}{\parbox[c]{1.6cm}{\centering Slope$^a$ (LIII)}} \\
\noalign{\smallskip}\hline\noalign{\smallskip}
\multicolumn{15}{c}{as a function of \RG} \\
\noalign{\smallskip}\hline\noalign{\smallskip}
  {[Y/H]}   & $-$0.053 & $\pm$ 0.003 &    0.43 & $\pm$ 0.03 & 0.14 & 429 & $-$0.033 & $\pm$ 0.007 & $-$0.062 & $\pm$ 0.012 & $-$0.044 & $\pm$ 0.004 & $-$0.061 & $\pm$ 0.003 \\
 {[La/H]}   & $-$0.020 & $\pm$ 0.003 &    0.13 & $\pm$ 0.03 & 0.14 & 424 &    0.002 & $\pm$ 0.005 & $-$0.045 & $\pm$ 0.012 & $-$0.019 & $\pm$ 0.005 & $-$0.031 & $\pm$ 0.004 \\
 {[Ce/H]}   & $-$0.024 & $\pm$ 0.003 &    0.20 & $\pm$ 0.03 & 0.14 & 421 &    0.008 & $\pm$ 0.007 & $-$0.043 & $\pm$ 0.012 & $-$0.021 & $\pm$ 0.004 & $-$0.034 & $\pm$ 0.003 \\
 {[Nd/H]}   & $-$0.025 & $\pm$ 0.003 &    0.24 & $\pm$ 0.03 & 0.13 & 430 &    0.006 & $\pm$ 0.006 & $-$0.046 & $\pm$ 0.013 & $-$0.006 & $\pm$ 0.004 & $-$0.037 & $\pm$ 0.003 \\
 {[Eu/H]}   & $-$0.030 & $\pm$ 0.004 &    0.28 & $\pm$ 0.03 & 0.16 & 420 & $-$0.013 & $\pm$ 0.005 & $-$0.066 & $\pm$ 0.013 & $-$0.021 & $\pm$ 0.004 & $-$0.042 & $\pm$ 0.005 \\[0.1cm]
 {[La/Fe]}  &    0.035 & $\pm$ 0.003 & $-$0.40 & $\pm$ 0.03 & 0.13 & 425 &    0.057 & $\pm$ 0.006 &    0.011 & $\pm$ 0.011 &    0.043 & $\pm$ 0.004 &    0.029 & $\pm$ 0.003 \\
 {[Ce/Fe]}  &    0.027 & $\pm$ 0.003 & $-$0.31 & $\pm$ 0.02 & 0.12 & 419 &    0.063 & $\pm$ 0.008 &    0.009 & $\pm$ 0.012 &    0.033 & $\pm$ 0.004 &    0.027 & $\pm$ 0.002 \\
 {[Nd/Fe]}  &    0.027 & $\pm$ 0.002 & $-$0.26 & $\pm$ 0.02 & 0.10 & 427 &    0.057 & $\pm$ 0.006 & $-$0.023 & $\pm$ 0.011 &    0.045 & $\pm$ 0.004 &    0.023 & $\pm$ 0.002 \\
 {[Eu/Fe]}  &    0.025 & $\pm$ 0.003 & $-$0.26 & $\pm$ 0.03 & 0.14 & 420 &    0.043 & $\pm$ 0.006 & $-$0.007 & $\pm$ 0.010 &    0.030 & $\pm$ 0.003 &    0.015 & $\pm$ 0.004 \\
\noalign{\smallskip}\hline\noalign{\smallskip}
\multicolumn{15}{c}{as a function of \logP} \\
\noalign{\smallskip}\hline\noalign{\smallskip}
  {[Y/H]}   &    0.20  & $\pm$ 0.03  & $-$0.20 & $\pm$ 0.03 & 0.17 & 430 &    0.28  & $\pm$ 0.06  &    0.27  & $\pm$ 0.07  &    0.12  & $\pm$ 0.03  &    0.13  & $\pm$ 0.03  \\
 {[La/H]}   &    0.10  & $\pm$ 0.02  & $-$0.13 & $\pm$ 0.02 & 0.14 & 424 &    0.06  & $\pm$ 0.04  &    0.16  & $\pm$ 0.07  &    0.04  & $\pm$ 0.03  &    0.11  & $\pm$ 0.03  \\
 {[Ce/H]}   &    0.06  & $\pm$ 0.02  & $-$0.07 & $\pm$ 0.02 & 0.13 & 417 & $-$0.10  & $\pm$ 0.07  &    0.10  & $\pm$ 0.07  &    0.00  & $\pm$ 0.03  &    0.04  & $\pm$ 0.03  \\
 {[Nd/H]}   &    0.10  & $\pm$ 0.02  & $-$0.07 & $\pm$ 0.02 & 0.14 & 430 & $-$0.03  & $\pm$ 0.05  &    0.13  & $\pm$ 0.07  &    0.04  & $\pm$ 0.03  &    0.12  & $\pm$ 0.02  \\
 {[Eu/H]}   &    0.15  & $\pm$ 0.03  & $-$0.11 & $\pm$ 0.03 & 0.17 & 418 &    0.08  & $\pm$ 0.04  &    0.26  & $\pm$ 0.08  &    0.09  & $\pm$ 0.03  &    0.15  & $\pm$ 0.04  \\[0.1cm]
 {[Ce/Fe]}  & $-$0.09  & $\pm$ 0.02  &    0.02 & $\pm$ 0.02 & 0.13 & 423 &  $-$0.55 & $\pm$ 0.10  &  $-$0.04 & $\pm$ 0.07  & $-$0.08  & $\pm$ 0.03  & $-$0.09  & $\pm$ 0.02  \\
\hline
\end{tabular}
}
\tablefoot{\tablefoottext{a}{In units of \dexkpc\ if in function of \RG, and
dex per logarithmic day if in function of \logP}.
Columns from 2 to 5 shows the results for all the different samples fitted
together. We also list the standard deviation ($\sigma$) of the residuals
and the number of data points ($N$) used in the fit. The slopes using only
the stars of our sample (TS: this study) and of previous studies (LEM, LII,
and LIII) are shown for comparison.}
\end{table*}

\clearpage
\begin{table*}[p]
\centering
\caption[]{Galactic Cepheids for which the abundances heavy elements was
           available in the literature.}
\label{table_liter}
{\scriptsize 
\begin{tabular}{r r@{}l r@{}l c r@{}l r@{}l c r@{}l r@{}l c r@{}l r@{}l c r@{}l r@{}l c}
\noalign{\smallskip}\hline\hline\noalign{\smallskip}
Name &
\multicolumn{2}{c}{[Y/H]$_{\rm lit}$} & \multicolumn{2}{c}{[Y/H]} & Ref. &
\multicolumn{2}{c}{[La/H]$_{\rm lit}$} & \multicolumn{2}{c}{[La/H]} & Ref. &
\multicolumn{2}{c}{[Ce/H]$_{\rm lit}$} & \multicolumn{2}{c}{[Ce/H]} & Ref. &
\multicolumn{2}{c}{[Nd/H]$_{\rm lit}$} & \multicolumn{2}{c}{[Nd/H]} & Ref. &
\multicolumn{2}{c}{[Eu/H]$_{\rm lit}$} & \multicolumn{2}{c}{[Eu/H]} & Ref. \\
\noalign{\smallskip}\hline\noalign{\smallskip}
       T\,Ant & $-$0&.06     & $-$0&.21     &   LIII &    0&.19     & $-$0&.12     &   LIII &	 0&.02     & $-$0&.15     &   LIII &	0&.04	  & $-$0&.11	 &   LIII &    0&.08	 & $-$0&.03	&   LIII \\
      BC\,Aql & $-$0&.11     & $-$0&.26     &   LIII & $-$0&.19     & $-$0&.50     &   LIII & $-$0&.35     & $-$0&.52     &   LIII & $-$0&.60	  & $-$0&.75	 &   LIII &    0&.10	 & $-$0&.01	&   LIII \\
      EV\,Aql &    0&.19     &    0&.04     &   LIII &    0&.34     &	 0&.03     &   LIII &	 0&.18     &	0&.01     &   LIII &	0&.19	  &    0&.04	 &   LIII &    0&.31	 &    0&.20	&   LIII \\
      FF\,Aql &    0&.27     &    0&.06     &    LII &    0&.23     & $-$0&.06     &	LII & $-$0&.14     & $-$0&.08     &    LII &	0&.12	  &    0&.03	 &    LII &    0&.14	 &    0&.04	&    LII \\
      FM\,Aql &    0&.31     &    0&.16     &   LIII &    0&.65     &	 0&.34     &   LIII &	 0&.29     &	0&.12     &   LIII &	0&.39	  &    0&.24	 &   LIII &    0&.22	 &    0&.11	&   LIII \\
      FN\,Aql &    0&.05     & $-$0&.10     &   LIII &    0&.13     & $-$0&.18     &   LIII &	 0&.02     & $-$0&.15     &   LIII &	0&.07	  & $-$0&.08	 &   LIII &    0&.05	 & $-$0&.06	&   LIII \\
      KL\,Aql &    0&.38     &    0&.23     &   LIII &    0&.39     &	 0&.08     &   LIII &	 0&.35     &	0&.18     &   LIII &	0&.36	  &    0&.21	 &   LIII &    0&.34	 &    0&.23	&   LIII \\
      SZ\,Aql &    0&.37     &    0&.16     &    LII &    0&.27     & $-$0&.02     &	LII & $-$0&.09     & $-$0&.03     &    LII &	0&.15	  &    0&.06	 &    LII &    0&.14	 &    0&.04	&    LII \\
      TT\,Aql &    0&.28     &    0&.13     &   LIII &    0&.33     &	 0&.02     &   LIII &	 0&.32     &	0&.15     &   LIII &	0&.25	  &    0&.10	 &   LIII &    0&.17	 &    0&.06	&   LIII \\
       U\,Aql &    0&.24     &    0&.09     &   LIII &    0&.36     &	 0&.05     &   LIII &	 0&.21     &	0&.04     &   LIII &	0&.26	  &    0&.11	 &   LIII &    0&.16	 &    0&.05	&   LIII \\
   V1162\,Aql &    0&.21     & $-$0&.00     &    LII &    0&.09     & $-$0&.20     &	LII & $-$0&.24     & $-$0&.18     &    LII &	0&.02	  & $-$0&.07	 &    LII & $-$0&.06	 & $-$0&.16	&    LII \\
   V1344\,Aql &    0&.16     &    0&.01     &   LIII &    0&.28     & $-$0&.03     &   LIII &	 0&.13     & $-$0&.04     &   LIII &	0&.17	  &    0&.02	 &   LIII &    0&.03	 & $-$0&.08	&   LIII \\
   V1359\,Aql &    0&.38     &    0&.23     &   LIII &    0&.24     & $-$0&.07     &   LIII &	 0&.25     &	0&.08     &   LIII &	0&.28	  &    0&.13	 &   LIII &    0&.12	 &    0&.01	&   LIII \\
    V336\,Aql &    0&.27     &    0&.12     &   LIII &    0&.30     & $-$0&.01     &   LIII &	 0&.15     & $-$0&.02     &   LIII &	0&.24	  &    0&.09	 &   LIII &    0&.21	 &    0&.10	&   LIII \\
    V493\,Aql &    0&.02     & $-$0&.13     &   LIII &    0&.25     & $-$0&.06     &   LIII & $-$0&.02     & $-$0&.19     &   LIII &	0&.05	  & $-$0&.10	 &   LIII & $-$0&.01	 & $-$0&.12	&   LIII \\
    V496\,Aql &    0&.14     & $-$0&.07     &    LII &    0&.09     & $-$0&.20     &	LII & $-$0&.23     & $-$0&.17     &    LII &	0&.02	  & $-$0&.07	 &    LII & $-$0&.01	 & $-$0&.11	&    LII \\
    V526\,Aql &    0&.43     &    0&.28     &   LIII &    0&.56     &	 0&.25     &   LIII &	 0&.36     &	0&.19     &   LIII &	0&.46	  &    0&.31	 &   LIII &    0&.24	 &    0&.13	&   LIII \\
    V600\,Aql &    0&.16     & $-$0&.05     &    LII &    0&.15     & $-$0&.14     &	LII & $-$0&.05     &	0&.01     &    LII &	0&.03	  & $-$0&.06	 &    LII &    0&.00	 & $-$0&.10	&    LII \\
    V733\,Aql &    0&.16     & $-$0&.05     &    LII &    0&.21     & $-$0&.08     &	LII & $-$0&.13     & $-$0&.07     &    LII &	0&.02	  & $-$0&.07	 &    LII & $-$0&.08	 & $-$0&.18	&    LII \\
    V916\,Aql &    0&.34     &    0&.19     &   LIII &    0&.32     &	 0&.01     &   LIII &	 0&.36     &	0&.19     &   LIII &	0&.27	  &    0&.12	 &   LIII &    0&.19	 &    0&.08	&   LIII \\
  $\eta$\,Aql &    0&.27     &    0&.06     &    LII &    0&.24     & $-$0&.05     &	LII & $-$0&.07     & $-$0&.01     &    LII &	0&.16	  &    0&.07	 &    LII &    0&.09	 & $-$0&.01	&    LII \\
    V340\,Ara &    0&.04     &    0&.04     &     TS & $-$0&.04     & $-$0&.04     &	 TS &	 0&.13     &	0&.13     &	TS &	0&.02	  &    0&.02	 &     TS &    0&.13	 &    0&.13	&     TS \\
      AN\,Aur &    0&.09     & $-$0&.06     &   LIII &    0&.27     & $-$0&.04     &   LIII &	 0&.18     &	0&.01     &   LIII &	0&.20	  &    0&.05	 &   LIII & $-$0&.04	 & $-$0&.15	&   LIII \\
      AO\,Aur & $-$0&.40     & $-$0&.52     &    LEM & $-$0&.21     & $-$0&.44     &	LEM & $-$0&.17     & $-$0&.32     &    LEM & $-$0&.22	  & $-$0&.39	 &    LEM & $-$0&.15	 & $-$0&.34	&    LEM \\
      AS\,Aur & $-$0&.20     & $-$0&.20     &     TS & $-$0&.16     & $-$0&.16     &	 TS & $-$0&.05     & $-$0&.22     &   LIII & $-$0&.06	  & $-$0&.06	 &     TS & $-$0&.03	 & $-$0&.03	&     TS \\
      AX\,Aur & $-$0&.35     & $-$0&.47     &    LEM & $-$0&.22     & $-$0&.45     &	LEM & $-$0&.08     & $-$0&.23     &    LEM & $-$0&.10	  & $-$0&.27	 &    LEM & $-$0&.31	 & $-$0&.50	&    LEM \\
      BK\,Aur &    0&.08     & $-$0&.04     &    LEM &    0&.26     &	 0&.03     &	LEM &	 0&.31     &	0&.16     &    LEM &	0&.28	  &    0&.11	 &    LEM &    0&.27	 &    0&.08	&    LEM \\
      CO\,Aur &    0&.08     & $-$0&.07     &   LIII &    0&.10     & $-$0&.21     &   LIII &	 0&.13     & $-$0&.04     &   LIII &	0&.14	  & $-$0&.01	 &   LIII &    0&.13	 &    0&.02	&   LIII \\
      CY\,Aur &    0&.02     & $-$0&.13     &   LIII &    0&.36     &	 0&.05     &   LIII &	 0&.13     & $-$0&.04     &   LIII &	0&.15	  &    0&.00	 &   LIII &    0&.19	 &    0&.08	&   LIII \\
      ER\,Aur & $-$0&.10     & $-$0&.25     &   LIII &    0&.08     & $-$0&.23     &   LIII & $-$0&.03     & $-$0&.20     &   LIII &	0&.01	  & $-$0&.14	 &   LIII & $-$0&.18	 & $-$0&.29	&   LIII \\
      EW\,Aur & $-$0&.55     & $-$0&.70     &   LIII & $-$0&.13     & $-$0&.44     &   LIII & $-$0&.54     & $-$0&.71     &   LIII & $-$0&.38	  & $-$0&.53	 &   LIII & $-$0&.24	 & $-$0&.35	&   LIII \\
      FF\,Aur & $-$0&.47     & $-$0&.62     &   LIII &    0&.00     & $-$0&.31     &   LIII & $-$0&.11     & $-$0&.28     &   LIII & $-$0&.33	  & $-$0&.48	 &   LIII &     & \ldots &     & \ldots & \ldots \\
      GT\,Aur &    0&.03     & $-$0&.12     &   LIII &    0&.14     & $-$0&.17     &   LIII &	 0&.12     & $-$0&.05     &   LIII &	0&.18	  &    0&.03	 &   LIII & $-$0&.02	 & $-$0&.13	&   LIII \\
      GV\,Aur & $-$0&.21     & $-$0&.36     &   LIII &    0&.08     & $-$0&.23     &   LIII & $-$0&.04     & $-$0&.21     &   LIII & $-$0&.05	  & $-$0&.20	 &   LIII & $-$0&.07	 & $-$0&.18	&   LIII \\
      IN\,Aur & $-$0&.26     & $-$0&.41     &   LIII & $-$0&.07     & $-$0&.38     &   LIII & $-$0&.05     & $-$0&.22     &   LIII & $-$0&.09	  & $-$0&.24	 &   LIII &    0&.06	 & $-$0&.05	&   LIII \\
      RT\,Aur &    0&.26     &    0&.11     &   LIII &    0&.29     & $-$0&.02     &   LIII &	 0&.16     & $-$0&.01     &   LIII &	0&.17	  &    0&.02	 &   LIII &    0&.19	 &    0&.08	&   LIII \\
      RX\,Aur &    0&.21     &    0&.06     &   LIII &    0&.49     &	 0&.18     &   LIII &	 0&.17     & $-$0&.00     &   LIII &	0&.33	  &    0&.18	 &   LIII &    0&.24	 &    0&.13	&   LIII \\
      SY\,Aur & $-$0&.07     & $-$0&.19     &    LEM &    0&.04     & $-$0&.19     &	LEM &	 0&.22     &	0&.07     &    LEM &	0&.17	  & $-$0&.00	 &    LEM &    0&.14	 & $-$0&.05	&    LEM \\
    V335\,Aur & $-$0&.21     & $-$0&.36     &   LIII &     & \ldots &     & \ldots & \ldots & $-$0&.20     & $-$0&.37     &   LIII & $-$0&.04	  & $-$0&.19	 &   LIII &     & \ldots &     & \ldots & \ldots \\
    V637\,Aur & $-$0&.09     & $-$0&.24     &   LIII &    0&.09     & $-$0&.22     &   LIII &	 0&.07     & $-$0&.10     &   LIII &	0&.11	  & $-$0&.04	 &   LIII & $-$0&.03	 & $-$0&.14	&   LIII \\
       Y\,Aur & $-$0&.36     & $-$0&.48     &    LEM & $-$0&.24     & $-$0&.47     &	LEM & $-$0&.21     & $-$0&.36     &    LEM & $-$0&.24	  & $-$0&.41	 &    LEM & $-$0&.11	 & $-$0&.30	&    LEM \\
      YZ\,Aur & $-$0&.34     & $-$0&.46     &    LEM & $-$0&.19     & $-$0&.42     &	LEM &	 0&.25     &	0&.10     &    LEM & $-$0&.17	  & $-$0&.32	 &   LIII & $-$0&.16	 & $-$0&.35	&    LEM \\
      AO\,CMa &    0&.08     &    0&.08     &     TS &    0&.13     &	 0&.13     &	 TS &	 0&.18     &	0&.18     &	TS &	0&.11	  &    0&.11	 &     TS &    0&.20	 &    0&.20	&     TS \\
      RW\,CMa & $-$0&.06     & $-$0&.06     &     TS & $-$0&.04     & $-$0&.04     &	 TS &     & \ldots &     & \ldots & \ldots &     & \ldots &     & \ldots & \ldots & $-$0&.08	 & $-$0&.08	&     TS \\
      RY\,CMa & $-$0&.08     & $-$0&.20     &    LEM &    0&.13     & $-$0&.10     &	LEM &	 0&.20     &	0&.05     &    LEM &	0&.19	  &    0&.02	 &    LEM &    0&.28	 &    0&.09	&    LEM \\
      RZ\,CMa & $-$0&.16     & $-$0&.28     &    LEM & $-$0&.07     & $-$0&.30     &	LEM &	 0&.09     & $-$0&.06     &    LEM &	0&.14	  & $-$0&.03	 &    LEM &    0&.09	 & $-$0&.10	&    LEM \\
      SS\,CMa &    0&.14     &    0&.14     &     TS &    0&.06     &	 0&.06     &	 TS &	 0&.13     &	0&.13     &	TS &	0&.09	  &    0&.09	 &     TS &    0&.10	 &    0&.10	&     TS \\
      TV\,CMa &    0&.17     &    0&.17     &     TS & $-$0&.04     & $-$0&.04     &	 TS & $-$0&.04     & $-$0&.04     &	TS & $-$0&.10	  & $-$0&.10	 &     TS &    0&.04	 &    0&.04	&     TS \\
      TW\,CMa & $-$0&.06     & $-$0&.06     &     TS &    0&.17     &	 0&.17     &	 TS &	 0&.14     &	0&.14     &	TS &	0&.02	  &    0&.02	 &     TS &    0&.06	 &    0&.06	&     TS \\
      VZ\,CMa &     & \ldots &     & \ldots & \ldots &    0&.33     &	 0&.04     &	LII & $-$0&.02     &	0&.04     &    LII &	0&.41	  &    0&.32	 &    LII &    0&.22	 &    0&.12	&    LII \\
      XZ\,CMa &     & \ldots &     & \ldots & \ldots &     & \ldots &     & \ldots & \ldots &     & \ldots &     & \ldots & \ldots &     & \ldots &     & \ldots & \ldots &     & \ldots &     & \ldots & \ldots \\
      AB\,Cam &    0&.05     & $-$0&.10     &   LIII &    0&.22     & $-$0&.09     &   LIII &	 0&.14     & $-$0&.03     &   LIII &	0&.21	  &    0&.06	 &   LIII & $-$0&.03	 & $-$0&.14	&   LIII \\
      AC\,Cam &    0&.04     & $-$0&.11     &   LIII & $-$0&.04     & $-$0&.35     &   LIII & $-$0&.08     & $-$0&.25     &   LIII &	0&.03	  & $-$0&.12	 &   LIII &    0&.71	 &    0&.60	&   LIII \\
      AD\,Cam & $-$0&.25     & $-$0&.40     &   LIII &    0&.08     & $-$0&.23     &   LIII &	 0&.05     & $-$0&.12     &   LIII & $-$0&.07	  & $-$0&.22	 &   LIII & $-$0&.16	 & $-$0&.27	&   LIII \\
      AM\,Cam & $-$0&.01     & $-$0&.16     &   LIII &    0&.21     & $-$0&.10     &   LIII &	 0&.15     & $-$0&.02     &   LIII &	0&.01	  & $-$0&.14	 &   LIII &    0&.08	 & $-$0&.03	&   LIII \\
      CK\,Cam &    0&.12     & $-$0&.03     &   LIII &    0&.19     & $-$0&.12     &   LIII &	 0&.11     & $-$0&.06     &   LIII &	0&.13	  & $-$0&.02	 &   LIII &    0&.12	 &    0&.01	&   LIII \\
      LO\,Cam &    0&.16     &    0&.01     &   LIII &    0&.30     & $-$0&.01     &   LIII &	 0&.20     &	0&.03     &   LIII &	0&.22	  &    0&.07	 &   LIII &    0&.30	 &    0&.19	&   LIII \\
      MN\,Cam &    0&.10     & $-$0&.05     &   LIII &    0&.24     & $-$0&.07     &   LIII &	 0&.13     & $-$0&.04     &   LIII &	0&.23	  &    0&.08	 &   LIII &    0&.06	 & $-$0&.05	&   LIII \\
      MQ\,Cam & $-$0&.11     & $-$0&.26     &   LIII &    0&.19     & $-$0&.12     &   LIII &	 0&.01     & $-$0&.16     &   LIII &	0&.10	  & $-$0&.05	 &   LIII & $-$0&.12	 & $-$0&.23	&   LIII \\
      OX\,Cam &    0&.10     & $-$0&.05     &   LIII &    0&.31     &	 0&.00     &   LIII &	 0&.25     &	0&.08     &   LIII &	0&.29	  &    0&.14	 &   LIII &    0&.04	 & $-$0&.07	&   LIII \\
      PV\,Cam & $-$0&.08     & $-$0&.23     &   LIII &    0&.13     & $-$0&.18     &   LIII &	 0&.03     & $-$0&.14     &   LIII &	0&.00	  & $-$0&.15	 &   LIII & $-$0&.02	 & $-$0&.13	&   LIII \\
      QS\,Cam &    0&.15     & $-$0&.00     &   LIII &    0&.43     &	 0&.12     &   LIII &	 0&.15     & $-$0&.02     &   LIII &	0&.22	  &    0&.07	 &   LIII &    0&.54	 &    0&.43	&   LIII \\
      RW\,Cam & $-$0&.06     & $-$0&.21     &   LIII &    0&.38     &	 0&.07     &   LIII & $-$0&.14     & $-$0&.31     &   LIII &	0&.05	  & $-$0&.10	 &   LIII &    0&.15	 &    0&.04	&   LIII \\
      RX\,Cam &    0&.16     &    0&.01     &   LIII &    0&.34     &	 0&.03     &   LIII &	 0&.17     & $-$0&.00     &   LIII &	0&.22	  &    0&.07	 &   LIII &    0&.12	 &    0&.01	&   LIII \\
      TV\,Cam &    0&.14     & $-$0&.01     &   LIII &    0&.29     & $-$0&.02     &   LIII &	 0&.24     &	0&.07     &   LIII &	0&.29	  &    0&.14	 &   LIII & $-$0&.08	 & $-$0&.19	&   LIII \\
    V359\,Cam & $-$0&.23     & $-$0&.38     &   LIII &    0&.23     & $-$0&.08     &   LIII & $-$0&.05     & $-$0&.22     &   LIII & $-$0&.06	  & $-$0&.21	 &   LIII & $-$0&.11	 & $-$0&.22	&   LIII \\
      AQ\,Car & $-$0&.02     & $-$0&.14     &    LEM &    0&.02     & $-$0&.21     &	LEM &	 0&.02     & $-$0&.13     &    LEM &	0&.25	  &    0&.08	 &    LEM &    0&.18	 &    0&.07	&   LIII \\
      CN\,Car &    0&.34     &    0&.19     &   LIII &    0&.45     &	 0&.14     &   LIII &	 0&.24     &	0&.07     &   LIII &	0&.28	  &    0&.13	 &   LIII &    0&.43	 &    0&.32	&   LIII \\
      CY\,Car &    0&.18     &    0&.03     &   LIII &    0&.25     & $-$0&.06     &   LIII &	 0&.09     & $-$0&.08     &   LIII &	0&.17	  &    0&.02	 &   LIII &    0&.07	 & $-$0&.04	&   LIII \\
      DY\,Car &    0&.18     &    0&.03     &   LIII &    0&.21     & $-$0&.10     &   LIII &	 0&.01     & $-$0&.16     &   LIII &	0&.25	  &    0&.10	 &   LIII &    0&.01	 & $-$0&.09	&    LII \\
      ER\,Car &    0&.17     &    0&.02     &   LIII &    0&.30     & $-$0&.01     &   LIII &	 0&.18     &	0&.01     &   LIII &	0&.20	  &    0&.05	 &   LIII &    0&.14	 &    0&.03	&   LIII \\
      FI\,Car &    0&.25     &    0&.10     &   LIII &    0&.52     &	 0&.21     &   LIII &	 0&.35     &	0&.18     &   LIII &	0&.44	  &    0&.29	 &   LIII &    0&.44	 &    0&.33	&   LIII \\
      FR\,Car &    0&.20     &    0&.05     &   LIII &    0&.30     & $-$0&.01     &   LIII &	 0&.17     & $-$0&.00     &   LIII &	0&.21	  &    0&.06	 &   LIII &    0&.34	 &    0&.23	&   LIII \\
      GH\,Car &    0&.35     &    0&.20     &   LIII &    0&.41     &	 0&.10     &   LIII &	 0&.26     &	0&.09     &   LIII &	0&.30	  &    0&.15	 &   LIII &    0&.47	 &    0&.36	&   LIII \\
      GX\,Car &    0&.23     &    0&.08     &   LIII &    0&.30     & $-$0&.01     &   LIII &	 0&.16     & $-$0&.01     &   LIII &	0&.26	  &    0&.11	 &   LIII &    0&.04	 & $-$0&.07	&   LIII \\
\hline\noalign{\smallskip}
\multicolumn{26}{r}{\it {\footnotesize continued on next page}} \\
\end{tabular}}
\tablefoot{The columns first give the original abundance estimate available
           in the literature and then the abundances rescaled according to
	   the zero-point differences listed in Table~\ref{table_diff_ab}.
	   The priority was given in the following order: we first adopt the
	   abundances provided by our group, this study (TS) and LEM, and
	   then those provided by the other studies, LIII, LII, and YON.}
\end{table*}
\addtocounter{table}{-1}
\begin{table*}[p]
\centering
\caption[]{continued.}
{\scriptsize 
\begin{tabular}{r r@{}l r@{}l c r@{}l r@{}l c r@{}l r@{}l c r@{}l r@{}l c r@{}l r@{}l c}
\noalign{\smallskip}\hline\hline\noalign{\smallskip}
Name &
\multicolumn{2}{c}{[Y/H]$_{\rm lit}$} & \multicolumn{2}{c}{[Y/H]} & Ref. &
\multicolumn{2}{c}{[La/H]$_{\rm lit}$} & \multicolumn{2}{c}{[La/H]} & Ref. &
\multicolumn{2}{c}{[Ce/H]$_{\rm lit}$} & \multicolumn{2}{c}{[Ce/H]} & Ref. &
\multicolumn{2}{c}{[Nd/H]$_{\rm lit}$} & \multicolumn{2}{c}{[Nd/H]} & Ref. &
\multicolumn{2}{c}{[Eu/H]$_{\rm lit}$} & \multicolumn{2}{c}{[Eu/H]} & Ref. \\
\noalign{\smallskip}\hline\noalign{\smallskip}
      HW\,Car &    0&.15     & $-$0&.00     &   LIII &    0&.21     & $-$0&.10     &   LIII &	 0&.11     & $-$0&.06     &   LIII &	0&.17	  &    0&.02	 &   LIII &    0&.04	 & $-$0&.07	&   LIII \\
      IO\,Car &    0&.28     &    0&.13     &   LIII &    0&.36     &	 0&.05     &   LIII &	 0&.18     &	0&.01     &   LIII &	0&.25	  &    0&.10	 &   LIII &    0&.21	 &    0&.10	&   LIII \\
      IT\,Car &    0&.17     &    0&.02     &   LIII &    0&.26     & $-$0&.05     &   LIII &	 0&.17     & $-$0&.00     &   LIII &	0&.22	  &    0&.07	 &   LIII &    0&.13	 &    0&.02	&   LIII \\
       L\,Car &    0&.29     &    0&.17     &    LEM &    0&.29     &	 0&.06     &	LEM &	 0&.42     &	0&.27     &    LEM &	0&.47	  &    0&.30	 &    LEM &    0&.61	 &    0&.42	&    LEM \\
      SX\,Car &    0&.22     &    0&.07     &   LIII &    0&.39     &	 0&.08     &   LIII &	 0&.06     & $-$0&.11     &   LIII &	0&.19	  &    0&.04	 &   LIII &    0&.03	 & $-$0&.08	&   LIII \\
       U\,Car &    0&.19     &    0&.04     &   LIII &    0&.51     &	 0&.20     &   LIII & $-$0&.09     & $-$0&.26     &   LIII &	0&.24	  &    0&.09	 &   LIII &    0&.16	 &    0&.05	&   LIII \\
      UW\,Car &    0&.17     &    0&.02     &   LIII &    0&.27     & $-$0&.04     &   LIII &	 0&.10     & $-$0&.07     &   LIII &	0&.07	  & $-$0&.08	 &   LIII &    0&.22	 &    0&.11	&   LIII \\
      UX\,Car & $-$0&.01     & $-$0&.13     &    LEM &    0&.11     & $-$0&.12     &	LEM &	 0&.25     &	0&.10     &    LEM &	0&.18	  &    0&.01	 &    LEM &    0&.14	 & $-$0&.05	&    LEM \\
      UY\,Car &    0&.23     &    0&.08     &   LIII &    0&.33     &	 0&.02     &   LIII &	 0&.10     & $-$0&.07     &   LIII &	0&.13	  & $-$0&.02	 &   LIII &    0&.13	 &    0&.03	&    LII \\
      UZ\,Car &    0&.16     &    0&.01     &   LIII &    0&.42     &	 0&.11     &   LIII &	 0&.08     & $-$0&.09     &   LIII &	0&.16	  &    0&.01	 &   LIII &    0&.03	 & $-$0&.08	&   LIII \\
       V\,Car &    0&.07     & $-$0&.05     &    LEM &    0&.26     &	 0&.03     &	LEM &	 0&.15     & $-$0&.00     &    LEM &	0&.16	  & $-$0&.01	 &    LEM &    0&.38	 &    0&.19	&    LEM \\
    V397\,Car & $-$0&.04     & $-$0&.16     &    LEM &    0&.06     & $-$0&.17     &	LEM &	 0&.06     & $-$0&.09     &    LEM &	0&.02	  & $-$0&.15	 &    LEM &    0&.12	 & $-$0&.07	&    LEM \\
      VY\,Car &    0&.20     &    0&.08     &    LEM &    0&.32     &	 0&.09     &	LEM &	 0&.39     &	0&.24     &    LEM &	0&.38	  &    0&.21	 &    LEM &    0&.39	 &    0&.20	&    LEM \\
      WW\,Car &    0&.05     & $-$0&.10     &   LIII &    0&.14     & $-$0&.17     &   LIII &	 0&.01     & $-$0&.16     &   LIII &	0&.03	  & $-$0&.12	 &   LIII &    0&.01	 & $-$0&.10	&   LIII \\
      WZ\,Car &    0&.01     & $-$0&.14     &   LIII &    0&.37     &	 0&.06     &   LIII &	 0&.03     & $-$0&.14     &   LIII &	0&.05	  & $-$0&.10	 &   LIII &    0&.26	 &    0&.15	&   LIII \\
      XX\,Car &    0&.38     &    0&.23     &   LIII &    0&.42     &	 0&.11     &   LIII &	 0&.25     &	0&.08     &   LIII &	0&.40	  &    0&.25	 &   LIII &    0&.26	 &    0&.15	&   LIII \\
      XY\,Car &    0&.15     & $-$0&.00     &   LIII &    0&.26     & $-$0&.05     &   LIII &	 0&.14     & $-$0&.03     &   LIII &	0&.15	  &    0&.00	 &   LIII &    0&.32	 &    0&.21	&   LIII \\
      XZ\,Car &    0&.23     &    0&.08     &   LIII &    0&.32     &	 0&.01     &   LIII &	 0&.17     & $-$0&.00     &   LIII &	0&.19	  &    0&.04	 &   LIII &    0&.38	 &    0&.27	&   LIII \\
      YZ\,Car &    0&.20     &    0&.05     &   LIII &    0&.30     & $-$0&.01     &   LIII &	 0&.16     & $-$0&.01     &   LIII &	0&.17	  &    0&.02	 &   LIII &    0&.05	 & $-$0&.06	&   LIII \\
      AP\,Cas &    0&.03     & $-$0&.12     &   LIII &    0&.18     & $-$0&.13     &   LIII &	 0&.10     & $-$0&.07     &   LIII &	0&.12	  & $-$0&.03	 &   LIII &    0&.04	 & $-$0&.07	&   LIII \\
      AS\,Cas & $-$0&.12     & $-$0&.27     &   LIII &    0&.35     &	 0&.04     &   LIII &	 0&.07     & $-$0&.10     &   LIII &	0&.07	  & $-$0&.08	 &   LIII &     & \ldots &     & \ldots & \ldots \\
      AW\,Cas &    0&.11     & $-$0&.04     &   LIII &    0&.23     & $-$0&.08     &   LIII &	 0&.13     & $-$0&.04     &   LIII &	0&.15	  &    0&.00	 &   LIII &    0&.08	 & $-$0&.03	&   LIII \\
      AY\,Cas &    0&.11     & $-$0&.04     &   LIII &    0&.23     & $-$0&.08     &   LIII &	 0&.13     & $-$0&.04     &   LIII &	0&.20	  &    0&.05	 &   LIII &    0&.10	 & $-$0&.01	&   LIII \\
      BF\,Cas &    0&.10     & $-$0&.05     &   LIII &    0&.14     & $-$0&.17     &   LIII &	 0&.01     & $-$0&.16     &   LIII &	0&.05	  & $-$0&.10	 &   LIII &     & \ldots &     & \ldots & \ldots \\
      BP\,Cas &    0&.18     &    0&.03     &   LIII &    0&.26     & $-$0&.05     &   LIII &	 0&.27     &	0&.10     &   LIII &	0&.27	  &    0&.12	 &   LIII &    0&.19	 &    0&.08	&   LIII \\
      BV\,Cas &    0&.14     & $-$0&.01     &   LIII &    0&.22     & $-$0&.09     &   LIII &	 0&.09     & $-$0&.08     &   LIII &	0&.16	  &    0&.01	 &   LIII &    0&.09	 & $-$0&.02	&   LIII \\
      BY\,Cas &    0&.25     &    0&.10     &   LIII &    0&.22     & $-$0&.09     &   LIII &	 0&.29     &	0&.12     &   LIII &	0&.23	  &    0&.08	 &   LIII &    0&.21	 &    0&.10	&   LIII \\
      CD\,Cas &    0&.21     &    0&.06     &   LIII &    0&.17     & $-$0&.14     &   LIII &	 0&.14     & $-$0&.03     &   LIII &	0&.24	  &    0&.09	 &   LIII &    0&.45	 &    0&.34	&   LIII \\
      CF\,Cas &    0&.12     & $-$0&.03     &   LIII &    0&.20     & $-$0&.11     &   LIII &	 0&.18     &	0&.01     &   LIII &	0&.16	  &    0&.01	 &   LIII &    0&.10	 & $-$0&.01	&   LIII \\
      CG\,Cas &    0&.18     &    0&.03     &   LIII &    0&.33     &	 0&.02     &   LIII &	 0&.14     & $-$0&.03     &   LIII &	0&.16	  &    0&.01	 &   LIII &    0&.03	 & $-$0&.08	&   LIII \\
      CH\,Cas &    0&.70     &    0&.49     &    LII &     & \ldots &     & \ldots & \ldots &	 0&.11     &	0&.17     &    LII &	0&.20	  &    0&.11	 &    LII &    0&.50	 &    0&.40	&    LII \\
      CT\,Cas & $-$0&.02     & $-$0&.17     &   LIII &    0&.31     &	 0&.00     &   LIII &	 0&.02     & $-$0&.15     &   LIII &	0&.07	  & $-$0&.08	 &   LIII &    0&.03	 & $-$0&.08	&   LIII \\
      CY\,Cas &    0&.21     & $-$0&.00     &    LII &    0&.23     & $-$0&.06     &	LII &	 0&.00     &	0&.06     &    LII &	0&.03	  & $-$0&.06	 &    LII &    0&.22	 &    0&.12	&    LII \\
      CZ\,Cas &    0&.14     & $-$0&.01     &   LIII &    0&.24     & $-$0&.07     &   LIII &	 0&.08     & $-$0&.09     &   LIII &	0&.15	  &    0&.00	 &   LIII &    0&.18	 &    0&.07	&   LIII \\
      DD\,Cas &    0&.24     &    0&.03     &    LII &    0&.09     & $-$0&.20     &	LII & $-$0&.02     &	0&.04     &    LII &	0&.06	  & $-$0&.03	 &    LII &    0&.09	 & $-$0&.01	&    LII \\
      DF\,Cas &    0&.06     & $-$0&.15     &    LII &     & \ldots &     & \ldots & \ldots &	 0&.07     &	0&.13     &    LII &	0&.38	  &    0&.29	 &    LII &    0&.28	 &    0&.18	&    LII \\
      DL\,Cas &    0&.21     & $-$0&.00     &    LII &    0&.12     & $-$0&.17     &	LII &	 0&.09     &	0&.15     &    LII &	0&.12	  &    0&.03	 &    LII &    0&.11	 &    0&.01	&    LII \\
      DW\,Cas &    0&.15     & $-$0&.00     &   LIII &    0&.31     &	 0&.00     &   LIII &	 0&.14     & $-$0&.03     &   LIII &	0&.15	  &    0&.00	 &   LIII &    0&.09	 & $-$0&.02	&   LIII \\
      EX\,Cas & $-$0&.03     & $-$0&.18     &   LIII &    0&.48     &	 0&.17     &   LIII &	 0&.19     &	0&.02     &   LIII &	0&.12	  & $-$0&.03	 &   LIII & $-$0&.29	 & $-$0&.40	&   LIII \\
      FM\,Cas & $-$0&.06     & $-$0&.27     &    LII & $-$0&.01     & $-$0&.30     &	LII & $-$0&.25     & $-$0&.19     &    LII & $-$0&.14	  & $-$0&.23	 &    LII & $-$0&.25	 & $-$0&.35	&    LII \\
      FO\,Cas & $-$0&.52     & $-$0&.67     &   LIII & $-$0&.30     & $-$0&.61     &   LIII & $-$0&.36     & $-$0&.53     &   LIII & $-$0&.35	  & $-$0&.50	 &   LIII &     & \ldots &     & \ldots & \ldots \\
      FW\,Cas & $-$0&.10     & $-$0&.25     &   LIII &    0&.22     & $-$0&.09     &   LIII &	 0&.11     & $-$0&.06     &   LIII &	0&.06	  & $-$0&.09	 &   LIII & $-$0&.10	 & $-$0&.21	&   LIII \\
      GL\,Cas &    0&.00     & $-$0&.15     &   LIII &    0&.29     & $-$0&.02     &   LIII &	 0&.14     & $-$0&.03     &   LIII &	0&.02	  & $-$0&.13	 &   LIII &    0&.11	 & $-$0&.00	&   LIII \\
      GM\,Cas & $-$0&.03     & $-$0&.18     &   LIII &    0&.26     & $-$0&.05     &   LIII &	 0&.04     & $-$0&.13     &   LIII &	0&.05	  & $-$0&.10	 &   LIII &    0&.26	 &    0&.15	&   LIII \\
      GO\,Cas &    0&.17     &    0&.02     &   LIII &    0&.37     &	 0&.06     &   LIII &	 0&.12     & $-$0&.05     &   LIII &	0&.16	  &    0&.01	 &   LIII &    0&.01	 & $-$0&.10	&   LIII \\
      HK\,Cas &    0&.37     &    0&.22     &   LIII &    0&.35     &	 0&.04     &   LIII &	 0&.31     &	0&.14     &   LIII &	0&.39	  &    0&.24	 &   LIII &     & \ldots &     & \ldots & \ldots \\
      IO\,Cas & $-$0&.48     & $-$0&.63     &   LIII &    0&.11     & $-$0&.20     &   LIII & $-$0&.53     & $-$0&.70     &   LIII & $-$0&.40	  & $-$0&.55	 &   LIII & $-$0&.30	 & $-$0&.41	&   LIII \\
      KK\,Cas &    0&.21     &    0&.06     &   LIII &    0&.33     &	 0&.02     &   LIII &	 0&.24     &	0&.07     &   LIII &	0&.33	  &    0&.18	 &   LIII & $-$0&.10	 & $-$0&.21	&   LIII \\
      LT\,Cas & $-$0&.42     & $-$0&.57     &   LIII & $-$0&.25     & $-$0&.56     &   LIII & $-$0&.25     & $-$0&.42     &   LIII & $-$0&.19	  & $-$0&.34	 &   LIII & $-$0&.17	 & $-$0&.28	&   LIII \\
      NP\,Cas &    0&.05     & $-$0&.10     &   LIII &    0&.39     &	 0&.08     &   LIII &	 0&.05     & $-$0&.12     &   LIII &	0&.17	  &    0&.02	 &   LIII &    0&.28	 &    0&.17	&   LIII \\
      NY\,Cas & $-$0&.59     & $-$0&.74     &   LIII &    0&.01     & $-$0&.30     &   LIII & $-$0&.55     & $-$0&.72     &   LIII & $-$0&.37	  & $-$0&.52	 &   LIII & $-$0&.49	 & $-$0&.60	&   LIII \\
      OP\,Cas &    0&.15     & $-$0&.00     &   LIII &    0&.14     & $-$0&.17     &   LIII &	 0&.10     & $-$0&.07     &   LIII &	0&.19	  &    0&.04	 &   LIII &    0&.03	 & $-$0&.08	&   LIII \\
      OZ\,Cas & $-$0&.03     & $-$0&.18     &   LIII &    0&.50     &	 0&.19     &   LIII &	 0&.21     &	0&.04     &   LIII &	0&.09	  & $-$0&.06	 &   LIII & $-$0&.08	 & $-$0&.19	&   LIII \\
      PW\,Cas &    0&.11     & $-$0&.04     &   LIII &    0&.28     & $-$0&.03     &   LIII &	 0&.22     &	0&.05     &   LIII &	0&.20	  &    0&.05	 &   LIII &    0&.01	 & $-$0&.10	&   LIII \\
      RS\,Cas &    0&.26     &    0&.11     &   LIII &    0&.24     & $-$0&.07     &   LIII &	 0&.36     &	0&.19     &   LIII &	0&.34	  &    0&.19	 &   LIII &    0&.26	 &    0&.15	&   LIII \\
      RW\,Cas &    0&.23     &    0&.02     &    LII &     & \ldots &     & \ldots & \ldots &	 0&.00     &	0&.06     &    LII &	0&.11	  &    0&.02	 &    LII &    0&.23	 &    0&.13	&    LII \\
      RY\,Cas &    0&.61     &    0&.40     &    LII &     & \ldots &     & \ldots & \ldots &	 0&.35     &	0&.41     &    LII &	0&.40	  &    0&.31	 &    LII &    0&.53	 &    0&.43	&    LII \\
      SU\,Cas &    0&.26     &    0&.05     &    LII &    0&.22     & $-$0&.07     &	LII & $-$0&.05     &	0&.01     &    LII &	0&.14	  &    0&.05	 &    LII &    0&.09	 & $-$0&.01	&    LII \\
      SW\,Cas &    0&.31     &    0&.10     &    LII &    0&.30     &	 0&.01     &	LII & $-$0&.12     & $-$0&.06     &    LII &	0&.16	  &    0&.07	 &    LII &    0&.15	 &    0&.05	&    LII \\
      SY\,Cas &    0&.11     & $-$0&.10     &    LII &    0&.42     &	 0&.13     &	LII &	 0&.01     &	0&.07     &    LII &	0&.14	  &    0&.05	 &    LII &    0&.10	 &    0&.00	&    LII \\
      SZ\,Cas &    0&.15     & $-$0&.00     &   LIII &    0&.46     &	 0&.15     &   LIII &	 0&.30     &	0&.13     &   LIII &	0&.31	  &    0&.16	 &   LIII &    0&.28	 &    0&.17	&   LIII \\
      TU\,Cas &    0&.17     & $-$0&.04     &    LII &    0&.24     & $-$0&.05     &	LII & $-$0&.05     &	0&.01     &    LII &	0&.08	  & $-$0&.01	 &    LII &    0&.11	 &    0&.01	&    LII \\
      UZ\,Cas &    0&.06     & $-$0&.09     &   LIII &    0&.15     & $-$0&.16     &   LIII &	 0&.15     & $-$0&.02     &   LIII &	0&.19	  &    0&.04	 &   LIII &    0&.03	 & $-$0&.08	&   LIII \\
   V1017\,Cas & $-$0&.15     & $-$0&.30     &   LIII &    0&.06     & $-$0&.25     &   LIII & $-$0&.02     & $-$0&.19     &   LIII &	0&.00	  & $-$0&.15	 &   LIII &    0&.02	 & $-$0&.09	&   LIII \\
   V1019\,Cas &    0&.18     &    0&.03     &   LIII &    0&.51     &	 0&.20     &   LIII &	 0&.18     &	0&.01     &   LIII &	0&.22	  &    0&.07	 &   LIII &    0&.36	 &    0&.25	&   LIII \\
   V1020\,Cas &    0&.30     &    0&.15     &   LIII &    0&.53     &	 0&.22     &   LIII &	 0&.37     &	0&.20     &   LIII &	0&.23	  &    0&.08	 &   LIII &    0&.19	 &    0&.08	&   LIII \\
   V1100\,Cas & $-$0&.06     & $-$0&.21     &   LIII &    0&.30     & $-$0&.01     &   LIII &	 0&.05     & $-$0&.12     &   LIII &	0&.08	  & $-$0&.07	 &   LIII &    0&.10	 & $-$0&.01	&   LIII \\
   V1154\,Cas & $-$0&.25     & $-$0&.40     &   LIII &    0&.28     & $-$0&.03     &   LIII & $-$0&.08     & $-$0&.25     &   LIII & $-$0&.08	  & $-$0&.23	 &   LIII & $-$0&.33	 & $-$0&.44	&   LIII \\
   V1206\,Cas &    0&.18     &    0&.03     &   LIII &    0&.38     &	 0&.07     &   LIII &	 0&.19     &	0&.02     &   LIII &	0&.27	  &    0&.12	 &   LIII &    0&.12	 &    0&.01	&   LIII \\
    V342\,Cas &    0&.11     & $-$0&.04     &   LIII &    0&.16     & $-$0&.15     &   LIII &	 0&.13     & $-$0&.04     &   LIII &	0&.19	  &    0&.04	 &   LIII &    0&.03	 & $-$0&.08	&   LIII \\
    V379\,Cas &    0&.16     & $-$0&.05     &    LII &    0&.30     &	 0&.01     &	LII & $-$0&.06     & $-$0&.00     &    LII &	0&.12	  &    0&.03	 &    LII &    0&.12	 &    0&.02	&    LII \\
    V395\,Cas &    0&.07     & $-$0&.08     &   LIII &    0&.30     & $-$0&.01     &   LIII &	 0&.09     & $-$0&.08     &   LIII &	0&.11	  & $-$0&.04	 &   LIII &    0&.26	 &    0&.15	&   LIII \\
    V407\,Cas &    0&.08     & $-$0&.07     &   LIII &    0&.19     & $-$0&.12     &   LIII &	 0&.05     & $-$0&.12     &   LIII &	0&.17	  &    0&.02	 &   LIII &    0&.25	 &    0&.14	&   LIII \\
    V556\,Cas &    0&.10     & $-$0&.05     &   LIII &    0&.25     & $-$0&.06     &   LIII &	 0&.19     &	0&.02     &   LIII &	0&.19	  &    0&.04	 &   LIII & $-$0&.12	 & $-$0&.23	&   LIII \\
    V636\,Cas &    0&.18     & $-$0&.03     &    LII &    0&.14     & $-$0&.15     &	LII & $-$0&.15     & $-$0&.09     &    LII &	0&.11	  &    0&.02	 &    LII &    0&.02	 & $-$0&.08	&    LII \\
\hline\noalign{\smallskip}
\multicolumn{26}{r}{\it {\footnotesize continued on next page}} \\
\end{tabular}}
\end{table*}
\addtocounter{table}{-1}
\begin{table*}[p]
\centering
\caption[]{continued.}
{\scriptsize 
\begin{tabular}{r r@{}l r@{}l c r@{}l r@{}l c r@{}l r@{}l c r@{}l r@{}l c r@{}l r@{}l c}
\noalign{\smallskip}\hline\hline\noalign{\smallskip}
Name &
\multicolumn{2}{c}{[Y/H]$_{\rm lit}$} & \multicolumn{2}{c}{[Y/H]} & Ref. &
\multicolumn{2}{c}{[La/H]$_{\rm lit}$} & \multicolumn{2}{c}{[La/H]} & Ref. &
\multicolumn{2}{c}{[Ce/H]$_{\rm lit}$} & \multicolumn{2}{c}{[Ce/H]} & Ref. &
\multicolumn{2}{c}{[Nd/H]$_{\rm lit}$} & \multicolumn{2}{c}{[Nd/H]} & Ref. &
\multicolumn{2}{c}{[Eu/H]$_{\rm lit}$} & \multicolumn{2}{c}{[Eu/H]} & Ref. \\
\noalign{\smallskip}\hline\noalign{\smallskip}
      VV\,Cas &    0&.06     & $-$0&.09     &   LIII &    0&.14     & $-$0&.17     &   LIII &	 0&.24     &	0&.07     &   LIII &	0&.21	  &    0&.06	 &   LIII &    0&.21	 &    0&.10	&   LIII \\
      VW\,Cas &    0&.30     &    0&.15     &   LIII &    0&.34     &	 0&.03     &   LIII &	 0&.41     &	0&.24     &   LIII &	0&.38	  &    0&.23	 &   LIII &    0&.28	 &    0&.17	&   LIII \\
      XY\,Cas &    0&.18     &    0&.03     &   LIII &    0&.35     &	 0&.04     &   LIII &	 0&.07     & $-$0&.10     &   LIII &	0&.13	  & $-$0&.02	 &   LIII &    0&.29	 &    0&.18	&   LIII \\
      AY\,Cen &    0&.17     &    0&.02     &   LIII &    0&.21     & $-$0&.10     &   LIII &	 0&.12     & $-$0&.05     &   LIII &	0&.15	  &    0&.00	 &   LIII &    0&.12	 &    0&.01	&   LIII \\
      AZ\,Cen &    0&.23     &    0&.08     &   LIII &    0&.27     & $-$0&.04     &   LIII &	 0&.14     & $-$0&.03     &   LIII &	0&.17	  &    0&.02	 &   LIII &    0&.06	 & $-$0&.05	&   LIII \\
      BB\,Cen &    0&.34     &    0&.19     &   LIII &    0&.35     &	 0&.04     &   LIII &	 0&.20     &	0&.03     &   LIII &	0&.26	  &    0&.11	 &   LIII &    0&.46	 &    0&.35	&   LIII \\
      KK\,Cen &    0&.31     &    0&.16     &   LIII &    0&.35     &	 0&.04     &   LIII &	 0&.25     &	0&.08     &   LIII &	0&.27	  &    0&.12	 &   LIII &    0&.29	 &    0&.18	&   LIII \\
      KN\,Cen &    0&.23     &    0&.23     &     TS &    0&.04     &	 0&.04     &	 TS & $-$0&.02     & $-$0&.02     &	TS &	0&.05	  &    0&.05	 &     TS &    0&.07	 &    0&.07	&     TS \\
      MZ\,Cen & $-$0&.08     & $-$0&.08     &     TS & $-$0&.22     & $-$0&.22     &	 TS &	 0&.16     & $-$0&.01     &   LIII &	0&.27	  &    0&.27	 &     TS & $-$0&.05	 & $-$0&.05	&     TS \\
      OO\,Cen &    0&.14     &    0&.14     &     TS &    0&.01     &	 0&.01     &	 TS &     & \ldots &     & \ldots & \ldots & $-$0&.04	  & $-$0&.04	 &     TS &    0&.05	 &    0&.05	&     TS \\
      QY\,Cen &    0&.28     &    0&.13     &   LIII &    0&.51     &	 0&.20     &   LIII &	 0&.12     & $-$0&.05     &   LIII &	0&.23	  &    0&.08	 &   LIII &    0&.60	 &    0&.49	&   LIII \\
      TX\,Cen &    0&.17     &    0&.17     &     TS &    0&.01     &	 0&.01     &	 TS &	 0&.07     &	0&.07     &	TS & $-$0&.05	  & $-$0&.05	 &     TS &    0&.07	 &    0&.07	&     TS \\
       V\,Cen &    0&.15     & $-$0&.00     &   LIII &    0&.31     &	 0&.00     &   LIII &	 0&.08     & $-$0&.09     &   LIII &	0&.14	  & $-$0&.01	 &   LIII &    0&.17	 &    0&.06	&   LIII \\
    V339\,Cen & $-$0&.09     & $-$0&.09     &     TS & $-$0&.25     & $-$0&.25     &	 TS & $-$0&.21     & $-$0&.21     &	TS & $-$0&.28	  & $-$0&.28	 &     TS & $-$0&.10	 & $-$0&.10	&     TS \\
    V378\,Cen &    0&.23     &    0&.08     &   LIII &    0&.28     & $-$0&.03     &   LIII &	 0&.12     & $-$0&.05     &   LIII &	0&.22	  &    0&.07	 &   LIII &    0&.44	 &    0&.33	&   LIII \\
    V381\,Cen &    0&.11     & $-$0&.04     &   LIII &    0&.20     & $-$0&.11     &   LIII &	 0&.03     & $-$0&.14     &   LIII &	0&.01	  & $-$0&.14	 &   LIII & $-$0&.06	 & $-$0&.17	&   LIII \\
    V419\,Cen &    0&.25     &    0&.10     &   LIII &    0&.50     &	 0&.19     &   LIII &	 0&.24     &	0&.07     &   LIII &	0&.28	  &    0&.13	 &   LIII &    0&.18	 &    0&.07	&   LIII \\
    V496\,Cen &    0&.20     &    0&.05     &   LIII &    0&.28     & $-$0&.03     &   LIII &	 0&.16     & $-$0&.01     &   LIII &	0&.15	  &    0&.00	 &   LIII &    0&.13	 &    0&.02	&   LIII \\
    V659\,Cen &    0&.20     &    0&.05     &   LIII &    0&.34     &	 0&.03     &   LIII &	 0&.14     & $-$0&.03     &   LIII &	0&.17	  &    0&.02	 &   LIII &    0&.10	 & $-$0&.01	&   LIII \\
    V737\,Cen &    0&.20     &    0&.05     &   LIII &    0&.21     & $-$0&.10     &   LIII &	 0&.12     & $-$0&.05     &   LIII &	0&.16	  &    0&.01	 &   LIII &    0&.15	 &    0&.04	&   LIII \\
      VW\,Cen &    0&.07     &    0&.07     &     TS & $-$0&.14     & $-$0&.14     &	 TS & $-$0&.19     & $-$0&.19     &	TS &	0&.23	  &    0&.23	 &     TS & $-$0&.05	 & $-$0&.05	&     TS \\
      XX\,Cen &    0&.28     &    0&.13     &   LIII &    0&.28     & $-$0&.03     &   LIII &	 0&.17     & $-$0&.00     &   LIII &	0&.25	  &    0&.10	 &   LIII &    0&.02	 & $-$0&.09	&   LIII \\
      AK\,Cep &    0&.16     &    0&.01     &   LIII &    0&.31     &	 0&.00     &   LIII &	 0&.13     & $-$0&.04     &   LIII &	0&.17	  &    0&.02	 &   LIII & $-$0&.19	 & $-$0&.30	&   LIII \\
      CN\,Cep &    0&.12     & $-$0&.03     &   LIII &    0&.19     & $-$0&.12     &   LIII &	 0&.05     & $-$0&.12     &   LIII &	0&.23	  &    0&.08	 &   LIII &    0&.24	 &    0&.13	&   LIII \\
      CP\,Cep &    0&.05     & $-$0&.16     &    LII &    0&.12     & $-$0&.17     &	LII & $-$0&.12     & $-$0&.06     &    LII &	0&.03	  & $-$0&.06	 &    LII &    0&.00	 & $-$0&.10	&    LII \\
      CR\,Cep &    0&.05     & $-$0&.16     &    LII &    0&.14     & $-$0&.15     &	LII & $-$0&.20     & $-$0&.14     &    LII &	0&.00	  & $-$0&.09	 &    LII & $-$0&.11	 & $-$0&.21	&    LII \\
      DR\,Cep & $-$0&.11     & $-$0&.26     &   LIII &    0&.43     &	 0&.12     &   LIII & $-$0&.31     & $-$0&.48     &   LIII &	0&.07	  & $-$0&.08	 &   LIII &    0&.07	 & $-$0&.04	&   LIII \\
      IR\,Cep &    0&.13     & $-$0&.08     &    LII &    0&.16     & $-$0&.13     &	LII &	 0&.02     &	0&.08     &    LII &	0&.09	  &    0&.00	 &    LII &    0&.12	 &    0&.02	&    LII \\
      IY\,Cep & $-$0&.05     & $-$0&.20     &   LIII &    0&.29     & $-$0&.02     &   LIII & $-$0&.08     & $-$0&.25     &   LIII &	0&.02	  & $-$0&.13	 &   LIII &    0&.35	 &    0&.24	&   LIII \\
      MU\,Cep &    0&.23     &    0&.08     &   LIII &    0&.70     &	 0&.39     &   LIII &	 0&.27     &	0&.10     &   LIII &	0&.27	  &    0&.12	 &   LIII &    0&.43	 &    0&.32	&   LIII \\
    V901\,Cep &    0&.10     & $-$0&.05     &   LIII &    0&.32     &	 0&.01     &   LIII &	 0&.08     & $-$0&.09     &   LIII &	0&.21	  &    0&.06	 &   LIII &     & \ldots &     & \ldots & \ldots \\
    V911\,Cep &    0&.10     & $-$0&.05     &   LIII &    0&.34     &	 0&.03     &   LIII &	 0&.34     &	0&.17     &   LIII &	0&.20	  &    0&.05	 &   LIII &    0&.45	 &    0&.34	&   LIII \\
$\delta$\,Cep &    0&.22     &    0&.07     &   LIII &    0&.43     &	 0&.12     &   LIII &	 0&.16     & $-$0&.01     &   LIII &	0&.24	  &    0&.09	 &   LIII &    0&.14	 &    0&.03	&   LIII \\
      AV\,Cir &    0&.32     &    0&.17     &   LIII &    0&.32     &	 0&.01     &   LIII &	 0&.21     &	0&.04     &   LIII &	0&.24	  &    0&.09	 &   LIII &    0&.13	 &    0&.02	&   LIII \\
      AX\,Cir &    0&.05     & $-$0&.10     &   LIII &    0&.26     & $-$0&.05     &   LIII &	 0&.06     & $-$0&.11     &   LIII &	0&.09	  & $-$0&.06	 &   LIII &    0&.06	 & $-$0&.05	&   LIII \\
      BP\,Cir &    0&.14     & $-$0&.01     &   LIII &    0&.22     & $-$0&.09     &   LIII &	 0&.03     & $-$0&.14     &   LIII &	0&.03	  & $-$0&.12	 &   LIII & $-$0&.15	 & $-$0&.26	&   LIII \\
      AD\,Cru &    0&.16     &    0&.01     &   LIII &    0&.26     & $-$0&.05     &   LIII &	 0&.14     & $-$0&.03     &   LIII &	0&.19	  &    0&.04	 &   LIII &    0&.40	 &    0&.29	&   LIII \\
      AG\,Cru &    0&.16     &    0&.01     &   LIII &    0&.26     & $-$0&.05     &   LIII &	 0&.01     & $-$0&.16     &   LIII &	0&.08	  & $-$0&.07	 &   LIII & $-$0&.03	 & $-$0&.14	&   LIII \\
      BG\,Cru &    0&.10     & $-$0&.05     &   LIII &    0&.47     &	 0&.16     &   LIII &	 0&.09     & $-$0&.08     &   LIII &	0&.28	  &    0&.13	 &   LIII &    0&.23	 &    0&.13	&    LII \\
       R\,Cru &    0&.24     &    0&.09     &   LIII &    0&.24     & $-$0&.07     &   LIII &	 0&.13     & $-$0&.04     &   LIII &	0&.18	  &    0&.03	 &   LIII &    0&.26	 &    0&.15	&   LIII \\
       S\,Cru &    0&.20     &    0&.05     &   LIII &    0&.32     &	 0&.01     &   LIII &	 0&.11     & $-$0&.06     &   LIII &	0&.11	  & $-$0&.04	 &   LIII & $-$0&.05	 & $-$0&.16	&   LIII \\
       T\,Cru &    0&.18     &    0&.03     &   LIII &    0&.29     & $-$0&.02     &   LIII &	 0&.10     & $-$0&.07     &   LIII &	0&.15	  &    0&.00	 &   LIII &    0&.01	 & $-$0&.10	&   LIII \\
      VW\,Cru &    0&.26     &    0&.11     &   LIII &    0&.31     &	 0&.00     &   LIII &	 0&.18     &	0&.01     &   LIII &	0&.25	  &    0&.10	 &   LIII &    0&.38	 &    0&.27	&   LIII \\
       X\,Cru &    0&.16     &    0&.01     &   LIII &    0&.29     & $-$0&.02     &   LIII &	 0&.14     & $-$0&.03     &   LIII &	0&.16	  &    0&.01	 &   LIII &    0&.00	 & $-$0&.11	&   LIII \\
      BZ\,Cyg &    0&.17     & $-$0&.04     &    LII &    0&.31     &	 0&.02     &	LII & $-$0&.07     & $-$0&.01     &    LII &	0&.15	  &    0&.06	 &    LII &    0&.12	 &    0&.02	&    LII \\
      CD\,Cyg &    0&.23     &    0&.08     &   LIII &    0&.40     &	 0&.09     &   LIII &	 0&.17     & $-$0&.00     &   LIII &	0&.23	  &    0&.08	 &   LIII &    0&.39	 &    0&.28	&   LIII \\
      DT\,Cyg &    0&.31     &    0&.10     &    LII &    0&.22     & $-$0&.07     &	LII & $-$0&.03     &	0&.03     &    LII &	0&.19	  &    0&.10	 &    LII &    0&.18	 &    0&.08	&    LII \\
      EP\,Cyg & $-$0&.09     & $-$0&.24     &   LIII &    0&.08     & $-$0&.23     &   LIII & $-$0&.08     & $-$0&.25     &   LIII &	0&.07	  & $-$0&.08	 &   LIII & $-$0&.09	 & $-$0&.20	&   LIII \\
      EU\,Cyg & $-$0&.22     & $-$0&.37     &   LIII &    0&.12     & $-$0&.19     &   LIII &	 0&.02     & $-$0&.15     &   LIII &	0&.03	  & $-$0&.12	 &   LIII & $-$0&.03	 & $-$0&.14	&   LIII \\
      EX\,Cyg &    0&.28     &    0&.13     &   LIII &    0&.35     &	 0&.04     &   LIII &	 0&.23     &	0&.06     &   LIII &	0&.38	  &    0&.23	 &   LIII &    0&.48	 &    0&.37	&   LIII \\
      EZ\,Cyg &    0&.35     &    0&.20     &   LIII &    0&.38     &	 0&.07     &   LIII &	 0&.55     &	0&.38     &   LIII &	0&.30	  &    0&.15	 &   LIII &    0&.25	 &    0&.14	&   LIII \\
      GH\,Cyg &    0&.32     &    0&.17     &   LIII &    0&.32     &	 0&.01     &   LIII &	 0&.20     &	0&.03     &   LIII &	0&.27	  &    0&.12	 &   LIII &    0&.36	 &    0&.25	&   LIII \\
      GI\,Cyg &    0&.23     &    0&.08     &   LIII &    0&.46     &	 0&.15     &   LIII &	 0&.38     &	0&.21     &   LIII &	0&.34	  &    0&.19	 &   LIII &    0&.26	 &    0&.15	&   LIII \\
      GL\,Cyg &    0&.10     & $-$0&.05     &   LIII &    0&.22     & $-$0&.09     &   LIII &	 0&.16     & $-$0&.01     &   LIII &	0&.16	  &    0&.01	 &   LIII &    0&.20	 &    0&.09	&   LIII \\
      IY\,Cyg &    0&.05     & $-$0&.10     &   LIII &    0&.33     &	 0&.02     &   LIII &	 0&.07     & $-$0&.10     &   LIII &	0&.18	  &    0&.03	 &   LIII & $-$0&.28	 & $-$0&.39	&   LIII \\
      KX\,Cyg &    0&.29     &    0&.14     &   LIII &    0&.51     &	 0&.20     &   LIII &	 0&.20     &	0&.03     &   LIII &	0&.33	  &    0&.18	 &   LIII &    0&.23	 &    0&.12	&   LIII \\
      MW\,Cyg &    0&.24     &    0&.03     &    LII &    0&.15     & $-$0&.14     &	LII &	 0&.04     &	0&.10     &    LII &	0&.13	  &    0&.04	 &    LII &    0&.06	 & $-$0&.04	&    LII \\
      SU\,Cyg &    0&.15     & $-$0&.06     &    LII &    0&.24     & $-$0&.05     &	LII & $-$0&.11     & $-$0&.05     &    LII &	0&.15	  &    0&.06	 &    LII &    0&.08	 & $-$0&.02	&    LII \\
      SZ\,Cyg &    0&.22     &    0&.01     &    LII &    0&.21     & $-$0&.08     &	LII &	 0&.00     &	0&.06     &    LII &	0&.03	  & $-$0&.06	 &    LII &    0&.15	 &    0&.05	&    LII \\
      TX\,Cyg &    0&.45     &    0&.24     &    LII &    0&.44     &	 0&.15     &	LII &	 0&.23     &	0&.29     &    LII &	0&.30	  &    0&.21	 &    LII &    0&.13	 &    0&.03	&    LII \\
   V1020\,Cyg &    0&.46     &    0&.31     &   LIII &    0&.55     &	 0&.24     &   LIII &	 0&.58     &	0&.41     &   LIII &	0&.40	  &    0&.25	 &   LIII &     & \ldots &     & \ldots & \ldots \\
   V1025\,Cyg &    0&.24     &    0&.09     &   LIII &    0&.24     & $-$0&.07     &   LIII &	 0&.15     & $-$0&.02     &   LIII &	0&.22	  &    0&.07	 &   LIII &    0&.33	 &    0&.22	&   LIII \\
   V1033\,Cyg &    0&.23     &    0&.08     &   LIII &    0&.34     &	 0&.03     &   LIII &	 0&.20     &	0&.03     &   LIII &	0&.25	  &    0&.10	 &   LIII &    0&.06	 & $-$0&.05	&   LIII \\
   V1046\,Cyg &    0&.34     &    0&.19     &   LIII &    0&.43     &	 0&.12     &   LIII &	 0&.28     &	0&.11     &   LIII &	0&.38	  &    0&.23	 &   LIII &    0&.33	 &    0&.22	&   LIII \\
   V1154\,Cyg &    0&.08     & $-$0&.13     &    LII &    0&.18     & $-$0&.11     &	LII & $-$0&.10     & $-$0&.04     &    LII &	0&.05	  & $-$0&.04	 &    LII & $-$0&.06	 & $-$0&.16	&    LII \\
   V1334\,Cyg &    0&.18     & $-$0&.03     &    LII &    0&.27     & $-$0&.02     &	LII & $-$0&.09     & $-$0&.03     &    LII &	0&.16	  &    0&.07	 &    LII &    0&.17	 &    0&.07	&    LII \\
   V1364\,Cyg &    0&.26     &    0&.11     &   LIII &    0&.32     &	 0&.01     &   LIII &	 0&.32     &	0&.15     &   LIII &	0&.25	  &    0&.10	 &   LIII &    0&.15	 &    0&.04	&   LIII \\
   V1397\,Cyg &    0&.07     & $-$0&.08     &   LIII &    0&.23     & $-$0&.08     &   LIII &	 0&.14     & $-$0&.03     &   LIII &	0&.16	  &    0&.01	 &   LIII &    0&.02	 & $-$0&.09	&   LIII \\
   V1726\,Cyg &    0&.14     & $-$0&.07     &    LII &     & \ldots &     & \ldots & \ldots &     & \ldots &     & \ldots & \ldots &	0&.24	  &    0&.15	 &    LII &    0&.31	 &    0&.21	&    LII \\
    V347\,Cyg &    0&.33     &    0&.18     &   LIII &    0&.36     &	 0&.05     &   LIII &	 0&.29     &	0&.12     &   LIII &	0&.36	  &    0&.21	 &   LIII &    0&.16	 &    0&.05	&   LIII \\
    V356\,Cyg &    0&.24     &    0&.09     &   LIII &    0&.34     &	 0&.03     &   LIII &	 0&.28     &	0&.11     &   LIII &	0&.20	  &    0&.05	 &   LIII &    0&.29	 &    0&.18	&   LIII \\
    V386\,Cyg &    0&.35     &    0&.14     &    LII &    0&.41     &	 0&.12     &	LII &	 0&.10     &	0&.16     &    LII &	0&.42	  &    0&.33	 &    LII &    0&.41	 &    0&.31	&    LII \\
    V396\,Cyg &    0&.00     & $-$0&.15     &   LIII &    0&.19     & $-$0&.12     &   LIII &	 0&.01     & $-$0&.16     &   LIII &	0&.05	  & $-$0&.10	 &   LIII &    0&.00	 & $-$0&.11	&   LIII \\
    V402\,Cyg &    0&.15     & $-$0&.06     &    LII &    0&.16     & $-$0&.13     &	LII &	 0&.03     &	0&.09     &    LII &	0&.01	  & $-$0&.08	 &    LII &    0&.01	 & $-$0&.09	&    LII \\
    V438\,Cyg &    0&.05     & $-$0&.10     &   LIII &    0&.19     & $-$0&.12     &   LIII & $-$0&.01     & $-$0&.18     &   LIII &	0&.12	  & $-$0&.03	 &   LIII & $-$0&.03	 & $-$0&.14	&   LIII \\
\hline\noalign{\smallskip}
\multicolumn{26}{r}{\it {\footnotesize continued on next page}} \\
\end{tabular}}
\end{table*}
\addtocounter{table}{-1}
\begin{table*}[p]
\centering
\caption[]{continued.}
{\scriptsize 
\begin{tabular}{r r@{}l r@{}l c r@{}l r@{}l c r@{}l r@{}l c r@{}l r@{}l c r@{}l r@{}l c}
\noalign{\smallskip}\hline\hline\noalign{\smallskip}
Name &
\multicolumn{2}{c}{[Y/H]$_{\rm lit}$} & \multicolumn{2}{c}{[Y/H]} & Ref. &
\multicolumn{2}{c}{[La/H]$_{\rm lit}$} & \multicolumn{2}{c}{[La/H]} & Ref. &
\multicolumn{2}{c}{[Ce/H]$_{\rm lit}$} & \multicolumn{2}{c}{[Ce/H]} & Ref. &
\multicolumn{2}{c}{[Nd/H]$_{\rm lit}$} & \multicolumn{2}{c}{[Nd/H]} & Ref. &
\multicolumn{2}{c}{[Eu/H]$_{\rm lit}$} & \multicolumn{2}{c}{[Eu/H]} & Ref. \\
\noalign{\smallskip}\hline\noalign{\smallskip}
    V459\,Cyg &    0&.40     &    0&.25     &   LIII &    0&.44     &	 0&.13     &   LIII &	 0&.28     &	0&.11     &   LIII &	0&.35	  &    0&.20	 &   LIII &    0&.24	 &    0&.13	&   LIII \\
    V492\,Cyg &    0&.24     &    0&.09     &   LIII &    0&.38     &	 0&.07     &   LIII &	 0&.25     &	0&.08     &   LIII &	0&.26	  &    0&.11	 &   LIII &    0&.19	 &    0&.08	&   LIII \\
    V495\,Cyg &    0&.23     &    0&.08     &   LIII &    0&.14     & $-$0&.17     &   LIII &	 0&.22     &	0&.05     &   LIII &	0&.26	  &    0&.11	 &   LIII &    0&.24	 &    0&.13	&   LIII \\
    V514\,Cyg &    0&.25     &    0&.10     &   LIII &    0&.35     &	 0&.04     &   LIII &	 0&.22     &	0&.05     &   LIII &	0&.29	  &    0&.14	 &   LIII &    0&.35	 &    0&.24	&   LIII \\
    V520\,Cyg &    0&.11     & $-$0&.04     &   LIII &    0&.12     & $-$0&.19     &   LIII &	 0&.23     &	0&.06     &   LIII &	0&.18	  &    0&.03	 &   LIII &    0&.13	 &    0&.02	&   LIII \\
    V532\,Cyg &    0&.12     & $-$0&.09     &    LII &    0&.09     & $-$0&.20     &	LII &	 0&.00     &	0&.06     &    LII & $-$0&.05	  & $-$0&.14	 &    LII & $-$0&.03	 & $-$0&.13	&    LII \\
    V538\,Cyg &    0&.08     & $-$0&.07     &   LIII &    0&.12     & $-$0&.19     &   LIII &	 0&.20     &	0&.03     &   LIII &	0&.18	  &    0&.03	 &   LIII &    0&.10	 & $-$0&.01	&   LIII \\
    V547\,Cyg &    0&.24     &    0&.09     &   LIII &    0&.26     & $-$0&.05     &   LIII &	 0&.24     &	0&.07     &   LIII &	0&.24	  &    0&.09	 &   LIII &    0&.18	 &    0&.07	&   LIII \\
    V609\,Cyg &    0&.36     &    0&.21     &   LIII &    0&.50     &	 0&.19     &   LIII &	 0&.37     &	0&.20     &   LIII &	0&.38	  &    0&.23	 &   LIII &    0&.53	 &    0&.42	&   LIII \\
    V621\,Cyg &    0&.07     & $-$0&.08     &   LIII &    0&.47     &	 0&.16     &   LIII &	 0&.17     & $-$0&.00     &   LIII &	0&.12	  & $-$0&.03	 &   LIII &    0&.41	 &    0&.30	&   LIII \\
    V924\,Cyg & $-$0&.18     & $-$0&.39     &    LII &     & \ldots &     & \ldots & \ldots &     & \ldots &     & \ldots & \ldots &	0&.04	  & $-$0&.05	 &    LII &     & \ldots &     & \ldots & \ldots \\
      VX\,Cyg &    0&.14     & $-$0&.07     &    LII &    0&.16     & $-$0&.13     &	LII & $-$0&.01     &	0&.05     &    LII &	0&.01	  & $-$0&.08	 &    LII &    0&.17	 &    0&.07	&    LII \\
      VY\,Cyg &    0&.17     & $-$0&.04     &    LII &    0&.21     & $-$0&.08     &	LII &	 0&.01     &	0&.07     &    LII &	0&.09	  &    0&.00	 &    LII &    0&.01	 & $-$0&.09	&    LII \\
      VZ\,Cyg &    0&.17     & $-$0&.04     &    LII &    0&.14     & $-$0&.15     &	LII &	 0&.01     &	0&.07     &    LII &	0&.08	  & $-$0&.01	 &    LII &    0&.11	 &    0&.01	&    LII \\
       X\,Cyg &    0&.28     &    0&.07     &    LII &    0&.23     & $-$0&.06     &	LII & $-$0&.07     & $-$0&.01     &    LII &	0&.12	  &    0&.03	 &    LII &    0&.11	 &    0&.01	&    LII \\
      EK\,Del & $-$1&.40     & $-$1&.55     &   LIII & $-$1&.28     & $-$1&.59     &   LIII & $-$1&.35     & $-$1&.52     &   LIII & $-$1&.39	  & $-$1&.54	 &   LIII & $-$0&.68	 & $-$0&.79	&   LIII \\
 $\beta$\,Dor &    0&.02     & $-$0&.19     &    LII &    0&.18     & $-$0&.11     &	LII &	 0&.05     &	0&.11     &    LII & $-$0&.05	  & $-$0&.14	 &    LII &    0&.04	 & $-$0&.06	&    LII \\
      AA\,Gem & $-$0&.18     & $-$0&.18     &     TS & $-$0&.16     & $-$0&.16     &	 TS &	 0&.34     &	0&.34     &	TS &	0&.02	  &    0&.02	 &     TS &    0&.08	 &    0&.08	&     TS \\
      AD\,Gem & $-$0&.31     & $-$0&.31     &     TS & $-$0&.17     & $-$0&.17     &	 TS &	 0&.26     &	0&.11     &    LEM & $-$0&.16	  & $-$0&.16	 &     TS & $-$0&.25	 & $-$0&.25	&     TS \\
      BB\,Gem & $-$0&.08     & $-$0&.23     &   LIII &    0&.23     & $-$0&.08     &   LIII &	 0&.05     & $-$0&.12     &   LIII &	0&.03	  & $-$0&.12	 &   LIII &    0&.01	 & $-$0&.10	&   LIII \\
      BW\,Gem & $-$0&.36     & $-$0&.36     &     TS & $-$0&.18     & $-$0&.18     &	 TS & $-$0&.06     & $-$0&.23     &   LIII & $-$0&.09	  & $-$0&.09	 &     TS & $-$0&.08	 & $-$0&.08	&     TS \\
      DX\,Gem & $-$0&.09     & $-$0&.09     &     TS &    0&.00     &	 0&.00     &	 TS &	 0&.08     & $-$0&.09     &   LIII &	0&.07	  &    0&.07	 &     TS & $-$0&.03	 & $-$0&.03	&     TS \\
      RZ\,Gem & $-$0&.15     & $-$0&.15     &     TS & $-$0&.27     & $-$0&.50     &	LEM & $-$0&.02     & $-$0&.17     &    LEM & $-$0&.19	  & $-$0&.36	 &    LEM & $-$0&.27	 & $-$0&.46	&    LEM \\
       W\,Gem & $-$0&.05     & $-$0&.20     &   LIII &    0&.12     & $-$0&.19     &   LIII &	 0&.02     & $-$0&.15     &   LIII &	0&.01	  & $-$0&.14	 &   LIII & $-$0&.04	 & $-$0&.15	&   LIII \\
 $\zeta$\,Gem &    0&.20     &    0&.05     &   LIII &    0&.33     &	 0&.02     &   LIII &	 0&.24     &	0&.07     &   LIII &	0&.22	  &    0&.07	 &   LIII &    0&.04	 & $-$0&.07	&   LIII \\
      BB\,Her &    0&.29     &    0&.14     &   LIII &    0&.26     & $-$0&.05     &   LIII &	 0&.14     & $-$0&.03     &   LIII &	0&.20	  &    0&.05	 &   LIII &    0&.27	 &    0&.16	&   LIII \\
      BG\,Lac &    0&.07     & $-$0&.08     &   LIII &    0&.25     & $-$0&.06     &   LIII &	 0&.09     & $-$0&.08     &   LIII &	0&.16	  &    0&.01	 &   LIII & $-$0&.03	 & $-$0&.14	&   LIII \\
      DF\,Lac &    0&.15     & $-$0&.00     &   LIII &    0&.24     & $-$0&.07     &   LIII &	 0&.14     & $-$0&.03     &   LIII &	0&.14	  & $-$0&.01	 &   LIII &    0&.12	 &    0&.01	&   LIII \\
      FQ\,Lac & $-$0&.74     & $-$0&.89     &   LIII &     & \ldots &     & \ldots & \ldots &	 0&.08     & $-$0&.09     &   LIII &	0&.08	  & $-$0&.07	 &   LIII &     & \ldots &     & \ldots & \ldots \\
      RR\,Lac &    0&.13     & $-$0&.02     &   LIII &    0&.26     & $-$0&.05     &   LIII &	 0&.08     & $-$0&.09     &   LIII &	0&.15	  &    0&.00	 &   LIII & $-$0&.01	 & $-$0&.12	&   LIII \\
       V\,Lac &    0&.22     &    0&.07     &   LIII &    0&.33     &	 0&.02     &   LIII &	 0&.15     & $-$0&.02     &   LIII &	0&.25	  &    0&.10	 &   LIII &    0&.33	 &    0&.22	&   LIII \\
    V411\,Lac &    0&.18     &    0&.03     &   LIII &    0&.48     &	 0&.17     &   LIII &	 0&.16     & $-$0&.01     &   LIII &	0&.24	  &    0&.09	 &   LIII &    0&.20	 &    0&.09	&   LIII \\
       X\,Lac &    0&.12     & $-$0&.03     &   LIII &    0&.25     & $-$0&.06     &   LIII &	 0&.17     & $-$0&.00     &   LIII &	0&.20	  &    0&.05	 &   LIII & $-$0&.10	 & $-$0&.21	&   LIII \\
       Y\,Lac &    0&.16     &    0&.01     &   LIII &    0&.46     &	 0&.15     &   LIII &	 0&.11     & $-$0&.06     &   LIII &	0&.14	  & $-$0&.01	 &   LIII &    0&.03	 & $-$0&.08	&   LIII \\
       Z\,Lac &    0&.18     &    0&.03     &   LIII &    0&.43     &	 0&.12     &   LIII &	 0&.28     &	0&.11     &   LIII &	0&.34	  &    0&.19	 &   LIII &    0&.03	 & $-$0&.08	&   LIII \\
      GH\,Lup &    0&.13     & $-$0&.02     &   LIII &    0&.23     & $-$0&.08     &   LIII &	 0&.10     & $-$0&.07     &   LIII &	0&.16	  &    0&.01	 &   LIII &    0&.02	 & $-$0&.09	&   LIII \\
    V473\,Lyr &    0&.05     & $-$0&.10     &   LIII &    0&.14     & $-$0&.17     &   LIII &	 0&.04     & $-$0&.13     &   LIII &	0&.05	  & $-$0&.10	 &   LIII & $-$0&.02	 & $-$0&.13	&   LIII \\
      AA\,Mon &    0&.09     & $-$0&.06     &   LIII &    0&.40     &	 0&.09     &   LIII &	 0&.03     & $-$0&.14     &   LIII &	0&.09	  & $-$0&.06	 &   LIII &    0&.16	 &    0&.05	&   LIII \\
      AC\,Mon &    0&.07     & $-$0&.08     &   LIII &    0&.21     & $-$0&.10     &   LIII &	 0&.17     & $-$0&.00     &   LIII &	0&.17	  &    0&.02	 &   LIII &    0&.14	 &    0&.03	&   LIII \\
      BE\,Mon &    0&.04     &    0&.04     &     TS &    0&.02     &	 0&.02     &	 TS &	 0&.11     & $-$0&.04     &    LEM &	0&.07	  &    0&.07	 &     TS &    0&.11	 &    0&.11	&     TS \\
      BV\,Mon & $-$0&.12     & $-$0&.24     &    LEM &    0&.24     &	 0&.01     &	LEM & $-$0&.02     & $-$0&.19     &   LIII & $-$0&.07	  & $-$0&.22	 &   LIII &    0&.27	 &    0&.08	&    LEM \\
      CS\,Mon &    0&.02     & $-$0&.13     &   LIII &    0&.34     &	 0&.03     &   LIII &	 0&.12     & $-$0&.05     &   LIII &	0&.13	  & $-$0&.02	 &   LIII &    0&.25	 &    0&.14	&   LIII \\
      CU\,Mon & $-$0&.14     & $-$0&.29     &   LIII &    0&.22     & $-$0&.09     &   LIII &	 0&.11     & $-$0&.06     &   LIII &	0&.08	  & $-$0&.07	 &   LIII &    0&.00	 & $-$0&.11	&   LIII \\
      CV\,Mon & $-$0&.07     & $-$0&.07     &     TS & $-$0&.16     & $-$0&.16     &	 TS & $-$0&.02     & $-$0&.02     &	TS & $-$0&.07	  & $-$0&.07	 &     TS & $-$0&.12	 & $-$0&.12	&     TS \\
      EE\,Mon & $-$0&.58     & $-$0&.73     &   LIII & $-$0&.02     & $-$0&.33     &   LIII & $-$0&.58     & $-$0&.75     &   LIII & $-$0&.60	  & $-$0&.75	 &   LIII &     & \ldots &     & \ldots & \ldots \\
      EK\,Mon & $-$0&.19     & $-$0&.31     &    LEM & $-$0&.05     & $-$0&.28     &	LEM &	 0&.18     &	0&.03     &    LEM &	0&.16	  & $-$0&.01	 &    LEM &    0&.01	 & $-$0&.18	&    LEM \\
      FG\,Mon & $-$0&.08     & $-$0&.23     &   LIII &    0&.12     & $-$0&.19     &   LIII & $-$0&.12     & $-$0&.29     &   LIII &	0&.07	  & $-$0&.08	 &   LIII & $-$0&.21	 & $-$0&.32	&   LIII \\
      FI\,Mon &    0&.05     & $-$0&.10     &   LIII &    0&.58     &	 0&.27     &   LIII & $-$0&.03     & $-$0&.20     &   LIII &	0&.04	  & $-$0&.11	 &   LIII &    0&.11	 &    0&.01	&    LII \\
      FT\,Mon & $-$0&.34     & $-$0&.34     &     TS & $-$0&.17     & $-$0&.17     &	 TS &	 0&.09     &	0&.09     &	TS &	0&.14	  &    0&.14	 &     TS & $-$0&.02	 & $-$0&.02	&     TS \\
      SV\,Mon & $-$0&.06     & $-$0&.06     &     TS & $-$0&.14     & $-$0&.14     &	 TS &	 0&.02     &	0&.02     &	TS &	0&.10	  &    0&.10	 &     TS & $-$0&.07	 & $-$0&.07	&     TS \\
       T\,Mon &    0&.30     &    0&.15     &   LIII &    0&.54     &	 0&.23     &   LIII &	 0&.16     & $-$0&.01     &   LIII &	0&.35	  &    0&.20	 &   LIII &    0&.36	 &    0&.25	&   LIII \\
      TW\,Mon & $-$0&.12     & $-$0&.12     &     TS &    0&.03     &	 0&.03     &	 TS &	 0&.00     & $-$0&.17     &   LIII &	0&.09	  &    0&.09	 &     TS &    0&.01	 &    0&.01	&     TS \\
      TX\,Mon &    0&.13     &    0&.13     &     TS &    0&.26     &	 0&.26     &	 TS &	 0&.20     &	0&.20     &	TS &	0&.23	  &    0&.23	 &     TS &    0&.14	 &    0&.14	&     TS \\
      TY\,Mon & $-$0&.09     & $-$0&.09     &     TS & $-$0&.01     & $-$0&.01     &	 TS &	 0&.02     & $-$0&.13     &    LEM &	0&.06	  &    0&.06	 &     TS & $-$0&.11	 & $-$0&.11	&     TS \\
      TZ\,Mon &    0&.00     &    0&.00     &     TS &    0&.06     &	 0&.06     &	 TS &	 0&.08     &	0&.08     &	TS &	0&.04	  &    0&.04	 &     TS &    0&.07	 &    0&.07	&     TS \\
      UY\,Mon & $-$0&.35     & $-$0&.47     &    LEM & $-$0&.12     & $-$0&.35     &	LEM & $-$0&.04     & $-$0&.19     &    LEM &	0&.01	  & $-$0&.16	 &    LEM & $-$0&.11	 & $-$0&.30	&    LEM \\
    V446\,Mon & $-$0&.34     & $-$0&.49     &   LIII & $-$0&.06     & $-$0&.37     &   LIII & $-$0&.04     & $-$0&.21     &   LIII & $-$0&.10	  & $-$0&.25	 &   LIII &     & \ldots &     & \ldots & \ldots \\
    V447\,Mon & $-$0&.33     & $-$0&.48     &   LIII & $-$0&.58     & $-$0&.89     &   LIII & $-$0&.14     & $-$0&.31     &   LIII & $-$0&.11	  & $-$0&.26	 &   LIII &    0&.03	 & $-$0&.08	&   LIII \\
    V465\,Mon & $-$0&.14     & $-$0&.14     &     TS &    0&.06     &	 0&.06     &	 TS &	 0&.22     &	0&.05     &   LIII &	0&.23	  &    0&.23	 &     TS &    0&.05	 &    0&.05	&     TS \\
    V484\,Mon &    0&.15     & $-$0&.00     &   LIII &    0&.48     &	 0&.17     &   LIII &	 0&.38     &	0&.21     &   LIII &	0&.21	  &    0&.06	 &   LIII &    0&.21	 &    0&.10	&   LIII \\
    V495\,Mon & $-$0&.26     & $-$0&.26     &     TS & $-$0&.06     & $-$0&.06     &	 TS & $-$0&.03     & $-$0&.03     &	TS & $-$0&.04	  & $-$0&.04	 &     TS & $-$0&.04	 & $-$0&.04	&     TS \\
    V504\,Mon &    0&.22     &    0&.07     &   LIII &    0&.41     &	 0&.10     &   LIII &	 0&.25     &	0&.08     &   LIII &	0&.35	  &    0&.20	 &   LIII & $-$0&.15	 & $-$0&.25	&    LII \\
    V508\,Mon & $-$0&.15     & $-$0&.15     &     TS &    0&.03     &	 0&.03     &	 TS &	 0&.13     & $-$0&.02     &    LEM &	0&.07	  &    0&.07	 &     TS & $-$0&.06	 & $-$0&.06	&     TS \\
    V510\,Mon & $-$0&.23     & $-$0&.23     &     TS & $-$0&.14     & $-$0&.14     &	 TS & $-$0&.07     & $-$0&.07     &	TS & $-$0&.11	  & $-$0&.11	 &     TS & $-$0&.04	 & $-$0&.04	&     TS \\
    V526\,Mon & $-$0&.04     & $-$0&.19     &   LIII &    0&.10     & $-$0&.21     &   LIII &	 0&.03     & $-$0&.14     &   LIII &	0&.05	  & $-$0&.10	 &   LIII &    0&.06	 & $-$0&.05	&   LIII \\
    V911\,Mon &    0&.13     & $-$0&.02     &   LIII &    0&.35     &	 0&.04     &   LIII &	 0&.37     &	0&.20     &   LIII &	0&.39	  &    0&.24	 &   LIII &    0&.28	 &    0&.17	&   LIII \\
      VZ\,Mon &    0&.05     & $-$0&.10     &   LIII &    0&.13     & $-$0&.18     &   LIII &	 0&.20     &	0&.03     &   LIII &	0&.04	  & $-$0&.11	 &   LIII &     & \ldots &     & \ldots & \ldots \\
      WW\,Mon & $-$0&.26     & $-$0&.38     &    LEM & $-$0&.23     & $-$0&.46     &	LEM & $-$0&.21     & $-$0&.36     &    LEM & $-$0&.14	  & $-$0&.31	 &    LEM & $-$0&.02	 & $-$0&.21	&    LEM \\
      XX\,Mon & $-$0&.07     & $-$0&.07     &     TS & $-$0&.11     & $-$0&.11     &	 TS &	 0&.11     &	0&.11     &	TS &	0&.44	  &    0&.44	 &     TS &    0&.08	 &    0&.08	&     TS \\
      YY\,Mon & $-$0&.51     & $-$0&.66     &   LIII & $-$0&.24     & $-$0&.55     &   LIII & $-$0&.65     & $-$0&.82     &   LIII & $-$0&.27	  & $-$0&.42	 &   LIII & $-$0&.76	 & $-$0&.87	&   LIII \\
       R\,Mus &    0&.24     &    0&.09     &   LIII &    0&.29     & $-$0&.02     &   LIII &	 0&.15     & $-$0&.02     &   LIII &	0&.20	  &    0&.05	 &   LIII &    0&.31	 &    0&.20	&   LIII \\
      RT\,Mus &    0&.23     &    0&.08     &   LIII &    0&.24     & $-$0&.07     &   LIII &	 0&.14     & $-$0&.03     &   LIII &	0&.19	  &    0&.04	 &   LIII & $-$0&.09	 & $-$0&.20	&   LIII \\
       S\,Mus &    0&.15     & $-$0&.00     &   LIII &    0&.23     & $-$0&.08     &   LIII &	 0&.11     & $-$0&.06     &   LIII &	0&.18	  &    0&.03	 &   LIII &    0&.14	 &    0&.03	&   LIII \\
      TZ\,Mus &    0&.18     &    0&.03     &   LIII &    0&.53     &	 0&.22     &   LIII &	 0&.09     & $-$0&.08     &   LIII &	0&.08	  & $-$0&.07	 &   LIII &    0&.17	 &    0&.06	&   LIII \\
      UU\,Mus &    0&.34     &    0&.19     &   LIII &    0&.48     &	 0&.17     &   LIII &	 0&.24     &	0&.07     &   LIII &	0&.27	  &    0&.12	 &   LIII &    0&.50	 &    0&.39	&   LIII \\
\hline\noalign{\smallskip}
\multicolumn{26}{r}{\it {\footnotesize continued on next page}} \\
\end{tabular}}
\end{table*}
\addtocounter{table}{-1}
\begin{table*}[p]
\centering
\caption[]{continued.}
{\scriptsize 
\begin{tabular}{r r@{}l r@{}l c r@{}l r@{}l c r@{}l r@{}l c r@{}l r@{}l c r@{}l r@{}l c}
\noalign{\smallskip}\hline\hline\noalign{\smallskip}
Name &
\multicolumn{2}{c}{[Y/H]$_{\rm lit}$} & \multicolumn{2}{c}{[Y/H]} & Ref. &
\multicolumn{2}{c}{[La/H]$_{\rm lit}$} & \multicolumn{2}{c}{[La/H]} & Ref. &
\multicolumn{2}{c}{[Ce/H]$_{\rm lit}$} & \multicolumn{2}{c}{[Ce/H]} & Ref. &
\multicolumn{2}{c}{[Nd/H]$_{\rm lit}$} & \multicolumn{2}{c}{[Nd/H]} & Ref. &
\multicolumn{2}{c}{[Eu/H]$_{\rm lit}$} & \multicolumn{2}{c}{[Eu/H]} & Ref. \\
\noalign{\smallskip}\hline\noalign{\smallskip}
      GU\,Nor & $-$0&.08     & $-$0&.08     &     TS & $-$0&.08     & $-$0&.08     &	 TS &	 0&.18     &	0&.01     &   LIII & $-$0&.23	  & $-$0&.23	 &     TS & $-$0&.07	 & $-$0&.07	&     TS \\
      IQ\,Nor & $-$0&.06     & $-$0&.06     &     TS & $-$0&.15     & $-$0&.15     &	 TS &     & \ldots &     & \ldots & \ldots & $-$0&.06	  & $-$0&.06	 &     TS & $-$0&.05	 & $-$0&.05	&     TS \\
      QZ\,Nor &    0&.25     &    0&.25     &     TS & $-$0&.07     & $-$0&.07     &	 TS &	 0&.03     &	0&.03     &	TS &	0&.12	  &    0&.12	 &     TS &    0&.16	 &    0&.16	&     TS \\
      RS\,Nor &    0&.15     &    0&.15     &     TS &    0&.01     &	 0&.01     &	 TS &     & \ldots &     & \ldots & \ldots &	0&.01	  &    0&.01	 &     TS &    0&.17	 &    0&.17	&     TS \\
       S\,Nor &    0&.21     &    0&.06     &   LIII &    0&.27     & $-$0&.04     &   LIII &	 0&.11     & $-$0&.06     &   LIII &	0&.19	  &    0&.04	 &   LIII &    0&.08	 & $-$0&.03	&   LIII \\
      SY\,Nor &    0&.06     &    0&.06     &     TS & $-$0&.10     & $-$0&.10     &	 TS &	 0&.08     &	0&.08     &	TS & $-$0&.04	  & $-$0&.04	 &     TS &    0&.08	 &    0&.08	&     TS \\
      TW\,Nor & $-$0&.10     & $-$0&.10     &     TS & $-$0&.19     & $-$0&.19     &	 TS & $-$0&.11     & $-$0&.11     &	TS & $-$0&.15	  & $-$0&.15	 &     TS &    0&.08	 &    0&.08	&     TS \\
       U\,Nor &    0&.23     &    0&.08     &   LIII &    0&.24     & $-$0&.07     &   LIII &	 0&.15     & $-$0&.02     &   LIII &	0&.25	  &    0&.10	 &   LIII &    0&.15	 &    0&.04	&   LIII \\
    V340\,Nor & $-$0&.31     & $-$0&.31     &     TS & $-$0&.13     & $-$0&.13     &	 TS &	 0&.06     & $-$0&.11     &   LIII & $-$0&.15	  & $-$0&.15	 &     TS & $-$0&.18	 & $-$0&.18	&     TS \\
      BF\,Oph &    0&.24     &    0&.09     &   LIII &    0&.25     & $-$0&.06     &   LIII &	 0&.14     & $-$0&.03     &   LIII &	0&.18	  &    0&.03	 &   LIII & $-$0&.02	 & $-$0&.13	&   LIII \\
       Y\,Oph &    0&.31     &    0&.10     &    LII &    0&.24     & $-$0&.05     &	LII & $-$0&.07     & $-$0&.01     &    LII &	0&.17	  &    0&.08	 &    LII &    0&.13	 &    0&.03	&    LII \\
      CR\,Ori & $-$0&.13     & $-$0&.28     &   LIII &    0&.17     & $-$0&.14     &   LIII &	 0&.06     & $-$0&.11     &   LIII &	0&.01	  & $-$0&.14	 &   LIII &    0&.02	 & $-$0&.09	&   LIII \\
      CS\,Ori & $-$0&.27     & $-$0&.27     &     TS &    0&.21     & $-$0&.08     &	LII &	 0&.14     & $-$0&.01     &    LEM & $-$0&.21	  & $-$0&.21	 &     TS & $-$0&.25	 & $-$0&.25	&     TS \\
      DF\,Ori & $-$0&.17     & $-$0&.32     &   LIII & $-$0&.02     & $-$0&.33     &   LIII & $-$0&.16     & $-$0&.33     &   LIII & $-$0&.05	  & $-$0&.20	 &   LIII & $-$0&.39	 & $-$0&.50	&   LIII \\
      GQ\,Ori &    0&.32     &    0&.11     &    LII &    0&.30     &	 0&.01     &	LII &	 0&.04     &	0&.10     &    LII &	0&.05	  & $-$0&.04	 &    LII &    0&.16	 &    0&.06	&    LII \\
      RS\,Ori &    0&.15     &    0&.15     &     TS &    0&.12     &	 0&.12     &	 TS &	 0&.37     &	0&.37     &	TS &	0&.25	  &    0&.25	 &     TS &    0&.21	 &    0&.21	&     TS \\
      AS\,Per &    0&.18     &    0&.03     &   LIII &    0&.32     &	 0&.01     &   LIII &	 0&.07     & $-$0&.10     &   LIII &	0&.10	  & $-$0&.05	 &   LIII &    0&.10	 &    0&.00	&    LII \\
      AW\,Per &    0&.21     &    0&.06     &   LIII &    0&.46     &	 0&.15     &   LIII &	 0&.20     &	0&.03     &   LIII &	0&.23	  &    0&.08	 &   LIII &    0&.00	 & $-$0&.11	&   LIII \\
      BM\,Per &    0&.31     &    0&.16     &   LIII &    0&.54     &	 0&.23     &   LIII &	 0&.27     &	0&.10     &   LIII &	0&.39	  &    0&.24	 &   LIII &    0&.36	 &    0&.25	&   LIII \\
      CI\,Per & $-$0&.32     & $-$0&.47     &   LIII & $-$0&.32     & $-$0&.63     &   LIII & $-$0&.45     & $-$0&.62     &   LIII & $-$0&.26	  & $-$0&.41	 &   LIII & $-$0&.55	 & $-$0&.66	&   LIII \\
      DW\,Per & $-$0&.05     & $-$0&.20     &   LIII &    0&.20     & $-$0&.11     &   LIII &	 0&.02     & $-$0&.15     &   LIII &	0&.10	  & $-$0&.05	 &   LIII &    0&.28	 &    0&.17	&   LIII \\
      GP\,Per & $-$1&.03     & $-$1&.18     &   LIII & $-$0&.72     & $-$1&.03     &   LIII & $-$0&.79     & $-$0&.96     &   LIII & $-$0&.79	  & $-$0&.94	 &   LIII & $-$0&.45	 & $-$0&.56	&   LIII \\
      HQ\,Per & $-$0&.35     & $-$0&.50     &   LIII & $-$0&.11     & $-$0&.42     &   LIII & $-$0&.20     & $-$0&.37     &   LIII & $-$0&.14	  & $-$0&.29	 &   LIII & $-$0&.18	 & $-$0&.29	&   LIII \\
      HZ\,Per & $-$0&.20     & $-$0&.35     &   LIII &    0&.26     & $-$0&.05     &   LIII & $-$0&.02     & $-$0&.19     &   LIII &	0&.04	  & $-$0&.11	 &   LIII &    0&.20	 &    0&.09	&   LIII \\
      MM\,Per &    0&.05     & $-$0&.10     &   LIII &    0&.18     & $-$0&.13     &   LIII &	 0&.11     & $-$0&.06     &   LIII &	0&.17	  &    0&.02	 &   LIII & $-$0&.01	 & $-$0&.12	&   LIII \\
      OT\,Per & $-$0&.05     & $-$0&.20     &   LIII &    0&.55     &	 0&.24     &   LIII &	 0&.12     & $-$0&.05     &   LIII &	0&.16	  &    0&.01	 &   LIII &    0&.46	 &    0&.35	&   LIII \\
      SV\,Per &    0&.19     &    0&.04     &   LIII &    0&.45     &	 0&.14     &   LIII &	 0&.21     &	0&.04     &   LIII &	0&.24	  &    0&.09	 &   LIII &    0&.16	 &    0&.05	&   LIII \\
      SX\,Per &    0&.05     & $-$0&.10     &   LIII &    0&.28     & $-$0&.03     &   LIII &	 0&.13     & $-$0&.04     &   LIII &	0&.19	  &    0&.04	 &   LIII &    0&.15	 &    0&.04	&   LIII \\
      UX\,Per &    0&.02     & $-$0&.13     &   LIII &    0&.20     & $-$0&.11     &   LIII &	 0&.08     & $-$0&.09     &   LIII &	0&.07	  & $-$0&.08	 &   LIII & $-$0&.11	 & $-$0&.22	&   LIII \\
      UY\,Per &    0&.23     &    0&.08     &   LIII &    0&.38     &	 0&.07     &   LIII &	 0&.29     &	0&.12     &   LIII &	0&.29	  &    0&.14	 &   LIII &    0&.28	 &    0&.17	&   LIII \\
    V440\,Per &    0&.27     &    0&.06     &    LII &    0&.31     &	 0&.02     &	LII & $-$0&.07     & $-$0&.01     &    LII &	0&.18	  &    0&.09	 &    LII &    0&.15	 &    0&.05	&    LII \\
    V891\,Per &    0&.09     & $-$0&.06     &   LIII &    0&.30     & $-$0&.01     &   LIII &	 0&.19     &	0&.02     &   LIII &	0&.22	  &    0&.07	 &   LIII &    0&.27	 &    0&.16	&   LIII \\
      VX\,Per &    0&.09     & $-$0&.06     &   LIII &    0&.45     &	 0&.14     &   LIII &	 0&.15     & $-$0&.02     &   LIII &	0&.19	  &    0&.04	 &   LIII & $-$0&.01	 & $-$0&.12	&   LIII \\
      VY\,Per &    0&.13     & $-$0&.02     &   LIII &    0&.29     & $-$0&.02     &   LIII &	 0&.19     &	0&.02     &   LIII &	0&.18	  &    0&.03	 &   LIII &    0&.05	 & $-$0&.06	&   LIII \\
      AD\,Pup &    0&.23     &    0&.11     &    LEM &    0&.42     &	 0&.19     &	LEM &	 0&.31     &	0&.16     &    LEM &	0&.50	  &    0&.33	 &    LEM &    0&.17	 & $-$0&.02	&    LEM \\
      AP\,Pup &    0&.11     & $-$0&.01     &    LEM &    0&.30     &	 0&.07     &	LEM &	 0&.35     &	0&.20     &    LEM &	0&.33	  &    0&.16	 &    LEM &    0&.32	 &    0&.13	&    LEM \\
      AQ\,Pup &    0&.48     &    0&.48     &     TS &    0&.05     &	 0&.05     &	 TS & $-$0&.13     & $-$0&.13     &	TS &	0&.03	  &    0&.03	 &     TS &    0&.12	 &    0&.12	&     TS \\
      AT\,Pup & $-$0&.19     & $-$0&.31     &    LEM & $-$0&.11     & $-$0&.34     &	LEM &	 0&.06     & $-$0&.09     &    LEM &	0&.06	  & $-$0&.11	 &    LEM & $-$0&.21	 & $-$0&.40	&    LEM \\
      BC\,Pup & $-$0&.35     & $-$0&.35     &     TS & $-$0&.15     & $-$0&.15     &	 TS &	 0&.07     &	0&.07     &	TS &	0&.37	  &    0&.37	 &     TS &    0&.02	 &    0&.02	&     TS \\
      BM\,Pup & $-$0&.05     & $-$0&.05     &     TS & $-$0&.11     & $-$0&.11     &	 TS &	 0&.19     &	0&.19     &	TS &	0&.06	  &    0&.06	 &     TS &    0&.04	 &    0&.04	&     TS \\
      BN\,Pup &    0&.30     &    0&.30     &     TS &    0&.09     &	 0&.09     &	 TS &	 0&.12     &	0&.12     &	TS &	0&.07	  &    0&.07	 &     TS &    0&.16	 &    0&.16	&     TS \\
      CE\,Pup &    0&.08     & $-$0&.07     &   LIII &    0&.25     & $-$0&.06     &   LIII &	 0&.11     & $-$0&.06     &   LIII &	0&.21	  &    0&.06	 &   LIII &    0&.15	 &    0&.04	&   LIII \\
      CK\,Pup & $-$0&.13     & $-$0&.13     &     TS & $-$0&.09     & $-$0&.09     &	 TS &	 0&.04     &	0&.04     &	TS & $-$0&.00	  & $-$0&.00	 &     TS &    0&.04	 &    0&.04	&     TS \\
      HW\,Pup & $-$0&.16     & $-$0&.16     &     TS & $-$0&.18     & $-$0&.18     &	 TS & $-$0&.10     & $-$0&.10     &	TS & $-$0&.15	  & $-$0&.15	 &     TS & $-$0&.07	 & $-$0&.07	&     TS \\
      LS\,Pup & $-$0&.11     & $-$0&.11     &     TS & $-$0&.01     & $-$0&.01     &	 TS & $-$0&.04     & $-$0&.04     &	TS &	0&.03	  &    0&.03	 &     TS &    0&.01	 &    0&.01	&     TS \\
      MY\,Pup &    0&.01     & $-$0&.11     &    LEM &    0&.17     & $-$0&.06     &	LEM &	 0&.25     &	0&.10     &    LEM &	0&.16	  & $-$0&.01	 &    LEM &    0&.16	 & $-$0&.03	&    LEM \\
      NT\,Pup &    0&.03     & $-$0&.12     &   LIII &    0&.21     & $-$0&.10     &   LIII &	 0&.09     & $-$0&.08     &   LIII &	0&.19	  &    0&.04	 &   LIII &    0&.14	 &    0&.03	&   LIII \\
      RS\,Pup &    0&.30     &    0&.18     &    LEM &    0&.24     &	 0&.01     &	LEM &	 0&.19     &	0&.04     &    LEM &	0&.39	  &    0&.22	 &    LEM &    0&.37	 &    0&.18	&    LEM \\
    V335\,Pup &    0&.26     &    0&.11     &   LIII &    0&.25     & $-$0&.06     &   LIII &	 0&.33     &	0&.16     &   LIII &	0&.29	  &    0&.14	 &   LIII &    0&.24	 &    0&.13	&   LIII \\
      VW\,Pup & $-$0&.35     & $-$0&.35     &     TS & $-$0&.38     & $-$0&.38     &	 TS &	 0&.08     &	0&.14     &    LII & $-$0&.46	  & $-$0&.46	 &     TS & $-$0&.18	 & $-$0&.18	&     TS \\
      VX\,Pup &    0&.04     & $-$0&.08     &    LEM &    0&.32     &	 0&.09     &	LEM &	 0&.35     &	0&.20     &    LEM &	0&.26	  &    0&.09	 &    LEM &    0&.23	 &    0&.04	&    LEM \\
      VZ\,Pup & $-$0&.01     & $-$0&.01     &     TS &    0&.09     &	 0&.09     &	 TS &	 0&.14     & $-$0&.01     &    LEM &	0&.18	  &    0&.18	 &     TS &    0&.06	 &    0&.06	&     TS \\
      WW\,Pup & $-$0&.30     & $-$0&.30     &     TS & $-$0&.23     & $-$0&.23     &	 TS &	 0&.00     &	0&.06     &    LII & $-$0&.44	  & $-$0&.44	 &     TS & $-$0&.07	 & $-$0&.07	&     TS \\
      WX\,Pup & $-$0&.01     & $-$0&.13     &    LEM &    0&.13     & $-$0&.10     &	LEM &	 0&.18     &	0&.03     &    LEM &	0&.24	  &    0&.07	 &    LEM & $-$0&.05	 & $-$0&.24	&    LEM \\
      WY\,Pup & $-$0&.43     & $-$0&.43     &     TS & $-$0&.18     & $-$0&.18     &	 TS &     & \ldots &     & \ldots & \ldots &	0&.13	  &    0&.13	 &     TS & $-$0&.03	 & $-$0&.03	&     TS \\
      WZ\,Pup & $-$0&.16     & $-$0&.16     &     TS & $-$0&.09     & $-$0&.09     &	 TS &	 0&.01     &	0&.01     &	TS & $-$0&.09	  & $-$0&.09	 &     TS & $-$0&.04	 & $-$0&.04	&     TS \\
       X\,Pup & $-$0&.15     & $-$0&.15     &     TS & $-$0&.02     & $-$0&.02     &	 TS & $-$0&.13     & $-$0&.13     &	TS &	0&.03	  &    0&.03	 &     TS & $-$0&.26	 & $-$0&.26	&     TS \\
      KQ\,Sco &    0&.03     &    0&.03     &     TS & $-$0&.08     & $-$0&.08     &	 TS & $-$0&.02     & $-$0&.02     &	TS & $-$0&.24	  & $-$0&.24	 &     TS & $-$0&.02	 & $-$0&.02	&     TS \\
      RV\,Sco &    0&.13     & $-$0&.02     &   LIII &    0&.26     & $-$0&.05     &   LIII &	 0&.10     & $-$0&.07     &   LIII &	0&.10	  & $-$0&.05	 &   LIII &    0&.27	 &    0&.16	&   LIII \\
      RY\,Sco & $-$0&.02     & $-$0&.02     &     TS & $-$0&.01     & $-$0&.01     &	 TS &	 0&.09     &	0&.09     &	TS &	0&.06	  &    0&.06	 &     TS &    0&.09	 &    0&.09	&     TS \\
    V470\,Sco &    0&.11     &    0&.11     &     TS & $-$0&.07     & $-$0&.07     &	 TS & $-$0&.12     & $-$0&.12     &	TS &	0&.03	  &    0&.03	 &     TS &    0&.16	 &    0&.16	&     TS \\
    V482\,Sco &    0&.27     &    0&.12     &   LIII &    0&.32     &	 0&.01     &   LIII &	 0&.20     &	0&.03     &   LIII &	0&.25	  &    0&.10	 &   LIII &    0&.35	 &    0&.24	&   LIII \\
    V500\,Sco & $-$0&.19     & $-$0&.19     &     TS & $-$0&.13     & $-$0&.13     &	 TS &	 0&.00     &	0&.00     &	TS & $-$0&.01	  & $-$0&.01	 &     TS & $-$0&.05	 & $-$0&.05	&     TS \\
    V636\,Sco &    0&.14     & $-$0&.01     &   LIII &    0&.28     & $-$0&.03     &   LIII &	 0&.14     & $-$0&.03     &   LIII &	0&.08	  & $-$0&.07	 &   LIII &    0&.02	 & $-$0&.09	&   LIII \\
    V950\,Sco &    0&.33     &    0&.18     &   LIII &    0&.35     &	 0&.04     &   LIII &	 0&.20     &	0&.03     &   LIII &	0&.26	  &    0&.11	 &   LIII &    0&.42	 &    0&.31	&   LIII \\
      BX\,Sct &    0&.32     &    0&.17     &   LIII &    0&.36     &	 0&.05     &   LIII &	 0&.24     &	0&.07     &   LIII &	0&.39	  &    0&.24	 &   LIII &    0&.53	 &    0&.42	&   LIII \\
      CK\,Sct &    0&.27     &    0&.12     &   LIII &    0&.37     &	 0&.06     &   LIII &	 0&.23     &	0&.06     &   LIII &	0&.26	  &    0&.11	 &   LIII &    0&.23	 &    0&.12	&   LIII \\
      CM\,Sct &    0&.21     &    0&.06     &   LIII &    0&.17     & $-$0&.14     &   LIII &	 0&.06     & $-$0&.11     &   LIII &	0&.13	  & $-$0&.02	 &   LIII &     & \ldots &     & \ldots & \ldots \\
      CN\,Sct &    0&.30     &    0&.15     &   LIII &    0&.19     & $-$0&.12     &   LIII &	 0&.13     & $-$0&.04     &   LIII &	0&.22	  &    0&.07	 &   LIII &    0&.05	 & $-$0&.06	&   LIII \\
      EV\,Sct &    0&.01     &    0&.01     &     TS &    0&.07     &	 0&.07     &	 TS &	 0&.23     &	0&.06     &   LIII &	0&.15	  &    0&.15	 &     TS &    0&.20	 &    0&.20	&     TS \\
      EW\,Sct &    0&.22     &    0&.01     &    LII &    0&.29     &	 0&.00     &	LII & $-$0&.07     & $-$0&.01     &    LII &	0&.17	  &    0&.08	 &    LII &    0&.06	 & $-$0&.04	&    LII \\
      RU\,Sct &    0&.18     &    0&.18     &     TS & $-$0&.09     & $-$0&.09     &	 TS & $-$0&.11     & $-$0&.11     &	TS & $-$0&.07	  & $-$0&.07	 &     TS &    0&.00	 &    0&.00	&     TS \\
      SS\,Sct &    0&.16     &    0&.01     &   LIII &    0&.48     &	 0&.17     &   LIII &	 0&.15     & $-$0&.02     &   LIII &	0&.20	  &    0&.05	 &   LIII & $-$0&.06	 & $-$0&.17	&   LIII \\
      TY\,Sct &    0&.38     &    0&.23     &   LIII &    0&.31     &	 0&.00     &   LIII &	 0&.16     & $-$0&.01     &   LIII &	0&.29	  &    0&.14	 &   LIII &    0&.14	 &    0&.03	&   LIII \\
      UZ\,Sct &    0&.19     &    0&.19     &     TS & $-$0&.01     & $-$0&.01     &	 TS &	 0&.16     &	0&.16     &	TS &	0&.05	  &    0&.05	 &     TS &    0&.17	 &    0&.17	&     TS \\
\hline\noalign{\smallskip}
\multicolumn{26}{r}{\it {\footnotesize continued on next page}} \\
\end{tabular}}
\end{table*}
\addtocounter{table}{-1}
\begin{table*}[p]
\centering
\caption[]{continued.}
{\scriptsize 
\begin{tabular}{r r@{}l r@{}l c r@{}l r@{}l c r@{}l r@{}l c r@{}l r@{}l c r@{}l r@{}l c}
\noalign{\smallskip}\hline\hline\noalign{\smallskip}
Name &
\multicolumn{2}{c}{[Y/H]$_{\rm lit}$} & \multicolumn{2}{c}{[Y/H]} & Ref. &
\multicolumn{2}{c}{[La/H]$_{\rm lit}$} & \multicolumn{2}{c}{[La/H]} & Ref. &
\multicolumn{2}{c}{[Ce/H]$_{\rm lit}$} & \multicolumn{2}{c}{[Ce/H]} & Ref. &
\multicolumn{2}{c}{[Nd/H]$_{\rm lit}$} & \multicolumn{2}{c}{[Nd/H]} & Ref. &
\multicolumn{2}{c}{[Eu/H]$_{\rm lit}$} & \multicolumn{2}{c}{[Eu/H]} & Ref. \\
\noalign{\smallskip}\hline\noalign{\smallskip}
    V367\,Sct & $-$0&.12     & $-$0&.12     &     TS & $-$0&.14     & $-$0&.14     &	 TS &	 0&.23     &	0&.23     &	TS &	0&.12	  &    0&.12	 &     TS & $-$0&.01	 & $-$0&.01	&     TS \\
       X\,Sct &    0&.04     &    0&.04     &     TS & $-$0&.07     & $-$0&.07     &	 TS &     & \ldots &     & \ldots & \ldots &	0&.03	  &    0&.03	 &     TS &    0&.08	 &    0&.08	&     TS \\
       Y\,Sct &    0&.30     &    0&.15     &   LIII &    0&.32     &	 0&.01     &   LIII &	 0&.20     &	0&.03     &   LIII &	0&.30	  &    0&.15	 &   LIII &    0&.14	 &    0&.03	&   LIII \\
       Z\,Sct & $-$0&.26     & $-$0&.26     &     TS & $-$0&.31     & $-$0&.31     &	 TS &	 0&.03     &	0&.03     &	TS & $-$0&.04	  & $-$0&.04	 &     TS & $-$0&.06	 & $-$0&.06	&     TS \\
      AA\,Ser &    0&.39     &    0&.39     &     TS &    0&.09     &	 0&.09     &	 TS &	 0&.04     &	0&.04     &	TS &	0&.15	  &    0&.15	 &     TS &    0&.30	 &    0&.30	&     TS \\
      BQ\,Ser &    0&.13     & $-$0&.08     &    LII &    0&.13     & $-$0&.16     &	LII & $-$0&.09     & $-$0&.03     &    LII &	0&.22	  &    0&.13	 &    LII &    0&.07	 & $-$0&.03	&    LII \\
      CR\,Ser &    0&.05     &    0&.05     &     TS & $-$0&.15     & $-$0&.15     &	 TS &	 0&.06     &	0&.06     &	TS &	0&.08	  &    0&.08	 &     TS &    0&.15	 &    0&.15	&     TS \\
      DV\,Ser &    0&.39     &    0&.24     &   LIII &    0&.43     &	 0&.12     &   LIII &	 0&.23     &	0&.06     &   LIII &	0&.27	  &    0&.12	 &   LIII &    0&.34	 &    0&.23	&   LIII \\
      DG\,Sge &    0&.21     &    0&.06     &   LIII &    0&.23     & $-$0&.08     &   LIII &	 0&.02     & $-$0&.15     &   LIII &	0&.14	  & $-$0&.01	 &   LIII &    0&.12	 &    0&.01	&   LIII \\
      GX\,Sge &    0&.37     &    0&.22     &   LIII &    0&.43     &	 0&.12     &   LIII &	 0&.28     &	0&.11     &   LIII &	0&.34	  &    0&.19	 &   LIII &    0&.33	 &    0&.22	&   LIII \\
      GY\,Sge &    0&.48     &    0&.33     &   LIII &    0&.56     &	 0&.25     &   LIII &	 0&.15     & $-$0&.02     &   LIII &	0&.40	  &    0&.25	 &   LIII &    0&.37	 &    0&.26	&   LIII \\
       S\,Sge &    0&.26     &    0&.05     &    LII &    0&.23     & $-$0&.06     &	LII & $-$0&.08     & $-$0&.02     &    LII &	0&.15	  &    0&.06	 &    LII &    0&.07	 & $-$0&.03	&    LII \\
      AP\,Sgr &    0&.28     &    0&.07     &    LII &    0&.02     & $-$0&.27     &	LII &	 0&.05     &	0&.11     &    LII &     & \ldots &     & \ldots & \ldots &    0&.14	 &    0&.04	&    LII \\
      AV\,Sgr &    0&.11     &    0&.11     &     TS & $-$0&.04     & $-$0&.04     &	 TS &	 0&.04     &	0&.04     &	TS & $-$0&.04	  & $-$0&.04	 &     TS &    0&.24	 &    0&.24	&     TS \\
      AY\,Sgr &    0&.07     &    0&.07     &     TS & $-$0&.05     & $-$0&.05     &	 TS & $-$0&.04     & $-$0&.04     &	TS &	0&.09	  &    0&.09	 &     TS &    0&.08	 &    0&.08	&     TS \\
      BB\,Sgr &    0&.18     & $-$0&.03     &    LII &    0&.22     & $-$0&.07     &	LII & $-$0&.09     & $-$0&.03     &    LII & $-$0&.03	  & $-$0&.12	 &    LII &    0&.04	 & $-$0&.06	&    LII \\
       U\,Sgr &    0&.19     & $-$0&.02     &    LII &    0&.18     & $-$0&.11     &	LII & $-$0&.03     &	0&.03     &    LII &	0&.09	  &    0&.00	 &    LII &    0&.07	 & $-$0&.03	&    LII \\
   V1954\,Sgr &    0&.07     &    0&.07     &     TS & $-$0&.11     & $-$0&.11     &	 TS &     & \ldots &     & \ldots & \ldots & $-$0&.07	  & $-$0&.07	 &     TS &    0&.04	 &    0&.04	&     TS \\
    V350\,Sgr &    0&.33     &    0&.12     &    LII &    0&.26     & $-$0&.03     &	LII &	 0&.05     &	0&.11     &    LII &	0&.16	  &    0&.07	 &    LII &    0&.15	 &    0&.05	&    LII \\
    V773\,Sgr & $-$0&.07     & $-$0&.07     &     TS & $-$0&.04     & $-$0&.04     &	 TS &     & \ldots &     & \ldots & \ldots & $-$0&.09	  & $-$0&.09	 &     TS & $-$0&.06	 & $-$0&.06	&     TS \\
      VY\,Sgr & $-$0&.01     & $-$0&.01     &     TS & $-$0&.11     & $-$0&.11     &	 TS & $-$0&.04     & $-$0&.04     &	TS & $-$0&.02	  & $-$0&.02	 &     TS &    0&.09	 &    0&.09	&     TS \\
       W\,Sgr &    0&.20     & $-$0&.01     &    LII &    0&.23     & $-$0&.06     &	LII & $-$0&.04     &	0&.02     &    LII &	0&.10	  &    0&.01	 &    LII &    0&.03	 & $-$0&.07	&    LII \\
      WZ\,Sgr &    0&.07     &    0&.07     &     TS & $-$0&.03     & $-$0&.03     &	 TS &	 0&.05     &	0&.05     &	TS &	0&.05	  &    0&.05	 &     TS &    0&.04	 &    0&.04	&     TS \\
      XX\,Sgr & $-$0&.09     & $-$0&.09     &     TS & $-$0&.10     & $-$0&.10     &	 TS &	 0&.10     &	0&.10     &	TS &	0&.04	  &    0&.04	 &     TS & $-$0&.03	 & $-$0&.03	&     TS \\
       Y\,Sgr &    0&.23     &    0&.02     &    LII &    0&.15     & $-$0&.14     &	LII & $-$0&.18     & $-$0&.12     &    LII &	0&.08	  & $-$0&.01	 &    LII &    0&.01	 & $-$0&.09	&    LII \\
      YZ\,Sgr &    0&.30     &    0&.09     &    LII &    0&.19     & $-$0&.10     &	LII & $-$0&.10     & $-$0&.04     &    LII &	0&.09	  &    0&.00	 &    LII &    0&.07	 & $-$0&.03	&    LII \\
      AE\,Tau & $-$0&.10     & $-$0&.25     &   LIII &    0&.04     & $-$0&.27     &   LIII &	 0&.00     & $-$0&.17     &   LIII & $-$0&.02	  & $-$0&.17	 &   LIII & $-$0&.12	 & $-$0&.23	&   LIII \\
      AV\,Tau & $-$0&.08     & $-$0&.20     &    LEM &    0&.08     & $-$0&.15     &	LEM &	 0&.11     & $-$0&.06     &   LIII &	0&.28	  &    0&.11	 &    LEM &    0&.16	 & $-$0&.03	&    LEM \\
      EF\,Tau & $-$0&.67     & $-$0&.88     &    LII & $-$0&.80     & $-$1&.09     &	LII & $-$0&.68     & $-$0&.62     &    LII & $-$0&.33	  & $-$0&.42	 &    LII & $-$0&.55	 & $-$0&.65	&    LII \\
      EU\,Tau &    0&.07     & $-$0&.14     &    LII &    0&.16     & $-$0&.13     &	LII & $-$0&.18     & $-$0&.12     &    LII & $-$0&.02	  & $-$0&.11	 &    LII &    0&.01	 & $-$0&.09	&    LII \\
      ST\,Tau & $-$0&.01     & $-$0&.13     &    LEM &    0&.09     & $-$0&.14     &	LEM &	 0&.37     &	0&.22     &    LEM &	0&.19	  &    0&.02	 &    LEM &    0&.12	 & $-$0&.07	&    LEM \\
      SZ\,Tau &    0&.21     & $-$0&.00     &    LII &    0&.25     & $-$0&.04     &	LII & $-$0&.03     &	0&.03     &    LII &	0&.11	  &    0&.02	 &    LII &    0&.14	 &    0&.04	&    LII \\
      LR\,TrA &    0&.35     &    0&.20     &   LIII &    0&.56     &	 0&.25     &   LIII &	 0&.16     & $-$0&.01     &   LIII &	0&.27	  &    0&.12	 &   LIII &    0&.30	 &    0&.19	&   LIII \\
       R\,TrA &    0&.29     &    0&.14     &   LIII &    0&.51     &	 0&.20     &   LIII &	 0&.29     &	0&.12     &   LIII &	0&.32	  &    0&.17	 &   LIII &    0&.47	 &    0&.36	&   LIII \\
       S\,TrA &    0&.28     &    0&.13     &   LIII &    0&.35     &	 0&.04     &   LIII &	 0&.17     & $-$0&.00     &   LIII &	0&.21	  &    0&.06	 &   LIII &    0&.30	 &    0&.19	&   LIII \\
      AE\,Vel &    0&.13     & $-$0&.02     &   LIII &    0&.31     &	 0&.00     &   LIII &	 0&.14     & $-$0&.03     &   LIII &	0&.20	  &    0&.05	 &   LIII &    0&.26	 &    0&.15	&   LIII \\
      AH\,Vel &    0&.11     & $-$0&.01     &    LEM &    0&.26     &	 0&.03     &	LEM &	 0&.33     &	0&.18     &    LEM &	0&.30	  &    0&.13	 &    LEM &    0&.30	 &    0&.11	&    LEM \\
      AX\,Vel &    0&.02     & $-$0&.10     &    LEM &    0&.28     &	 0&.05     &	LEM &	 0&.44     &	0&.29     &    LEM &	0&.16	  & $-$0&.01	 &    LEM &    0&.14	 & $-$0&.05	&    LEM \\
      BG\,Vel &    0&.08     & $-$0&.04     &    LEM &    0&.24     &	 0&.01     &	LEM &	 0&.35     &	0&.20     &    LEM &	0&.27	  &    0&.10	 &    LEM &    0&.21	 &    0&.02	&    LEM \\
      CS\,Vel &    0&.21     &    0&.06     &   LIII &    0&.38     &	 0&.07     &   LIII &	 0&.20     &	0&.03     &   LIII &	0&.27	  &    0&.12	 &   LIII &    0&.08	 & $-$0&.03	&   LIII \\
      CX\,Vel &    0&.18     &    0&.03     &   LIII &    0&.21     & $-$0&.10     &   LIII &	 0&.07     & $-$0&.10     &   LIII &	0&.16	  &    0&.01	 &   LIII &    0&.05	 & $-$0&.05	&    LII \\
      DK\,Vel &    0&.30     &    0&.15     &   LIII &    0&.39     &	 0&.08     &   LIII &	 0&.19     &	0&.02     &   LIII &	0&.27	  &    0&.12	 &   LIII &    0&.22	 &    0&.11	&   LIII \\
      DR\,Vel &    0&.13     &    0&.01     &    LEM &    0&.14     & $-$0&.09     &	LEM &	 0&.29     &	0&.14     &    LEM &	0&.17	  & $-$0&.00	 &    LEM & $-$0&.04	 & $-$0&.23	&    LEM \\
      EX\,Vel &    0&.14     & $-$0&.01     &   LIII &    0&.28     & $-$0&.03     &   LIII &	 0&.13     & $-$0&.04     &   LIII &	0&.21	  &    0&.06	 &   LIII &    0&.23	 &    0&.12	&   LIII \\
      EZ\,Vel & $-$0&.02     & $-$0&.02     &     TS & $-$0&.01     & $-$0&.01     &	 TS &	 0&.13     &	0&.13     &	TS & $-$0&.08	  & $-$0&.08	 &     TS &    0&.04	 &    0&.04	&     TS \\
      FG\,Vel &    0&.05     & $-$0&.10     &   LIII &    0&.27     & $-$0&.04     &   LIII &	 0&.02     & $-$0&.15     &   LIII &	0&.14	  & $-$0&.01	 &   LIII &    0&.05	 & $-$0&.06	&   LIII \\
      FN\,Vel &    0&.11     & $-$0&.04     &   LIII &    0&.27     & $-$0&.04     &   LIII &	 0&.11     & $-$0&.06     &   LIII &	0&.17	  &    0&.02	 &   LIII &    0&.27	 &    0&.16	&   LIII \\
      RY\,Vel &    0&.19     &    0&.07     &    LEM &    0&.07     & $-$0&.16     &	LEM &	 0&.15     & $-$0&.00     &    LEM &	0&.17	  & $-$0&.00	 &    LEM &    0&.50	 &    0&.31	&    LEM \\
      RZ\,Vel &    0&.34     &    0&.22     &    LEM &    0&.27     &	 0&.04     &	LEM &	 0&.25     &	0&.10     &    LEM &	0&.54	  &    0&.37	 &    LEM &    0&.37	 &    0&.18	&    LEM \\
      ST\,Vel & $-$0&.04     & $-$0&.16     &    LEM &    0&.12     & $-$0&.11     &	LEM &	 0&.17     &	0&.02     &    LEM &	0&.16	  & $-$0&.01	 &    LEM &    0&.20	 &    0&.01	&    LEM \\
      SV\,Vel &    0&.20     &    0&.05     &   LIII &    0&.27     & $-$0&.04     &   LIII &	 0&.13     & $-$0&.04     &   LIII &	0&.19	  &    0&.04	 &   LIII &    0&.17	 &    0&.06	&   LIII \\
      SW\,Vel & $-$0&.03     & $-$0&.15     &    LEM &    0&.10     & $-$0&.13     &	LEM &	 0&.04     & $-$0&.11     &    LEM & $-$0&.02	  & $-$0&.19	 &    LEM &    0&.08	 & $-$0&.11	&    LEM \\
      SX\,Vel &    0&.01     & $-$0&.11     &    LEM &    0&.15     & $-$0&.08     &	LEM &	 0&.19     &	0&.04     &    LEM &	0&.05	  & $-$0&.12	 &    LEM &    0&.19	 & $-$0&.00	&    LEM \\
       T\,Vel &    0&.30     &    0&.18     &    LEM &    0&.49     &	 0&.26     &	LEM &	 0&.63     &	0&.48     &    LEM &	0&.39	  &    0&.22	 &    LEM &    0&.45	 &    0&.26	&    LEM \\
       V\,Vel & $-$0&.13     & $-$0&.25     &    LEM &    0&.07     & $-$0&.16     &	LEM &	 0&.24     &	0&.09     &    LEM &	0&.19	  &    0&.02	 &    LEM & $-$0&.04	 & $-$0&.23	&    LEM \\
      XX\,Vel &    0&.27     &    0&.12     &   LIII &    0&.30     & $-$0&.01     &   LIII &	 0&.20     &	0&.03     &   LIII &	0&.27	  &    0&.12	 &   LIII &    0&.48	 &    0&.37	&   LIII \\
      AS\,Vul &    0&.27     &    0&.12     &   LIII &    0&.34     &	 0&.03     &   LIII &	 0&.20     &	0&.03     &   LIII &	0&.39	  &    0&.24	 &   LIII &    0&.26	 &    0&.15	&   LIII \\
      DG\,Vul &    0&.26     &    0&.11     &   LIII &    0&.21     & $-$0&.10     &   LIII &	 0&.13     & $-$0&.04     &   LIII &	0&.24	  &    0&.09	 &   LIII &    0&.35	 &    0&.24	&   LIII \\
       S\,Vul &    0&.23     &    0&.08     &   LIII &    0&.35     &	 0&.04     &   LIII &	 0&.08     & $-$0&.09     &   LIII &	0&.21	  &    0&.06	 &   LIII &    0&.23	 &    0&.12	&   LIII \\
      SV\,Vul &    0&.24     &    0&.03     &    LII &    0&.20     & $-$0&.09     &	LII & $-$0&.13     & $-$0&.07     &    LII &	0&.06	  & $-$0&.03	 &    LII &    0&.04	 & $-$0&.06	&    LII \\
       T\,Vul &    0&.15     & $-$0&.06     &    LII &    0&.24     & $-$0&.05     &	LII & $-$0&.08     & $-$0&.02     &    LII &	0&.14	  &    0&.05	 &    LII &    0&.08	 & $-$0&.02	&    LII \\
       U\,Vul &    0&.35     &    0&.20     &   LIII &    0&.39     &	 0&.08     &   LIII &	 0&.29     &	0&.12     &   LIII &	0&.32	  &    0&.17	 &   LIII &    0&.21	 &    0&.10	&   LIII \\
       X\,Vul &    0&.20     & $-$0&.01     &    LII &    0&.15     & $-$0&.14     &	LII & $-$0&.06     & $-$0&.00     &    LII &	0&.13	  &    0&.04	 &    LII &    0&.07	 & $-$0&.03	&    LII \\
\hline
\end{tabular}}
\end{table*}